\documentclass[pdflatex,sn-mathphys-num]{sn-jnl}

\usepackage{graphicx}%
\usepackage{multirow}%
\usepackage{amsmath,amssymb,amsfonts}%
\usepackage{amsthm}%
\usepackage{mathrsfs}%
\usepackage[title]{appendix}%
\usepackage{xcolor}%
\usepackage{textcomp}%
\usepackage{manyfoot}%
\usepackage{booktabs}%
\usepackage{algorithm}%
\usepackage{algorithmicx}%
\usepackage{algpseudocode}%
\usepackage{listings}%
\usepackage{subcaption}
\usepackage{setspace}
\usepackage{fontsize}

\theoremstyle{thmstyleone}%

\theoremstyle{thmstyletwo}%

\theoremstyle{thmstylethree}%

\begin{document}
\large

\title[AI Buckingham]{AI Buckingham: Agentic Discovery of Dimensionless Scaling Laws Across Physical Regimes}

\author[1,2]{\fnm{Ao} \sur{Cheng}}\email{chengao23@mails.ucas.ac.cn}
\author*[1,2]{\fnm{Lei} \sur{Zhang}}\email{zhanglei@imech.ac.cn}
\author*[1,2]{\fnm{Guowei} \sur{He}}\email{hgw@lnm.imech.ac.cn}

\affil[1]{State Key Laboratory of Nonlinear Mechanics, Institute of Mechanics, Chinese Academy of Sciences, Beijing 100190, China}
\affil[2]{School of Engineering Sciences, University of Chinese Academy of Sciences, Beijing 100049, China}


\abstract{\begin{large}
\begin{spacing}{1.1}

Classical dimensional analysis defines admissible dimensionless representations but does not determine which combinations capture the dominant physical dependence, whether it changes across physical regimes, or when dependent-variable rescaling is required. 
We present AI Buckingham, an agentic framework for autonomous dimensionless discovery from natural-language descriptions of physical problems and observational data. It couples reasoning and planning by a large language model (LLM) with predefined symbolic and numerical tools. 
Literature-based knowledge retrieval supports dimensional checking, gradient-based clustering identifies physical regimes, and mutual information provides an objective, reparameterization-invariant criterion for identifying dominant dimensionless groups. Explicit numerical criteria determine the group count and assess simplification. Across ten physical systems spanning fluid dynamics, magnetohydrodynamics, thermal convection and additive manufacturing, AI Buckingham recovers established similarity representations and resolves regime-dependent scalings. For the longstanding challenge of predicting turbulent drag on rough surfaces, it identifies informative reparameterizations of surface statistics that improve equivalent sand-grain roughness estimates for engineering-oriented drag modelling.

\end{spacing}
\end{large}}


\maketitle

\section{Introduction}\label{sec1}

Dimensional analysis, anchored by the Buckingham $\Pi$ theorem~\cite{buckingham1914}, expresses physical laws as relationships among a reduced set of dimensionless variables, independent of the choice of measurement units. Over the past century, it has produced canonical quantities such as the Reynolds number ($Re$) and Mach number ($Ma$), which organize the governing physics of fluid flows and underpin scale-model experiments and cross-scale extrapolation~\cite{barenblatt1996scaling}. Yet the Buckingham $\Pi$ theorem has two fundamental limitations: the admissible dimensionless groups are not unique, and the theorem provides no ranking of their relevance~\cite{constantine2017data}. The central challenge is therefore not merely to construct dimensionally admissible groups, but to uncover which combinations govern the physical system, which are redundant, and how many are needed to capture its essential behavior. 
Classical dimensional reasoning therefore relies heavily on domain expertise to identify relevant variables, construct a consistent dimensional description, and recognize the physically informative representation. This expertise barrier becomes increasingly restrictive in high-dimensional systems for which governing equations or established similarity arguments are unavailable.

Recent advances in data-driven methods have begun to address these limitations. They can identify dimensionless relationships without requiring governing equations, while enforcing dimensional consistency as a hard constraint. Gradient-based sensitivity measures and information-theoretic criteria quantify the relevance of candidate groups. Global optimization enables searches over the admissible dimensionless space when manual exploration becomes impractical in high dimensions. Recent approaches have focused primarily on two complementary tasks.

The first is \textbf{dominant dimensionless-group identification}: among the infinitely many admissible dimensionless combinations, which groups govern the physical system? Constantine~\textit{et al.}~\cite{constantine2017data} proposed active-subspace analysis, using gradient sensitivity to produce unique, relevance-ordered directions in the admissible dimensionless space. Gradient-based sensitivity is not reparameterization-invariant, however: rescaling the input variables alters the computed ranking, so the result depends on arbitrary scaling choices rather than intrinsic physical dependence. 
Other approaches incorporate dimensional constraints into model fitting to identify dimensionless groups and learn their functional relationship with the dependent variable~\cite{bakarji2022dimensionally,xie2022data}. Symbolic regression methods~\cite{schmidt2009distilling} have discovered nonlinear dimensionless forms beyond simple power-law monomials~\cite{udrescu2020ai,gan2025hierarchical}. 
Information-theoretic approaches replace gradient sensitivity with mutual information (MI); because MI is invariant under arbitrary invertible reparameterizations, the identified groups are \textit{objective}, unaffected by arbitrary choices of variable scaling or coordinate transformation~\cite{kraskov2004estimating,zhang2025mutual}. 
In \textbf{Type I (complete) similarity}~\cite{barenblatt1996scaling,barenblatt2003scaling}, the dependence on certain dimensionless groups vanishes in the relevant asymptotic limit, leaving a reduced relation involving only the remaining relevant groups.

The second is \textbf{scaling-transformation discovery}. In \textbf{Type II (incomplete) similarity}, data collapse requires appropriate dimensionless groups together with a non-trivial rescaling of the dependent variable that cannot be determined from the Buckingham $\Pi$ theorem alone~\cite{zeldovich1956motion,barenblatt1972self}. Classical treatments obtain such scalings from governing equations or asymptotic analysis, as illustrated by Guderley's implosion~\cite{guderley1942starke} and Goldenfeld's rough-pipe scaling~\cite{goldenfeld2006roughness}. A related example is van Driest's density-weighted velocity transformation for compressible turbulence, based on mixing-length theory~\cite{vanDriest1951turbulent}. Sanfins~\textit{et al.}~\cite{sanfins2025similarity} introduced Barenet to learn similarity relations and extract the associated incomplete-similarity exponents from its trained weights. Watanabe~\textit{et al.}~\cite{watanabe2025datadriven} used a linear bottleneck in log space to infer similarity exponents directly from data, while the nonlinear part of the network learned the associated scaling function. Both approaches start from specified dimensionless variables and a prescribed power-law similarity ansatz; neither automates the full analysis pipeline from the raw physical quantities to the final dimensionless scaling relation.

A further challenge is to discover when and where the dominant physical mechanisms change, and to uncover the distinct dimensionless representations governing each regime. 
In such \textbf{multiple-physical-regime systems}, a single global dimensionless representation can obscure distinct local scalings. Zhang~\textit{et al.}~\cite{zhang2024clustering} introduced clustering dimensionless learning, which partitions data based on gradient similarity, with projection directions and weights determined by the eigenvectors and eigenvalues of each cluster's gradient covariance matrix. Yuan and Lozano-Dur\'an~\cite{yuan2025dimensionless} identified physical regimes a posteriori by partitioning the space spanned by globally learned dimensionless groups according to local mutual-information-based prediction scores. However, because the local assessment is restricted to globally learned groups and a prescribed number of regions, it may overlook regimes governed by other locally dominant groups. Integrating the choice between single- and multiple-regime descriptions with the determination of similarity type and dominant groups remains an open component of the discovery problem.

Yet a unified framework capable of autonomously identifying physical regimes, uncovering their dominant dimensionless representations, and distinguishing different types of similarity remains lacking. 
In the present work, we introduce AI Buckingham, an agentic framework that transforms dimensional analysis from a sequence of expert-guided operations into an autonomous scientific discovery process. It integrates LLM-based reasoning and planning with literature-grounded retrieval, gradient-based regime analysis, and MI-based similarity learning to address both Type I and Type II similarity, as well as systems with multiple physical regimes. Specifically, our contributions are as follows.

\textbf{First, we establish end-to-end agentic orchestration} that integrates otherwise fragmented analysis steps into a coordinated autonomous workflow. In response to natural-language instructions, AI Buckingham selects analysis workflows, configures numerical policies, and coordinates LLM-based reasoning with literature retrieval and dedicated symbolic and numerical tools~\cite{cheng2026re}. 
\textbf{Second, we provide a unified treatment} of physical regimes and similarity representations built on four complementary mechanisms: (i) literature-grounded constraints for dimensional consistency; (ii) unsupervised, gradient-based preliminary partitioning of physical regimes; (iii) an objective mutual-information-based criterion for identifying dominant dimensionless groups; and (iv) sparsification and rationalization to enforce parsimony in physical representations. 
\textbf{Third, we demonstrate the agentic framework across ten physical systems}: six Type~I cases, two Type~II cases and two multiple-physical-regime cases. Beyond recovering established similarity laws, the framework reveals a distinct transition regime in the Cess eddy-viscosity model, whose dominant dimensionless group combines inner- and outer-layer scaling. More importantly, for rough-wall turbulence, it uncovers two previously unreported dimensionless combinations of roughness statistics that capture the combined effects of surface morphology and improve surrogate predictions of equivalent sand-grain roughness height.

\section{Results}\label{sec2}

\begin{figure}
    \centering
\begin{subfigure}[b]{0.99\textwidth}
\centering
\includegraphics[width = \textwidth]{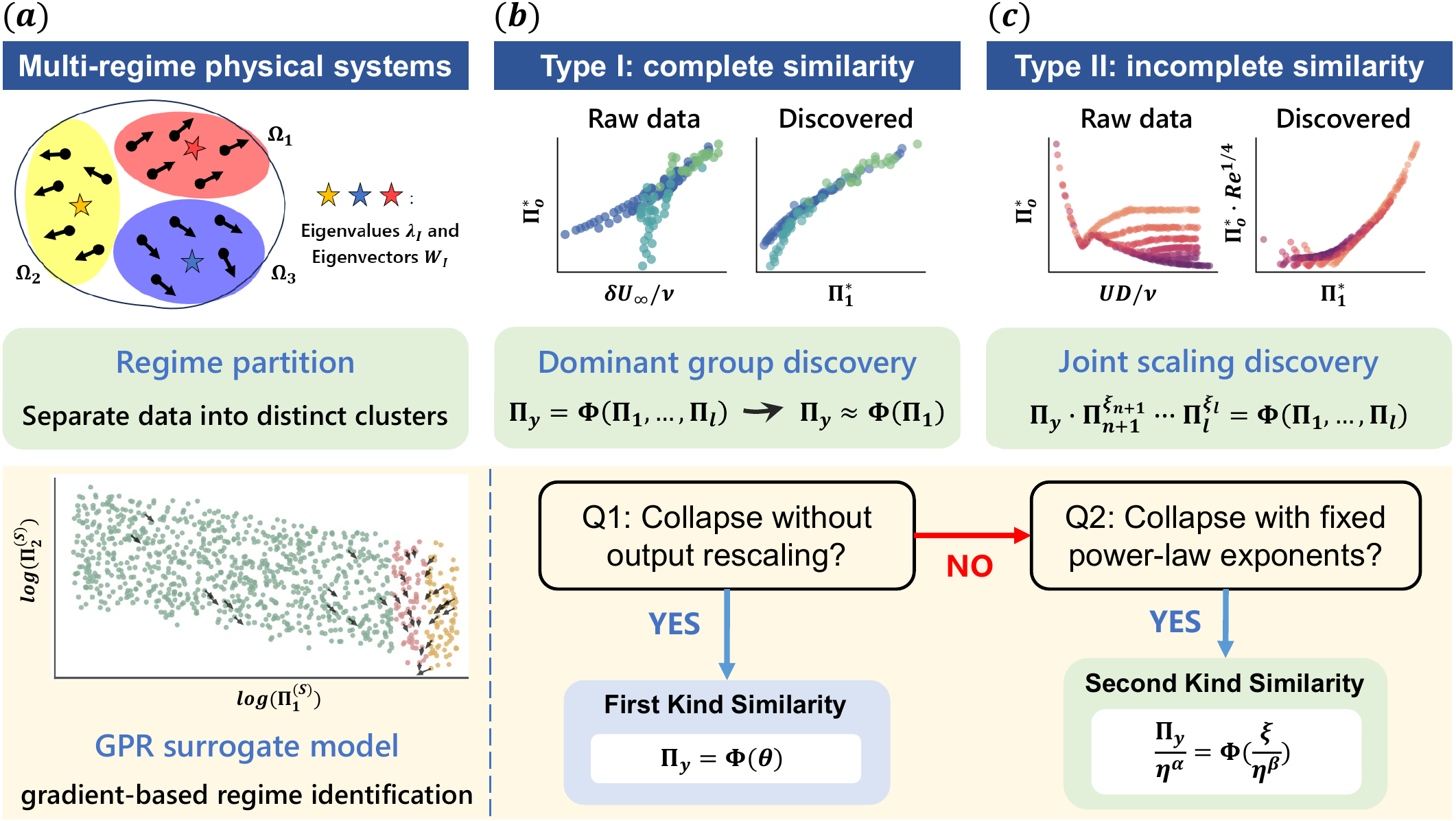}
\end{subfigure}
\caption{Overview of the physical similarity settings considered in this work. (a) Multi-regime physical systems, in which different regions of parameter space are governed by distinct dominant physical mechanisms. Clustering based on gradients from a Gaussian-process regression (GPR) surrogate identifies the corresponding physical regimes before dimensionless learning is performed independently within each cluster. (b) Type I similarity, where a reduced set of dimensionless similarity variables suffices for data collapse, without additional rescaling of the dependent variable. (c) Type II similarity, where data collapse additionally requires a power-law rescaling of the dependent variable.}
\label{fig:similarity_classes_}
\end{figure}

\begin{figure}
    \centering
\begin{subfigure}[b]{0.99\textwidth}
\centering
\includegraphics[width = \textwidth]{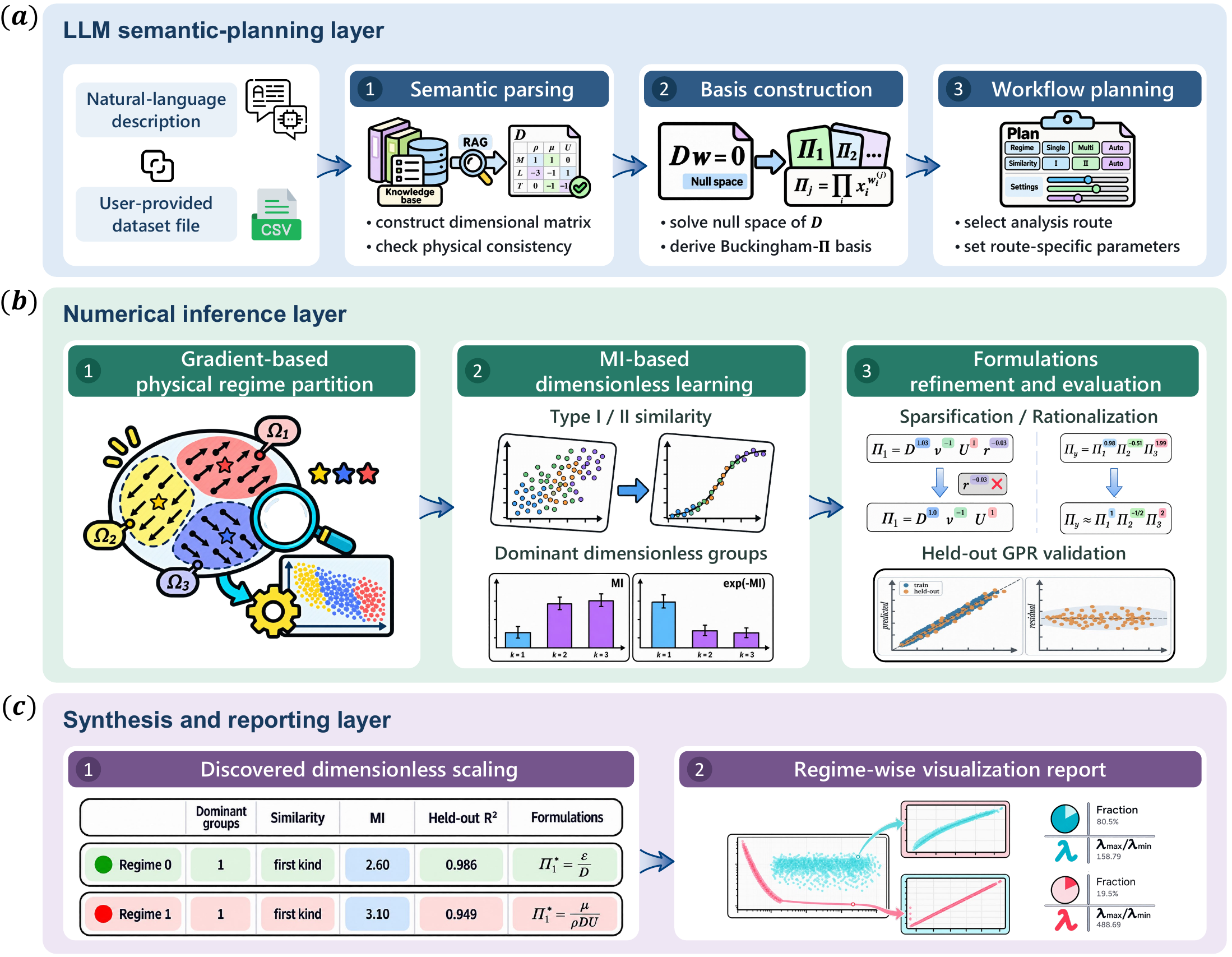}
\end{subfigure}
\caption{Schematic of the AI Buckingham workflow. (a) The semantic-planning layer parses the problem specification, checks variable dimensions, constructs the Buckingham--$\Pi$ basis, and selects the analysis route. (b) The numerical-inference layer performs regime and similarity-type screening when required, MI-based dimensionless learning with adaptive dimensionality selection, sparsification and rationalization, and held-out validation. (c) The synthesis-and-reporting layer summarizes the discovered scaling relations and regime-wise data collapse.}
\label{fig:agent_schem}
\end{figure}

Dimensional analysis defines the space of dimensionally admissible combinations, but leaves the physically relevant similarity representation to be discovered. The challenge extends beyond identifying dominant dimensionless groups: it requires determining whether distinct physical regimes exist, which groups govern each regime, and whether a non-trivial rescaling of the dependent variable is needed to achieve data collapse (Fig.~\ref{fig:similarity_classes_}). These questions are intrinsically coupled, as the dominant dimensionless groups may change across physical regimes, and the appropriate similarity representation may differ between complete and incomplete similarity.


To address this coupled discovery problem, we introduce AI Buckingham, an agentic framework that integrates LLM-based reasoning and planning, literature-grounded retrieval, and dedicated symbolic and numerical tools to enable autonomous dimensionless discovery (Fig.~\ref{fig:agent_schem}). Starting from natural-language problem descriptions and datasets, AI Buckingham identifies the governing physical quantities, validates their dimensional assignments, and constructs the admissible Buckingham--$\Pi$ space through symbolic null-space analysis. Unsupervised, gradient-based clustering then assesses whether regime partitioning is warranted and identifies distinct physical regimes. Within each regime, mutual information provides an objective, reparameterization-invariant criterion for discovering dominant dimensionless groups, distinguishing Type~I from Type~II similarity, and adaptively determining the required number of groups. Sparsification and rationalization further simplify the resulting representations into compact, physically interpretable scaling laws.

To support this workflow across a broad range of physical disciplines, we compile two extensible, literature-based knowledge bases covering fluid and solid mechanics, heat and mass transfer, electromagnetism, acoustics, and optics. They contain 247 physical-quantity entries and 251 dimensionless-relation entries, supporting dimensional consistency checks and physical interpretation of the discovered scaling laws, respectively.

\subsection{Type I similarity: complete similarity}
The Buckingham theorem provides only the functional form $\Pi_y = \Phi(\Pi_1, \Pi_2, \ldots, \Pi_n)$, where $\Pi_y$ is a dimensionless representation of the dependent variable and $\Pi_1, \ldots, \Pi_n$ are the $n$ dimensionless groups constructed from the governing parameters. Type I similarity (complete similarity)~\cite{barenblatt1996scaling} occurs when the dimensionless function $\Phi(\Pi_1, \ldots, \Pi_n)$ converges to a finite, non-zero limit as a subset of its arguments $\Pi_{k+1}, \ldots, \Pi_n$ approach zero or infinity, thereby reducing to a function $\Phi^{(0)}(\Pi_1, \ldots, \Pi_k)$ of fewer effective variables~\cite{sanfins2025similarity}.

\subsubsection{Magnetohydrodynamic power generator}

\begin{figure}[!t]
\centering
\includegraphics[width=\textwidth]{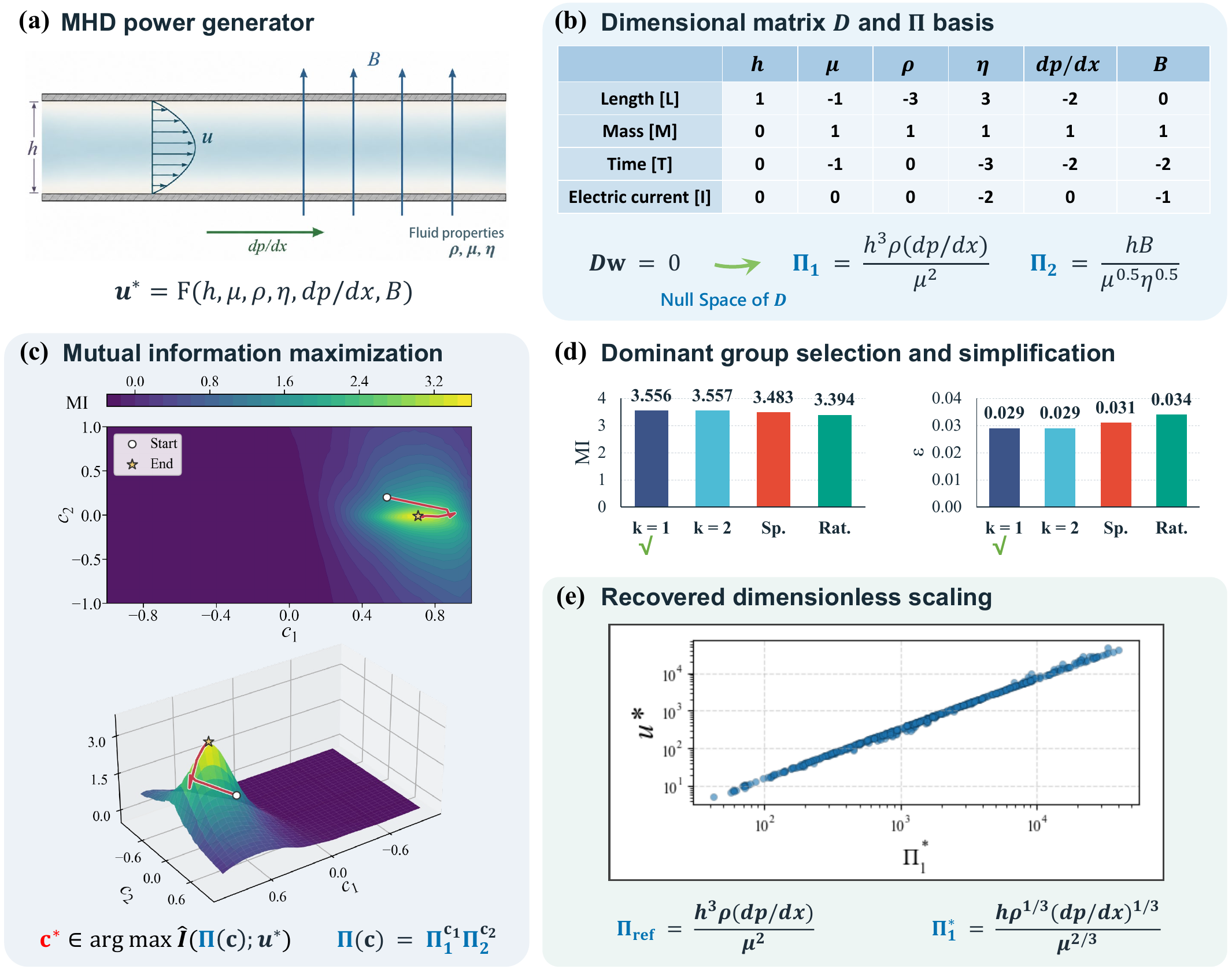}
\caption{Autonomous discovery of the dominant dimensionless group for a magnetohydrodynamic power generator. \textbf{(a)} Pressure-driven MHD flow in a square duct under a transverse magnetic field. \textbf{(b)} Dimensional matrix $\mathbf D\in\mathbb R^{4\times6}$ and two Buckingham--$\Pi$ basis groups constructed from the natural-language prompt. \textbf{(c)} MI optimization landscape, shown as a 2-D MI contour plot (upper) and the corresponding 3-D response surface (lower). \textbf{(d)} MI and normalized information score $\varepsilon_k$ for one- and two-group representations and the subsequent sparsification (Sp.) and rationalization (Rat.) stages. \textbf{(e)} Data collapse of $u^*$ against the discovered dominant dimensionless group $\Pi_1^*$.}
\label{fig:mhd_case}
\end{figure}

Magnetohydrodynamic (MHD) generators convert thermal energy directly into electrical power without moving mechanical components~\cite{thring1965magnetohydrodynamic,hammond1972magnetohydrodynamic}. In pressure-driven flow of an electrically conducting fluid through a square duct under a transverse magnetic field (Fig.~\ref{fig:mhd_case}a), the mean velocity $\overline u$ is governed by the balance of pressure, viscous and electromagnetic forces. This case tests whether electromagnetic effects contribute measurably to the dominant scaling of the dimensionless mean velocity $u^*=\overline u\rho h/\mu$ over the sampled parameter range. AI Buckingham receives the following natural-language prompt:

\begin{quote}
\textit{Perform dimensional analysis for a magnetohydrodynamics (MHD) generator to derive the dimensionless form of the average flow velocity (denoted as $u^*$). This analysis uses ONLY the following 6 physical quantities as input parameters: the side length of the square duct ($h$), the dynamic viscosity ($\mu$), the fluid density ($\rho$), the electrical resistivity of the fluid ($\eta$), the applied pressure gradient ($\mathrm{d}p/\mathrm{d}x$), and the magnetic field magnitude ($B$).}
\end{quote}

With $p=6$ and $\operatorname{rank}(\mathbf D)=4$, the admissible Buckingham--$\Pi$ space has dimension $n=2$ (Fig.~\ref{fig:mhd_case}b). The dataset consists of $N=483$ numerical samples from steady-state MHD duct flow simulations reported by Glaws~\textit{et al.}~\cite{glaws2016dimension}, covering a range of fluid properties, pressure gradients, and electromagnetic field strengths. The covariance matrix adaptation evolution strategy (CMA-ES) searches the Buckingham--$\Pi$ coefficients to maximize MI between the candidate dimensionless representation and $u^*$ (Fig.~\ref{fig:mhd_case}c). Adding a second dimensionless group produces a negligible reduction in the normalized information score, indicating that a single dimensionless group suffices to capture the predictive information (Fig.~\ref{fig:mhd_case}d):
\begin{equation}
\Pi_1^* = \frac{h^{1.000} \rho^{0.335} \eta^{0.003} (\mathrm{d}p/\mathrm{d}x)^{0.335}}{\mu^{0.668} B^{0.005}}.
\end{equation}
The negligibly small exponents of $\eta$ and $B$ suggest weak dependence of the learned representation on these quantities over the sampled parameter range. These quantities are masked out through zero-exponent constraints, and the representation is re-optimized in the restricted Buckingham--$\Pi$ space. Once $\eta$ and $B$ are eliminated, the remaining four variables ($h, \mu, \rho, \mathrm{d}p/\mathrm{d}x$) span three fundamental dimensions ($M, L, T$), leaving a one-dimensional null space. Maximum-exponent normalization (corresponding to a homeomorphic power reparameterization) with respect to the physical variables is combined with rationalization of the basis coefficients, yielding:
\begin{equation}
\Pi_1^* = \frac{h \rho^{1/3} (\mathrm{d}p/\mathrm{d}x)^{1/3}}{\mu^{2/3}}.
\end{equation}
The simplified dimensionless representation is mathematically equivalent to the Poiseuille-type scaling $h^3\rho(\mathrm{d}p/\mathrm{d}x)/\mu^2$. Figure~\ref{fig:mhd_case}e shows the resulting collapse of the dependent variable $u^*$ onto a single curve.

\subsubsection{Skin friction under pressure gradients}

\begin{figure}[!t]
\centering
\includegraphics[width=\textwidth]{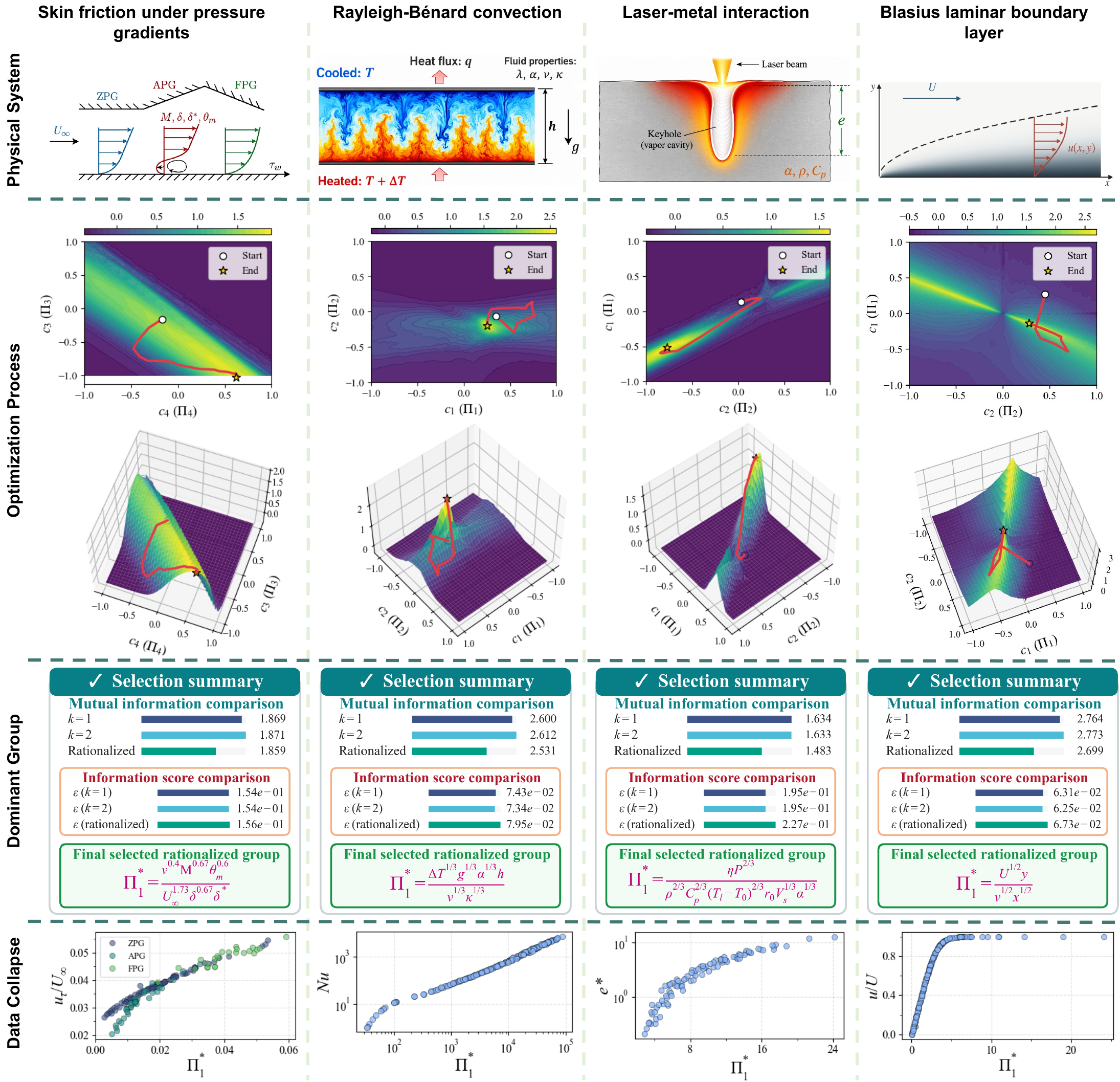}
\caption{Autonomous discovery of Type~I similarity in four representative systems. Each column shows the physical configuration, the MI optimization landscape, the comparison of one- and two-group representations, the rationalized dominant group, and the resulting data collapse.}
\label{fig:typeI_4cases}
\end{figure}

Accurate prediction of the friction velocity $u_\tau$ in turbulent boundary layers (TBLs) under streamwise pressure gradients is critical for aerodynamic design of wings, turbine blades, and nozzles. In zero-pressure-gradient (ZPG) TBLs, the normalized friction velocity $u_\tau/U_\infty$ is commonly correlated with a Reynolds number. When a pressure gradient is imposed, a favorable pressure gradient (FPG) accelerates the outer flow and tends to increase $u_\tau$ relative to a comparable ZPG case, while an adverse pressure gradient (APG) decelerates the flow and tends to reduce $u_\tau$, potentially leading to separation~\cite{zhou2025wall,zhang2025knowledge}. 

AI Buckingham receives six physical quantities characterizing the outer flow, fluid properties, and integral boundary-layer state: $\mathbf{q} = \left[ U_\infty, \, \nu, \, M, \, \delta, \, \delta^*, \, \theta_m \right]$, where $U_\infty$ is the free-stream velocity, $\nu$ is the kinematic viscosity, $M = \int_0^\delta u^2\,\mathrm{d}y$ is the wall-normal integral of the squared mean streamwise velocity, $\delta$ is the boundary-layer thickness at 99\% of the free-stream velocity, $\delta^* = \int_0^\delta (1 - u/U_\infty)\,\mathrm{d}y$ is the displacement thickness, and $\theta_m = \int_0^\delta (u/U_\infty)(1 - u/U_\infty)\,\mathrm{d}y$ is the momentum thickness. The dependent variable is the dimensionless friction velocity $u_\tau/U_\infty$, where $u_\tau = \sqrt{\tau_w/\rho_w}$.

With $p=6$ and $\operatorname{rank}(\mathbf D)=2$, the admissible Buckingham--$\Pi$ space has dimension $n=4$. Based on the normalized information-score comparison, AI Buckingham selects a single dominant group combining three thickness-based Reynolds numbers, $\mathrm{Re}_\delta=U_\infty\delta/\nu$, $\mathrm{Re}_{\delta^\ast}=U_\infty\delta^\ast/\nu$ and $\mathrm{Re}_{\theta}=U_\infty\theta_m/\nu$, with $M/(U_\infty\nu)$. The last parameter compares the velocity scale $M/\nu$ with $U_\infty$; Dixit~\textit{et al.}~\cite{dixit2022universal} motivated this scale through an argument based on the transfer of kinetic energy from the mean flow to large turbulent eddies. The latter two thickness-based Reynolds numbers can be reparameterized by introducing the classic shape factor $\delta^\ast/\theta_m$, giving
\begin{equation}
\Pi_1^\ast =
\mathrm{Re}_{\delta}^{-2/3}
\mathrm{Re}_{\delta^\ast}^{-2/5}
\left(\frac{\delta^*}{\theta_m}\right)^{-3/5}
\left(\frac{M}{U_\infty\nu}\right)^{2/3}.
\label{eq:tbl_discovered_group}
\end{equation}

The discovered formulation gives a remarkably tighter data collapse across ZPG, APG and FPG flows than Reynolds-number scaling alone (Fig.~\ref{fig:typeI_4cases}; \textcolor{blue}{Supplementary Fig. 5}). The dataset comprises $N=144$ experimental measurements and simulation results compiled by Dixit~\textit{et al.}~\cite{dixit2024generalized}, covering ZPG channel, pipe, and TBL flows, adverse- and favorable-pressure-gradient TBLs, and the pressure side of an airfoil.

\subsubsection{Rayleigh--B\'{e}nard convection}
Rayleigh-B\'{e}nard convection (RBC) is the canonical system for thermally driven turbulence, governing phenomena from atmospheric convection to stellar interiors and industrial heat exchangers~\cite{ahlers2009heat}. Buoyancy drives warm fluid upward from the bottom plate and cool fluid downward from the top, organizing a large-scale circulation. The Rayleigh number characterizes the competition between destabilizing buoyancy and the stabilizing effects of viscous dissipation and thermal diffusion.

AI Buckingham receives seven physical quantities characterizing the fluid properties and thermal configuration: $\mathbf{q} = \left[ h, \Delta T, \lambda, g, \alpha, \nu, \kappa \right]$, where $h$ is the cell height, $\Delta T$ is the temperature difference, $\lambda$ is the thermal conductivity, $g$ is the gravitational acceleration, $\alpha$ is the thermal expansion coefficient, $\nu$ is the kinematic viscosity, and $\kappa$ is the thermal diffusivity. The dependent variable is the Nusselt number $Nu$, which quantifies the enhancement of heat transfer relative to pure conduction. 
With $p=7$ and $\operatorname{rank}(\mathbf D)=4$, the admissible Buckingham--$\Pi$ space has dimension $n=3$. The agent recovers a single dominant group $\Pi_1^*=Ra^{1/3}$, an invertible reparameterization of the classical Rayleigh number:
\begin{equation}
Ra = \frac{g \alpha \Delta T h^3}{\nu \kappa}.
\end{equation}
The $N=182$ experimental measurements compiled from the literature by Xie~\textit{et al.}~\cite{xie2022data} collapse onto a single $Nu$--$Ra^{1/3}$ curve.

\subsubsection{Laser-metal interaction in additive manufacturing}
In laser powder-bed fusion, a focused laser beam scans across a metal powder bed, locally melting the material to build parts layer by layer. At sufficiently high energy input, recoil pressure from metal vaporization depresses the melt pool surface into a deep, narrow cavity known as a keyhole. Excessive keyhole depth $e$ can lead to instability and keyhole collapse, producing porosity defects that compromise part integrity.

AI Buckingham receives seven physical quantities characterizing the process parameters and thermophysical properties: $\mathbf{q} = \left[ \eta P, V_s, r_0, \alpha, \rho, C_p, T_l - T_0 \right]$, where $\eta P$ is the effective laser power (the product of laser power and absorption efficiency, treated as a single variable), $V_s$ is the scan speed, $r_0$ is the beam radius, $\alpha$ is the thermal diffusivity, $\rho$ is the material density, $C_p$ is the specific heat capacity, and $T_l-T_0$ is the temperature difference between melting and ambient conditions. The dependent variable is the normalized keyhole aspect ratio $e^* = e/r_0$. 
With $p=7$ and $\operatorname{rank}(\mathbf D)=4$, the admissible Buckingham--$\Pi$ space has dimension $n=3$. The discovered single dominant group, $\Pi_1^* = \left(\pi Ke\right)^{2/3}$, is mathematically equivalent to the Keyhole number:
\begin{equation}
Ke = \frac{\eta P}{(T_l - T_0) \pi \rho C_p \sqrt{\alpha V_s r_0^3}}.
\end{equation}
Xie~\textit{et al.}~\cite{xie2022data} identified $Ke$ through data-driven dimensional analysis to collapse the conduction-to-keyhole transition across diverse materials and process conditions. The dataset consists of $N=90$ high-speed X-ray imaging measurements of keyhole dynamics from laser powder-bed fusion experiments, covering a range of laser powers, scan speeds, and beam diameters across multiple material systems.

\subsubsection{Blasius laminar boundary layer}
The Blasius boundary layer serves as an exact theoretical benchmark for complete similarity. It describes steady, two-dimensional laminar flow over a semi-infinite flat plate parallel to a uniform free stream.

AI Buckingham receives four physical quantities characterizing the flow field and fluid properties: $\mathbf{q} = \left[ U, \nu, x, y \right]$, where $U$ is the free-stream velocity, $\nu$ is the kinematic viscosity, $x$ is the streamwise distance from the leading edge, and $y$ is the wall-normal distance. The dependent variable is the normalized streamwise velocity $u/U$. 
With $p=4$ and $\operatorname{rank}(\mathbf D)=2$, the admissible Buckingham--$\Pi$ space has dimension $n=2$. The recovered single dominant group is the Blasius similarity variable~\cite{blasius1907grenzschichten}:
\begin{equation}
\eta = y \sqrt{\frac{U}{\nu x}}.
\end{equation}

Plotting $u/U$ against $\eta$ collapses the velocity profiles sampled at arbitrary streamwise locations onto the theoretical curve $u/U = f'(\eta)$. The $N=200$ spatial data points were generated from numerical solutions of the Blasius equation $f^{\prime\prime\prime}+ff^{\prime\prime}/2=0$.

\subsubsection{Sand-grain roughness parameterization}

\begin{figure}[!t]
\centering
\includegraphics[width=\textwidth]{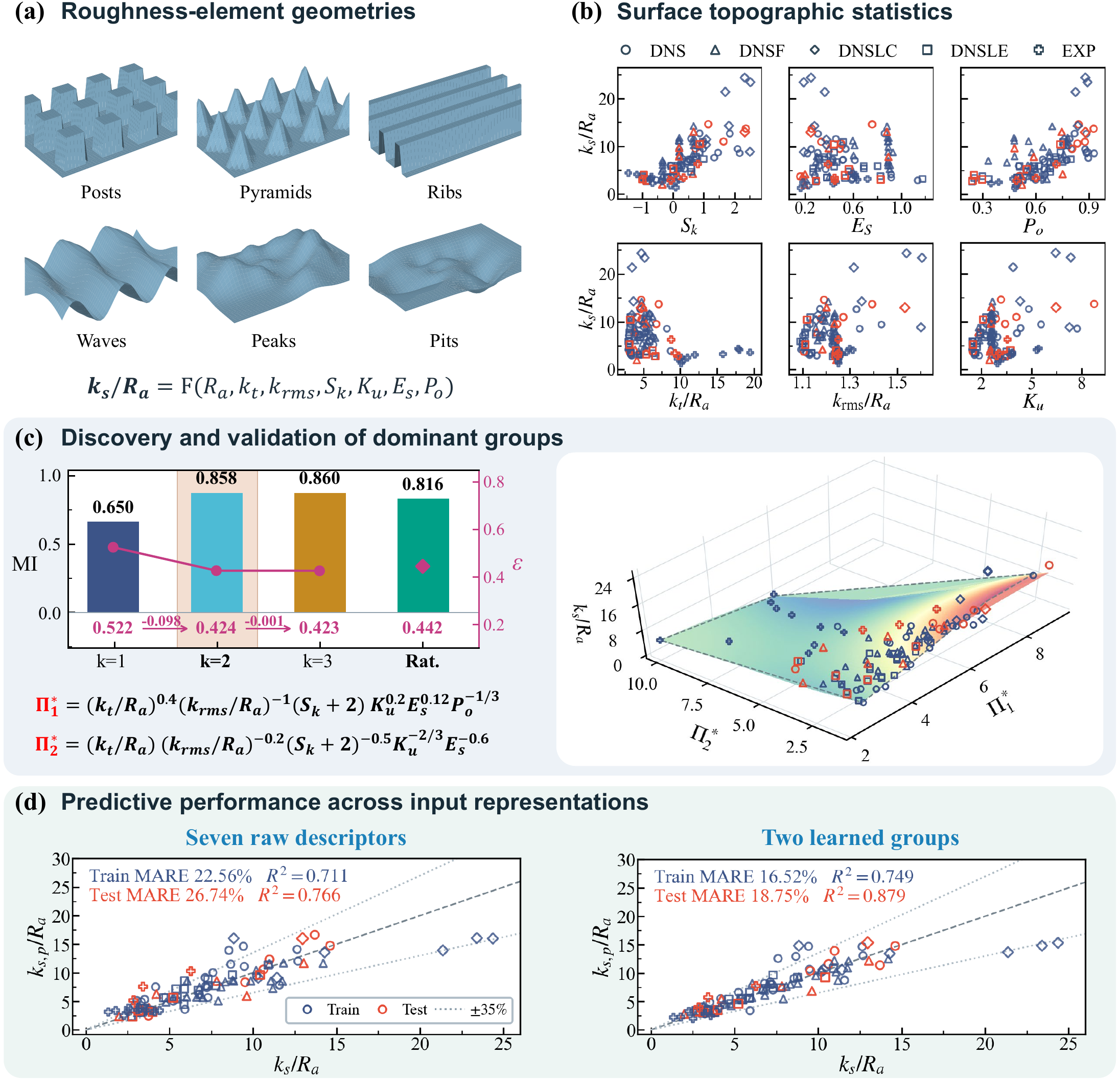}
\caption{Autonomous dimensionless discovery for equivalent sand-grain roughness. \textbf{(a)} Illustrative roughness-element geometries. \textbf{(b)} Scatter plots of $k_s/R_a$ against six dimensionless surface statistics. \textbf{(c)} MI and $\varepsilon$ for group-count selection and rationalization (left), and the GPR response surface in the learned dimensionless coordinates (right). The dashed boundary marks the convex hull of the training coordinates, with extrapolation outside. \textbf{(d)} GPR predictions $k_{s,p}/R_a$ against reference values $k_s/R_a$ using seven original surface statistics (left) and two rationalized groups (right). Both models use the same 83 training and 21 test samples. Marker shapes denote data sources, and dotted lines indicate $\pm35\%$ relative-error bounds.}
\label{fig:sandgrain_two_groups}
\end{figure}

Beyond recovering known single-group similarity laws, a critical test of AI Buckingham is whether it can discover \textit{new} dimensionless parameterizations that improve predictive accuracy in engineering practice. Predicting the equivalent sand-grain roughness height from surface topography in rough-wall turbulent flows exemplifies this challenge.

The equivalent sand-grain roughness height $k_s$ is a central parameter in rough-wall turbulence, providing a hydraulically equivalent measure of surface roughness for predicting skin friction. It is defined as the height of uniform Nikuradse sand grains that produce the same friction coefficient as the surface under consideration in the fully rough regime~\cite{nikuradse1950laws}, and is closely related to the roughness function $\Delta U^+$~\cite{yang2023prediction}. However, predicting $k_s$ from general surface topographies remains challenging, as the geometric statistics and, more importantly, their combinations that best capture drag-relevant effects remain unclear~\cite{chung2021predicting,nair2024rough}.


AI Buckingham receives seven commonly used surface statistics to identify their dominant dimensionless combinations (Fig.~\ref{fig:sandgrain_two_groups}a): $\mathbf q=[R_a,k_t,k_{\mathrm{rms}},S_k,K_u,E_S,P_o]$, where the entries denote arithmetic mean roughness, peak-to-valley height, root-mean-square height, skewness, kurtosis, effective slope and porosity, respectively. The dependent variable is $k_s/R_a$. With $p=7$ physical quantities and $\operatorname{rank}(\mathbf D)=1$, the admissible Buckingham--$\Pi$ space has dimension $n=6$. The translation $S_k\mapsto S_k+2$ is applied to provide a positive base for the product-of-powers representation of the similarity variables.

Applying AI Buckingham to 104 rough surfaces, we uncover two previously unreported, rationalized dimensionless groups that combine the seven conventional surface statistics into a compact representation of their effects on equivalent sand-grain roughness:
\begin{equation}
\begin{aligned}
\Pi_1^*
&=
\left(\frac{k_t}{R_a}\right)^{2/5}
\left(\frac{k_{\mathrm{rms}}}{R_a}\right)^{-1}
(S_k+2)K_u^{1/5}E_S^{0.12}P_o^{-1/3},
\\
\Pi_2^*
&=
\left(\frac{k_t}{R_a}\right)
\left(\frac{k_{\mathrm{rms}}}{R_a}\right)^{-1/5}
(S_k+2)^{-1/2}K_u^{-2/3}E_S^{-3/5}.
\end{aligned}
\end{equation}

 Beyond providing a more parsimonious description, these newly discovered groups also improve predictive accuracy: a GPR surrogate using the two groups achieves lower MARE on both training and test sets than one using all seven original statistics (Fig.~\ref{fig:sandgrain_two_groups}c,d). The dataset was compiled by Cheng~\textit{et al.}~\cite{cheng2026drag} from multiple numerical and experimental studies.

\subsection{Type II similarity: incomplete similarity}
Type II similarity (incomplete similarity)~\cite{barenblatt1996scaling,barenblatt1972self} arises when no such finite, non-zero limit exists. Instead, as $\Pi_{k+1}, \ldots, \Pi_n \to 0$ or $\infty$, $\Phi$ exhibits power-law asymptotic behaviour that cannot be obtained from dimensional analysis alone~\cite{sanfins2025similarity}. The dependent variable obeys the generalized homogeneity relation:
\begin{equation}
\Pi_y \cdot \Pi_{k+1}^{\xi_{k+1}} \cdots \Pi_n^{\xi_n} = \Phi^{(1)}\bigl(\Pi_1 \Pi_{k+1}^{\xi^{(1)}_{k+1}} \cdots \Pi_n^{\xi^{(1)}_n},\; \ldots,\; \Pi_k \Pi_{k+1}^{\xi^{(k)}_{k+1}} \cdots \Pi_n^{\xi^{(k)}_n}\bigr),
\end{equation}
where the exponents $\xi_j$ and $\xi_j^{(i)}$ are learned jointly with the similarity variables.

\subsubsection{Velocity scaling in compressible wall-bounded turbulence}

\begin{figure}[!t]
\centering
\includegraphics[width=\textwidth]{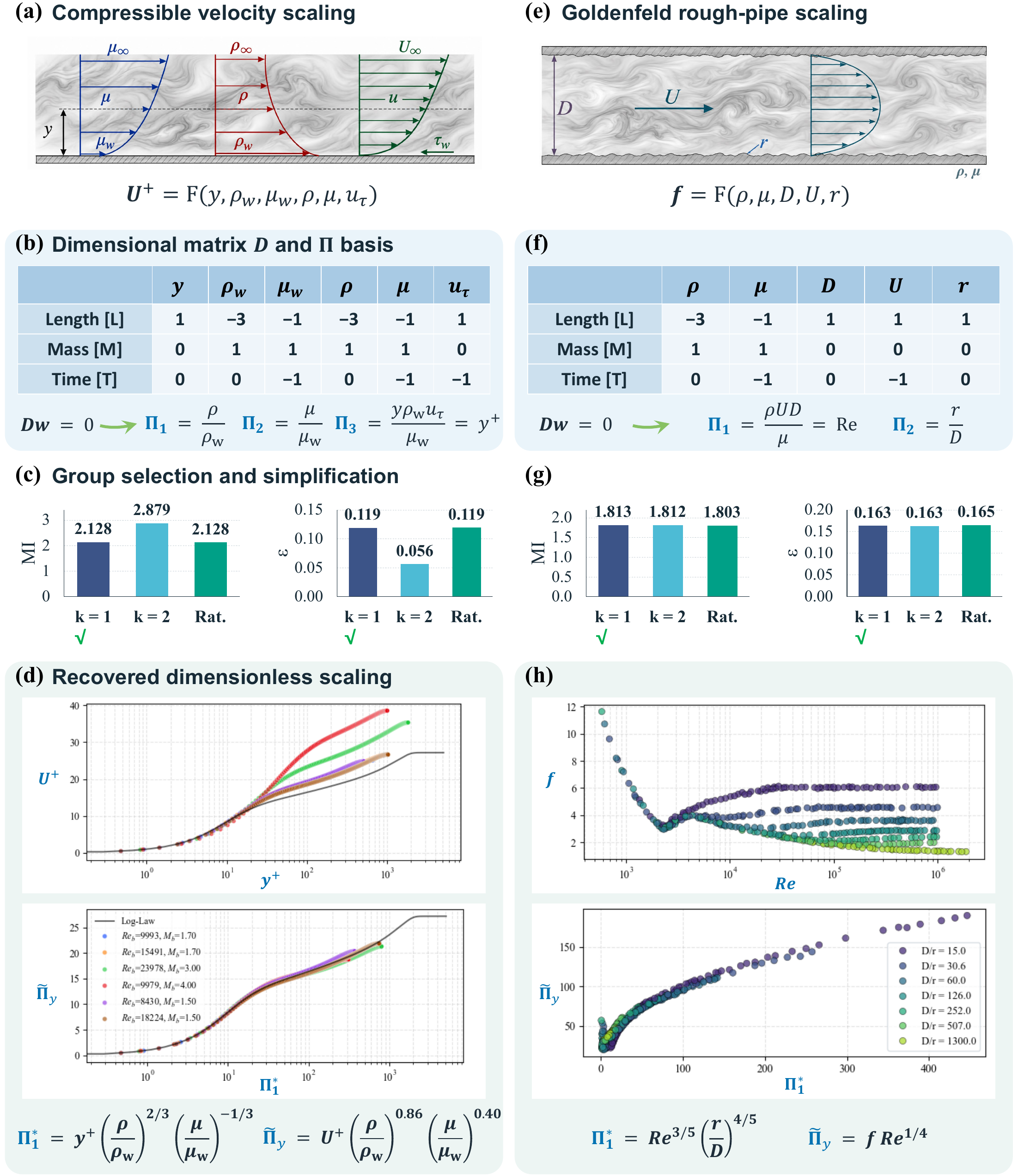}
\caption{Autonomous discovery of Type II similarity for compressible wall-bounded turbulence \textbf{(a--d)} and Goldenfeld's rough-pipe scaling \textbf{(e--h)}. \textbf{(a,e)} Schematics of the physical configurations. \textbf{(b,f)} Dimensional matrices $\mathbf D$ and Buckingham--$\Pi$ basis groups obtained from their null spaces. \textbf{(c,g)} MI and $\varepsilon$ for group-count selection and rationalization. \textbf{(d)} Upper panel: conventional wall-unit profiles, $U^+$ versus $y^+$. Lower panel: the learned representation, $\widetilde{\Pi}_y$ versus $\Pi_1^*$. \textbf{(h)} Upper panel: friction factor $f$ versus Reynolds number $Re$ at different relative roughnesses. Lower panel: the rescaled representation, $fRe^{1/4}$ versus $Re^{3/5}(r/D)^{4/5}$.}
\label{fig:cases_type_II}
\end{figure}

In incompressible wall-bounded turbulence, mean-velocity profiles obey the universal law of the wall $U^+=f(y^+)$, with collapse across Reynolds numbers in the logarithmic region. 
In compressible flows, however, the strong near-wall temperature gradients cause the density $\rho(y)$ and viscosity $\mu(y)$ to vary substantially in the wall-normal direction, disrupting this incompressible scaling. 
The classical remedy, due to van Driest, is to apply a density-weighted transformation intended to map the compressible mean velocity profile onto the incompressible law of the wall~\cite{vanDriest1951turbulent}. 
Here compressible wall-bounded velocity scaling is treated as a Type~II discovery problem, with power-law corrections learned for both $y^+$ and $U^+$.

AI Buckingham receives six physical quantities characterizing the local fluid state and wall conditions: $\mathbf q=[y,\rho_w,\mu_w,\rho,\mu,u_\tau]$, where $y$ is the wall-normal distance, $\rho_w$ and $\mu_w$ are the wall density and dynamic viscosity, $\rho$ and $\mu$ are the local density and dynamic viscosity at distance $y$, and $u_\tau=\sqrt{\tau_w/\rho_w}$ is the friction velocity. The dependent variable is the normalized mean streamwise velocity $U^+=U/u_\tau$. 
With $p=6$ and $\operatorname{rank}(\mathbf D)=3$, the admissible Buckingham--$\Pi$ space has dimension $n=3$ (Fig.~\ref{fig:cases_type_II}b). 
To retain comparability with the incompressible log-law, the prompt explicitly constrains the exponent of wall units to 1 in the similarity coordinate and 0 in the output-rescaling factor, thereby removing the arbitrary power-rescaling freedom and ensuring compressibility corrections are learned through the density and viscosity ratios. 
Although two groups yield higher MI, the reduction in $\varepsilon$ remains below the prescribed parsimony threshold, so a single dominant group is selected and rationalized, yielding
\begin{equation}
U^+
\left(\frac{\rho}{\rho_w}\right)^{0.86}
\left(\frac{\mu}{\mu_w}\right)^{0.40}
=
\Phi^{(1)}
\left[
y^+
\left(\frac{\rho}{\rho_w}\right)^{2/3}
\left(\frac{\mu}{\mu_w}\right)^{-1/3}
\right].
\end{equation}
In conventional wall units, profiles at different Mach numbers deviate from the incompressible reference beyond $y^+\approx10^2$. The learned transformation collapses these profiles onto a common curve over the sampled range (Fig.~\ref{fig:cases_type_II}d). The dataset contains $N=1,\!109$ samples from direct numerical simulations (DNS) of compressible channel and pipe flows, covering bulk Mach numbers $M_b\in[1.5,4.0]$ and bulk Reynolds numbers $Re_b\in[8430,23978]$, compiled from Trettel and Larsson~\cite{trettel2016mean} and Modesti and Pirozzoli~\cite{modesti2019direct}.

The learned transformation has an algebraic connection to the Trettel--Larsson semi-local coordinate~\cite{trettel2016mean}. Writing $R=\rho/\rho_w$ and $M=\mu/\mu_w$, the coordinates satisfy $\Pi_1^*=y_{\mathrm{TL}}^+R^{1/6}M^{2/3}$, where $y_{\mathrm{TL}}^+=y^+R^{1/2}M^{-1}$. Here the exponents are learned without imposing the logarithmic-layer and inner-layer stress-balance conditions used in the Trettel--Larsson derivation. Unlike their integral velocity transformation, the learned velocity rescaling is a local algebraic product requiring neither profile derivatives nor integration. Quantitative comparisons are provided in \textcolor{blue}{Supplementary Section~5.3}.

\subsubsection{Goldenfeld's roughness-induced criticality in pipe flow}
Goldenfeld~\cite{goldenfeld2006roughness} demonstrated that the friction factor $f$ in rough-pipe turbulent flow obeys Widom scaling---the same mathematical structure that describes ferromagnetic phase transitions. In this analogy, the inverse Reynolds number $Re^{-1}$ corresponds to reduced temperature $t=|T-T_c|/T_c$, with $Re\to\infty$ representing the critical limit, while the relative roughness $r/D$ plays the role of an external field $H$. Just as the magnetization $M(t,H)$ near the Curie point satisfies $M(t,H)=|t|^\beta\mathcal{M}(H/|t|^{\beta\delta})$, the friction factor admits the scaling form:
\begin{equation}
f\cdot Re^{1/4}=\Phi\bigl(Re^{3/4}(r/D)\bigr),
\end{equation}
where $\Phi(z)$ is a scaling function with the asymptotic limits $z\to0$ corresponding to the Blasius smooth-pipe regime ($f\propto Re^{-1/4}$) and $z\to\infty$ to the Strickler fully rough regime ($f\propto(r/D)^{1/3}$). Matching these asymptotic laws determines the exponents $1/4$ and $3/4$, providing a reference for Type~II similarity discovery.

AI Buckingham receives five physical quantities characterizing the pipe geometry and fluid state: $\mathbf q=[D,U,\rho,\mu,r]$, where $D$ is the pipe diameter, $U$ is the mean flow velocity, $\rho$ is the fluid density, $\mu$ is the dynamic viscosity, and $r$ is the sand-grain roughness height. The dependent variable is the dimensionless friction factor $f$. With $p=5$ and $\operatorname{rank}(\mathbf D)=3$, the admissible Buckingham--$\Pi$ space has dimension $n=2$. To fix the parameterization of the dependent-variable rescaling, the prompt explicitly constrains the exponent of $r/D$ in the output-rescaling factor to zero. A single dominant group together with the rescaling factor is recovered and rationalized, yielding
\begin{equation}
f \cdot Re^{1/4} = \Phi^{(1)}\left( Re^{3/5} \left( r/D \right)^{4/5} \right).
\end{equation}
Remarkably, the learned scaling argument can be rewritten as $[Re^{3/4}(r/D)]^{4/5}$ and is therefore equivalent to Goldenfeld's theoretical group under an invertible power reparameterization. Figure~\ref{fig:cases_type_II}h shows the resulting data collapse. The dataset consists of $N=414$ Nikuradse measurements~\cite{nikuradse1950laws} in sand-grain-roughened pipes, covering $Re\in[5.84\times10^2,1.88\times10^6]$ and seven roughness ratios $D/r\in[15,1300]$.


\subsection{Multi-regime physical systems}

\begin{figure}[!t]
\centering
\includegraphics[width=\textwidth]{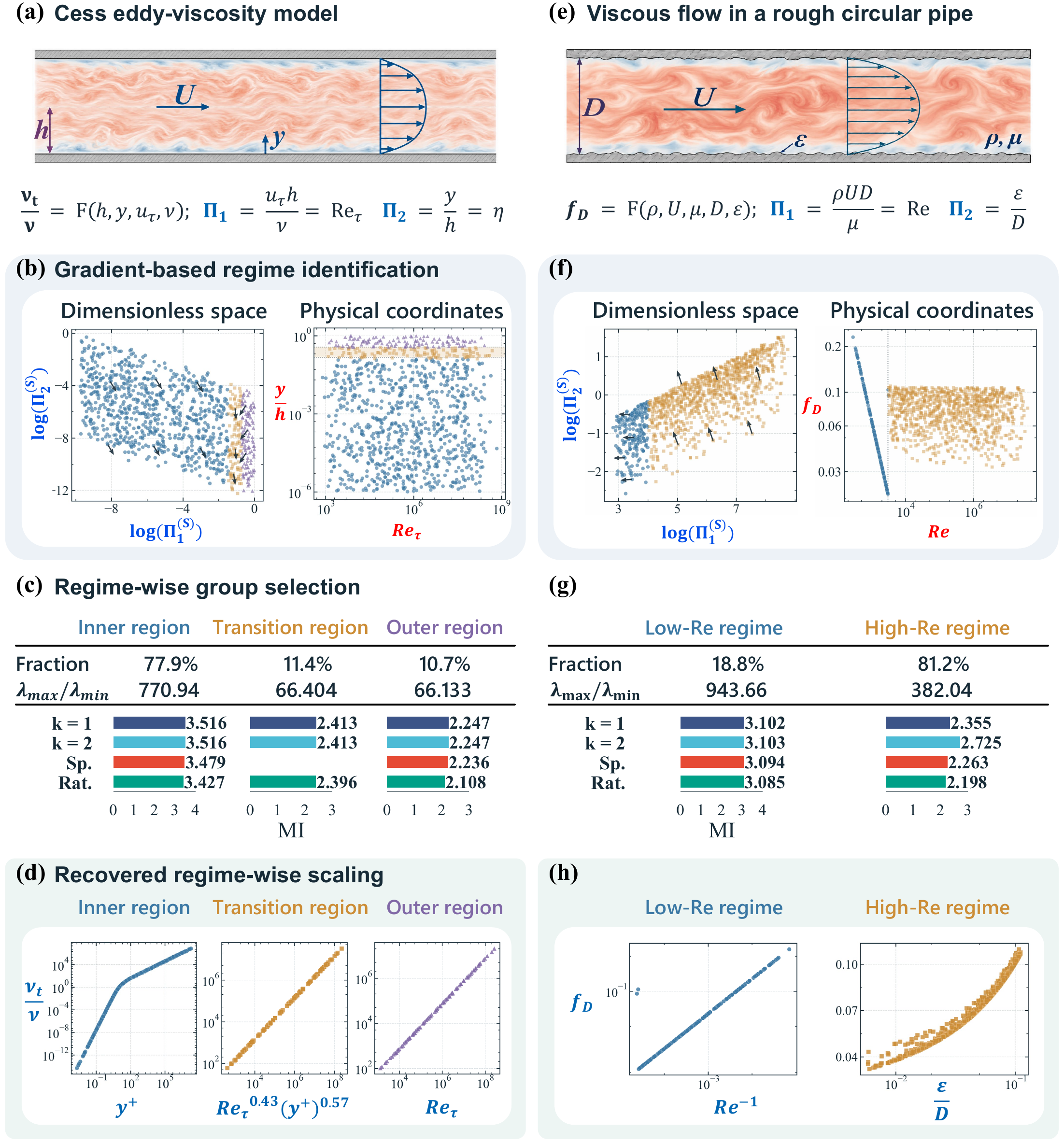}
\caption{Autonomous identification of physical regimes and regime-wise dominant scaling for the Cess eddy-viscosity model \textbf{(a--d)} and viscous flow in a rough circular pipe \textbf{(e--h)}. 
\textbf{(a,e)} Schematics of the physical configurations, with the corresponding functional relations and Buckingham--$\Pi$ basis groups. 
\textbf{(b,f)} Cluster assignments in the orthonormalized logarithmic Buckingham--$\Pi$ space (left) and physical coordinates (right). Arrows denote normalized GPR posterior-mean gradient directions. 
\textbf{(c,g)} Cluster fractions, eigenvalue ratios $\lambda_{\max}/\lambda_{\min}$ and MI comparisons for group selection and simplification. \textbf{(d,h)} Regime-wise data collapse using the discovered dominant dimensionless groups.}
\label{fig:multi_regime_}
\end{figure}

Beyond similarity discovery in the preceding single-physical-regime cases, AI Buckingham can also identify distinct regimes and recover their local similarity representations. This capability is assessed using two systems with multiple physical regimes: the Cess eddy-viscosity model and flow in a rough circular pipe, representing space-extended and parameter-dependent systems, respectively.

\subsubsection{Cess eddy-viscosity model in turbulent channel flows}
In turbulent channel flows, the Cess model is widely used for the eddy viscosity~\cite{cess1958survey,reynolds1972mechanics,wu2021space,wu2023composition,smits2011high},
\begin{equation}
\frac{\nu_t(y)}{\nu} = \frac{1}{2} \sqrt{ 1+ \frac{\kappa^2 Re_\tau^2}{9} \left[1-\left(1-\frac{y}{h}\right)^2\right]^2 \left[1+2\left(1-\frac{y}{h}\right)^2\right]^2 \left(1-e^{-y^+/A}\right)^2 } - \frac{1}{2},
\label{eq:cess}
\end{equation}
where $\kappa$ is the von K\'arm\'an constant and $A$ is a dimensionless damping constant. The dominant physical dependence varies continuously across the wall-normal direction, providing a benchmark for identifying distinct local similarity representations.

The dimensionless dependent variable $\nu_t/\nu$ is governed by four physical quantities, $\mathbf{q}=[h,y,u_\tau,\nu]$, where $h$ is the channel half-height, $y$ is the wall-normal distance, $u_\tau$ is the friction velocity, and $\nu$ is the fluid kinematic viscosity. With $p=4$ and $\operatorname{rank}(\mathbf D)=2$, the admissible Buckingham--$\Pi$ space has dimension $n=2$, with basis dimensionless parameters $Re_\tau=u_\tau h/\nu$ and $\eta=y/h$. The inner wall coordinate $y^+=Re_\tau\eta$ is an equivalent dimensionless variable within this space.
AI Buckingham autonomously identifies three spatially distinct regimes, corresponding to the near-wall, intermediate transition and outer regions, each described by a single dominant dimensionless group (Fig.~\ref{fig:multi_regime_}b,c). 
Near the wall, the recovered group is $\Pi_1^*=y^+=yu_\tau/\nu$, whereas toward the channel centre it becomes $\Pi_1^*=Re_\tau=hu_\tau/\nu$. In the intermediate region, the group is $\Pi_1^*=h^{0.43}y^{0.57}u_\tau/\nu=Re_\tau^{0.43}(y^+)^{0.57}$, retaining contributions from both the inner and outer similarity variables and characterizing the transition between the two limiting scaling mechanisms.

The dataset consists of $N=1000$ samples generated from Eq.~\ref{eq:cess}, covering $Re_\tau\in[7.85\times10^2,6.83\times10^8]$, $\eta\in[1.03\times10^{-6},0.998]$ and $y^+\in[2.05\times10^{-3},1.71\times10^8]$. Plotting $\nu_t/\nu$ against the locally identified group yields a one-dimensional collapse within each regime (Fig.~\ref{fig:multi_regime_}d). 
AI Buckingham identifies this transition regime by clustering samples using an eigenvalue-weighted measure of local gradient similarity. In contrast, IT--$\pi$~\cite{yuan2025dimensionless} fails to identify this regime as a distinct physical regime: its a posteriori prediction scores, evaluated over prescribed partitions of the space spanned by globally learned dimensionless groups, do not isolate the transition region (see \textcolor{blue}{Supplementary Section~5.4}). 

\subsubsection{Viscous flow in a rough circular pipe}
Viscous flow in a rough circular pipe (Fig.~\ref{fig:multi_regime_}e) exhibits different friction scalings as the Reynolds number and relative roughness vary~\cite{constantine2017data,xu2022artificial,zhang2024clustering}. The dimensionless friction factor $f_D$ is governed by five physical quantities, $\mathbf q=[\rho,U,\mu,D,\epsilon]$, where $\rho$ is the fluid density, $U$ is the mean velocity, $\mu$ is the dynamic viscosity, $D$ is the pipe diameter, and $\epsilon$ is the pipe wall roughness. With $p=5$ and $\operatorname{rank}(\mathbf D)=3$, the admissible Buckingham--$\Pi$ space has dimension $n=2$, with basis dimensionless parameters $Re=\rho U D/\mu$ and $\epsilon/D$.

The dataset is generated from a piecewise empirical relation with an imposed branch switch at $Re=3000$. For $Re\leq3000$, the friction factor follows the laminar relation $f_D=64/Re$; for $Re>3000$, it is implicitly determined by the Colebrook equation~\cite{white2016fluid,colebrook1937experiments},
\begin{equation}
\frac{1}{\sqrt{f_D}}=-2\log_{10}\left(\frac{\epsilon}{3.7D}+\frac{2.51}{Re\sqrt{f_D}}\right).
\label{eq:colebrook}
\end{equation}
With Eq.~\ref{eq:colebrook} solved numerically by Newton iteration, the resulting dataset consists of $N=1200$ samples, covering $Re\in[2.78\times10^2,3.95\times10^7]$ and $\epsilon/D\in[5.86\times10^{-3},1.10\times10^{-1}]$.

AI Buckingham identifies two regimes, each described by a single dominant dimensionless group (Fig.~\ref{fig:multi_regime_}f,g). When mapped to the physical parameter space, the partition closely follows the imposed division at $Re=3000$. In the low-Reynolds-number regime, the recovered group $\Pi_1^*=\mu/(\rho U D)=Re^{-1}$ captures the classical Reynolds-number dependence of laminar pipe flow. In the high-Reynolds-number regime, $\Pi_1^*=\epsilon/D$ indicates that relative roughness dominates as the flow approaches the fully rough asymptote. The corresponding regime-wise data collapse is shown in Fig.~\ref{fig:multi_regime_}h. Near the imposed discontinuity at $Re=3000$, a smooth GPR surrogate cannot represent the abrupt change exactly, making the local gradient estimates and the resulting partition sensitive to a small number of samples close to the branch boundary.

\section{Discussion}\label{sec3}
AI Buckingham brings dimensional specification, physical-regime identification, and similarity discovery into a unified, hierarchical framework for autonomous dimensionless learning. Its central contribution lies in coordinating these previously fragmented tasks through agentic orchestration, coupling LLM-based physical reasoning with explicit numerical decision criteria. This enables a unified treatment of both single- and multiple-regime systems, while accommodating Type~I and Type~II similarity within the same discovery workflow.

At the algorithmic level, eigenvalue-weighted gradient clustering identifies distinct regimes by grouping samples with similar local physical dependences~\cite{zhang2024clustering}. 
Within each regime, MI provides an objective, reparameterization-invariant criterion for discovering and selecting dominant dimensionless groups without prescribing their functional relationship with the output. In the existing dimensionless learning methods, identifying dominant dimensionless groups does not necessarily yield compact and physically interpretable representations. AI Buckingham further incorporates sparsification and rationalization to remove weakly contributing physical variables and simplify the learned exponents, respectively, with both operations controlled by explicit information-loss criteria.
These complementary mechanisms enable AI Buckingham to go beyond recovering established similarity laws and uncovering regime-specific representations.
In compressible wall-bounded turbulence, the learned local algebraic transformation applies compressibility corrections through density and viscosity ratios, thereby mapping the mean-velocity profiles onto the incompressible law of the wall. These power-law corrections are learned from data without imposing the logarithmic-layer and inner-layer stress-balance conditions underlying the Trettel--Larsson (TL) transformation~\cite{trettel2016mean}. Moreover, the resulting mapping requires neither profile differentiation nor integration and yields lower mean errors relative to the incompressible reference than TL and IT--$\pi$ for the examined profiles. 
In the Cess eddy-viscosity model, AI Buckingham autonomously identifies a distinct transition regime whose dominant group retains both inner and outer scaling contributions, a regime that is not isolated by the a posteriori identification employed in IT--$\pi$. 
For drag modelling in rough-wall turbulence, it further uncovers two previously unreported dimensionless combinations of roughness statistics. These compact representations capture the combined effects of surface morphology on equivalent sand-grain roughness and improve surrogate predictions compared with using all seven original surface statistics.

Existing methods typically address selected aspects of dimensionless discovery, including dominant-group identification~\cite{constantine2017data,xu2022artificial,zhang2025mutual,yuan2025dimensionless}, scaling-transformation discovery~\cite{sanfins2025similarity,watanabe2025datadriven} and multiple-physical-regime analysis~\cite{zhang2024clustering,yuan2025dimensionless}. These advances remain fragmented, with dimensional assignment and coordination of the analysis steps still relying heavily on expert guidance. As an agentic framework, AI Buckingham organizes the analysis into an autonomous workflow for discovering dimensionless scaling laws across physical regimes: the agent establishes dimensional assignments, plans workflow templates and configures numerical policies, reducing the manual intervention needed to connect successive stages. An additional contribution is the development and public release of two extensible, literature-based knowledge bases: 247 physical-quantity records support dimensional assignment and consistency checking, while 251 dimensionless-relation records provide physical context for interpreting newly discovered scaling laws. Together, these resources provide a foundation for extending autonomous dimensionless discovery across a broad range of physical disciplines. 

Several limitations remain to be addressed in future work. Errors in the GPR surrogate can propagate into local gradient estimates and alter the inferred regime partition, particularly in sparsely sampled regions or near abrupt changes in the response~\cite{zhang2024clustering,zhang2025mutual}.  Partitioning reduces the samples available for each MI estimate, making comparisons between group counts more sensitive to finite-sample errors as the joint dimension increases~\cite{kraskov2004estimating,yuan2025dimensionless}. Beyond these estimation issues, restricting dimensionless groups and Type~II rescaling factors to products of powers limits the direct discovery of similarity variables involving sums or more general nonlinear combinations. To relax this restriction, dimensionally constrained symbolic regression could extend the search to more general similarity variables and yield explicit relations in the learned coordinates~\cite{schmidt2009distilling,brunton2016discovering,cheng2026physics}. For broader applications, similarity representations inferred from measured and reconstructed flow fields~\cite{wan2026lagrangian} could be compared, and the learned dimensionless groups could be evaluated as inputs to turbulence models with regime-specific experts~\cite{liu2026factorized}.

\section{Methods}\label{sec4}
\subsection{Dimensional analysis and Buckingham-$\Pi$ formulation}
Given a target dimensionless response $\Pi_y$ and $p$ physical variables $\mathbf{x}=(x_1,\ldots,x_p)$, AI Buckingham first constructs the dimensional matrix $\mathbf{D}\in\mathbb{R}^{l\times p}$, whose columns contain the exponents of the input variables with respect to the $l$ relevant fundamental dimensions. The null space of $\mathbf{D}$ has dimension $n=p-\operatorname{rank}(\mathbf{D})$, and its basis $\{\mathbf{w}^{(1)},\ldots,\mathbf{w}^{(n)}\}$ is computed symbolically using SymPy. 

Each null-space basis vector $\mathbf{w}^{(q)}\in\mathbb{R}^{p}$ defines a basis dimensionless parameter $\Pi^{(q)} = \prod_{i=1}^{p}x_i^{w_i^{(q)}}$, satisfying the dimensional homogeneity constraint $\mathbf{D}\mathbf{w}^{(q)}=\mathbf{0}$. Any admissible dimensionless group can then be parameterized by $\mathbf{w}(\boldsymbol{\gamma}) = \sum_{q=1}^{n}\gamma_q\mathbf{w}^{(q)}$ or equivalently,
\[
\Pi(\boldsymbol{\gamma})
=
\prod_{i=1}^{p}
x_i^{w_i(\boldsymbol{\gamma})}
=
\prod_{q=1}^{n}
\left(\Pi^{(q)}\right)^{\gamma_q},
\]
where $\boldsymbol{\gamma}=(\gamma_1,\ldots,\gamma_n)^T$ contains the coefficients to be optimized and specifies a direction in the $n$-dimensional Buckingham--$\Pi$ space.

\subsection{Gradient-based identification of multiple physical regimes}\label{sec:cluster_analysis}
Identifying dominant dimensionless parameters in systems with multiple physical regimes conventionally requires substantial physical insight and prior knowledge. A single set of dominant groups may be insufficient when different subsets of the data follow distinct scaling behaviors. AI Buckingham therefore employs gradient-based clustering to partition the samples into clusters associated with different physical regimes~\cite{zhang2024clustering}.

Let $\{\Pi^{(S)}_1,\ldots,\Pi^{(S)}_n\}$ denote the dimensionless parameters corresponding to the orthonormalized null-space basis vectors. Orthonormalization provides orthogonal coordinates for evaluating angular similarities between gradient directions. We then define the logarithmic dimensionless coordinates $\mathbf{z} = \left[ \log \Pi^{(S)}_1, \ldots, \log \Pi^{(S)}_n \right]^T$. The logarithmic representation accommodates dimensionless quantities spanning several orders of magnitude and converts power-law relations into linear directions in $\log$-$\Pi$ space.

A Gaussian-process regression surrogate $g(\mathbf{z})\simeq\log \left({\Pi}_y\right)$ is fitted between the logarithmic dimensionless coordinates and the transformed response, thereby evaluating the posterior-mean gradient $\nabla_{\mathbf{z}}g(\mathbf{z}^{(i)})$ at each sample. 
The GPR formulation and analytical posterior-mean gradients are derived in \textcolor{blue}{Supplementary Section 1.2}.
For a candidate cluster $\Omega_I$ containing $N_I$ samples, the gradient directions are normalized and used to form the covariance matrix
\[
\mathbf{C}_I = \frac{1}{N_I} \sum_{\mathbf{z}^{(i)} \in \Omega_I} \left( \frac{\nabla_{\mathbf{z}} g(\mathbf{z}^{(i)})}{\|\nabla_{\mathbf{z}} g(\mathbf{z}^{(i)})\|_2} \right) \left( \frac{\nabla_{\mathbf{z}} g(\mathbf{z}^{(i)})}{\|\nabla_{\mathbf{z}} g(\mathbf{z}^{(i)})\|_2} \right)^T,
\]
whose eigendecomposition is $\mathbf{C}_I = \mathbf{V}_I \mathbf{\Lambda}_I \mathbf{V}_I^T$. Let $\lambda_j^{(I)}$ and $\mathbf{v}_j^{(I)}$ denote the corresponding eigenvalues and eigenvectors, and let $\boldsymbol{\lambda}_I=(\lambda_1^{(I)},\ldots,\lambda_n^{(I)})^T$. The similarity between a sample gradient and cluster $\Omega_I$ is evaluated using the eigenvalue-weighted projection
\[
Sim
\left(
\nabla_{\mathbf{z}}g(\mathbf{z}),\Omega_I
\right)
=
\sum_{j=1}^{n}
\left(
\frac{\lambda_j^{(I)}}
{\|\boldsymbol{\lambda}_I\|_2}
\right)^2
\left[
\nabla_{\mathbf{z}}g(\mathbf{z})
\cdot
\mathbf{v}_j^{(I)}
\right]^2.
\]

Samples are iteratively assigned to the cluster that maximizes this similarity, followed by recomputation of the gradient covariance and its eigenpairs for each updated cluster, until the assignments converge. When AI Buckingham is requested to autonomously identify distinct physical regimes from data, candidate $K$-cluster partitions are evaluated, with $K\in\{2,3,4\}$ by default. A partition is considered admissible only when every cluster $\Omega_I$ satisfies $\lambda_1^{(I)}/\lambda_n^{(I)} \geq E, N_I/N \geq p_{\min}, N_I \geq N_{\min}$, with $E=50$, $p_{\min}=0.05$, and $N_{\min}=60$ by default.

The admissibility criteria support the existence of dominant physical dependences within each cluster, but do not guarantee that these dependences differ between clusters. To bridge single- and multiple-regime treatments, AI Buckingham employs a directional-separation criterion based on the leading covariance eigenvectors $\mathbf{v}_1^{(A)}$ and $\mathbf{v}_1^{(B)}$ of the candidate $K=2$ partition. Their directional alignment in the orthonormalized logarithmic Buckingham--$\Pi$ coordinates $\mathbf{z}$ is quantified by the absolute cosine similarity $s_{\mathrm{dir}}=\left|(\mathbf{v}_1^{(A)})^{T}\mathbf{v}_1^{(B)}\right|$. This directional requirement is evaluated independently of partition admissibility. A multiple-regime treatment requires at least one admissible candidate partition and $s_{\mathrm{dir}}<\tau_{\mathrm{dir}}$, with $\tau_{\mathrm{dir}}=0.90$ by default. A single-regime treatment is adopted when no candidate partition is admissible or $s_{\mathrm{dir}}\geq\tau_{\mathrm{dir}}$.

The eigenvalue-ratio criterion provides a lower bound on the number of clusters by requiring a dominant physical dependence within each cluster, while the cluster-fraction and sample-count criteria provide an upper bound by limiting excessive subdivision. Within the admissible range, smaller $K$ identifies the principal physical regimes, whereas larger $K$ yields progressively more detailed and refined partitions. The regime-analysis policies and partition-selection logic are detailed in \textcolor{blue}{Supplementary Section 3.1}. After the cluster partition is determined, the MI-based dimensionless-learning procedure is applied independently to each cluster.

\subsection{Mutual-information-based dimensionless learning}
MI quantifies the dependence between a dimensionless representation and the response $\Pi_y$, providing an objective, model-free relevance criterion that is invariant under invertible reparameterizations~\cite{zhang2025mutual,yuan2025dimensionless}.

For a candidate representation containing $k$ dimensionless groups, define $\boldsymbol{\Pi}_k(\boldsymbol{\Gamma}_k)=\left[\Pi(\boldsymbol{\gamma}_1),\ldots,\Pi(\boldsymbol{\gamma}_k)\right]$ and $\boldsymbol{\Gamma}_k=\left[\boldsymbol{\gamma}_1,\ldots,\boldsymbol{\gamma}_k\right]\in\mathbb{R}^{n\times k}$, where each $\boldsymbol{\gamma}_j\in\mathbb{R}^{n}$ specifies one direction in the Buckingham--$\Pi$ space and hence defines an admissible dimensionless group $\Pi(\boldsymbol{\gamma}_j)$. For Type I similarity, the dominant groups are obtained by maximizing their joint MI with the dimensionless response,
\[
\boldsymbol{\Gamma}_k^*
=
\arg\max_{\boldsymbol{\Gamma}_k}
\widehat{I}
\left(
\boldsymbol{\Pi}_k(\boldsymbol{\Gamma}_k);
\Pi_y
\right).
\]
For Type II similarity, an additional coefficient vector $\boldsymbol{\gamma}_o$ defines the multiplicative output-scaling factor through $\widetilde{\Pi}_y(\boldsymbol{\gamma}_o)=\Pi_y\Pi(\boldsymbol{\gamma}_o)$, and the similarity variables and output-scaling exponents are optimized jointly,
\[
\left(
\boldsymbol{\Gamma}_k^*,
\boldsymbol{\gamma}_o^*
\right)
=
\arg\max_{\boldsymbol{\Gamma}_k,\boldsymbol{\gamma}_o}
\widehat{I}
\left(
\boldsymbol{\Pi}_k(\boldsymbol{\Gamma}_k);
\widetilde{\Pi}_y(\boldsymbol{\gamma}_o)
\right).
\]
Here, $\widehat{I}$ is estimated directly from the samples using the Kraskov nearest-neighbour estimator~\cite{kraskov2004estimating} or, alternatively, a histogram-based estimator. The resulting non-convex optimization problems are solved primarily using CMA-ES with multiple restarts. 
When similarity type is not prescribed a priori, AI Buckingham compares the Type I and Type II formulations at $k=1$. Type I is selected when omitting dependent-variable rescaling causes a relative loss in MI no greater than $\tau_{\mathrm{MI}}$, with $\tau_{\mathrm{MI}}=0.15$ by default; otherwise, Type II is selected.

For $k\geq2$, a cosine-similarity penalty is used by default to prevent distinct optimized groups from collapsing onto nearly identical directions in the Buckingham--$\Pi$ space. For two coefficient vectors $\boldsymbol{\gamma}_i$ and $\boldsymbol{\gamma}_j$, their absolute cosine similarity is
\[
c_{ij}
=
\frac{
\left|
\boldsymbol{\gamma}_i^T\boldsymbol{\gamma}_j
\right|
}{
\|\boldsymbol{\gamma}_i\|_2
\|\boldsymbol{\gamma}_j\|_2
}.
\]
Candidate solutions with $c_{ij}>c_{\max}$ are penalized, with $c_{\max}=0.7$ by default. For Type II similarity, the same criterion can also be applied between the dimensionless-group directions and the output-scaling direction.

To determine the number of dominant dimensionless groups, AI Buckingham evaluates candidate representations sequentially from $k=1$. For each $k$, let $\widehat{I}_k$ denote the optimized MI and define the normalized information score $\varepsilon_k=\exp[-\max(\widehat{I}_k,0)]$, with smaller values indicating stronger mutual dependence between the dimensionless representation and the response~\cite{yuan2025dimensionless}. The contribution of an additional dimensionless group is quantified by the absolute score reduction $\Delta\varepsilon_k=\varepsilon_{k-1}-\varepsilon_k$. The search proceeds when $\Delta\varepsilon_k\geq\Delta\varepsilon_{\min}$, with $\Delta\varepsilon_{\min}=0.08$ by default; otherwise, the preceding representation is selected, giving $k^*=k-1$. Uncertainty in the estimated MI and normalized information score is quantified from variability across CMA-ES restarts and data subsamples. 
Details of the MI criterion, estimators and similarity-type screening are provided in \textcolor{blue}{Supplementary Sections~1.3--1.4 and~3.2}.

\subsection{Simplification and validation of dimensionless representations}
\label{sec:sparse_rationalize}
\textbf{Sparsification.}
A physical variable is considered for removal when $\max_{1\leq j\leq k^*}\left|\widetilde{w}_i^{(j)}\right|<\delta_1$, where $\delta_1=0.05$ and $\widetilde{w}_i^{(j)}=w_i^{(j)}/\max_{1\leq r\leq p}|w_r^{(j)}|$. For Type II similarity, the physical exponents of the output-scaling factor are additionally included in the sparsification criterion. Sparsification is not applied when fixed coefficients in the Buckingham--$\Pi$ basis are imposed.

Rather than directly setting these exponents to zero, AI Buckingham constructs a restricted null-space basis by masking the candidate physical variables through zero-exponent constraints. The selected $k^*$ dimensionless parameters are then re-optimized in this restricted space while preserving dimensional homogeneity. The sparsified solution is retained only when the increase in normalized information score does not exceed $\delta_2=0.05$.

\textbf{Rationalization.}
The optimized Buckingham-$\Pi$ basis coefficients are subsequently approximated by simple rational values using a bounded-denominator continued-fraction method. For each coefficient $\gamma$, a rational approximation $a/q$ with $q\leq Q_{\max}=5$ is adopted when $\left|\gamma-\frac{a}{q}\right|\leq\delta_3=0.05$; otherwise, a decimal approximation is retained. The rationalized dimensionless parameters are then re-evaluated using the same MI estimator and are accepted only when the relative decrease in MI does not exceed $\delta_4=10\%$.

\textbf{Collapse validation.}
The data-collapse quality of the discovered representation is assessed using a separate non-parametric GPR model. The model is fitted in the learned dimensionless coordinates using 80\% of the samples and evaluated on the remaining 20\%. A representation passes collapse validation when the mean test-set coefficient of determination satisfies $\overline{R^2}\geq0.85$. The simplification and validation procedures are detailed in \textcolor{blue}{Supplementary Sections~3.4--3.5}.

\subsection{LLM-based physical reasoning and workflow planning}
The LLM interprets the natural-language problem description and dataset preview to identify the dependent variable and assign physical meanings and candidate SI dimensions to the governing quantities. These assignments are checked against retrieved physical definitions, with conflicts returned for revision before basis construction.

Given the problem description, dataset summary and constructed Buckingham--$\Pi$ basis, the LLM selects a predefined workflow template and configures numerical policies and parameters subject to user constraints. System prompts define the planning rules, while predefined schemas constrain the plan to permitted fields and options. 
LangGraph~\cite{langgraph_2025} coordinates these interactions and plan execution. Deterministic control logic translates the validated plan into calls to predefined tools, with data-dependent decisions governed by the criteria described above. The LLM then interprets the numerical results with support from retrieved dimensionless relations.

\subsection{Knowledge-base construction and retrieval augmentation}
Two Markdown knowledge bases covering physical quantities and dimensionless relations were compiled offline through a Codex-assisted literature-curation procedure~\cite{OpenAI2025Codex}. Codex retrieved and cross-checked authoritative references on SI units, established dimensional-analysis resources and domain-specific literature, and standardized quantity names, symbols and aliases while preserving source provenance and notation ambiguities. Physical-quantity entries contain definitions and SI dimensional vectors; dimensionless-relation entries contain defining expressions and physical interpretations. Both retain names, notation, domains and sources.

Record embeddings are generated with GLM-Embedding-3, normalized to unit length and cached locally. Candidate entries are ranked by a hybrid relevance score $s=w_ls_l+w_ss_s$, where $s_l$ combines name, symbol and alias matching with contextual token overlap, and $s_s$ is the cosine similarity in embedding space. For dimensional checking, queries include the dataset column name, inferred physical meaning and problem context but exclude the candidate SI exponents. The default weights $(w_l,w_s)=(0.65,0.35)$ emphasize precise quantity identification. Retrieval for physical interpretation uses $(0.20,0.80)$, emphasizing semantic similarity because discovered expressions may not contain the conventional names or symbols used in the literature. Workflow implementation and retrieval procedures are detailed in \textcolor{blue}{Supplementary Section~2}.


\backmatter

\bmhead{Data availability}
The datasets and numerical results supporting this study are available at~\url{https://github.com/ChengAo21/ai-buckingham}.

\bmhead{Code availability}
The source code, executed notebooks and reproduction scripts are available in the same repository. A Streamlit-based interactive web application is available at~\url{https://ai-buckingham.streamlit.app/}.

\bmhead{Acknowledgements}
This work was supported by NSFC Excellence Research Group Program for ``Multiscale Problems in Nonlinear Mechanics'' (No.\ 12588201) and by the CAS Project for Young Scientists in Basic Research (No.\ YSBR-136). L.\ Zhang was also supported by the National Natural Science Foundation of China (NSFC) under Grant No.\ 12202451. We thank Dr.\ Zhideng Zhou for valuable discussions on the sand-grain roughness case study.

\section*{Author contributions}
A.C.\ designed and implemented the agent framework, conducted the validation experiments, and drafted the manuscript. L.Z.\ developed the mutual-information-based dimensionless learning and gradient-based multi-regime identification methodologies, supervised the research, and edited the manuscript. G.-W.H.\ conceived and directed the overall research program and contributed to the physical interpretation of results. All authors discussed the results and approved the final manuscript.

\section*{Competing interests}
The authors declare no competing interests.


\bibliography{sn-bibliography}

\end{document}


\maketitle
\tableofcontents

\section{Theoretical and Computational Foundations}\label{sec:theo_comp_foundation}

\subsection{The Buckingham $\Pi$ Theorem, admissible dimensionless space and similarity types}

The Buckingham $\Pi$ theorem provides the foundation for expressing physical relationships independently of the choice of units~\cite{buckingham1914,bridgman1922,barenblatt1996scaling}. Consider a physical system with a dimensionless dependent variable denoted by $\Pi_y$ and described by $p$ quantities $\mathbf{x}=(x_1,x_2,\ldots,x_p)$, where each quantity $x_i$ possesses a physical dimension expressible as a product of $l$ fundamental dimensions (for mechanical systems, $l = 3$: mass $M$, length $L$, time $T$; for thermo-mechanical systems, $l = 4$ with the addition of temperature $\Theta$; for electro-mechanical systems, $l = 4$ with electric current $I$). The dimensional form of $x_i$ is written as
\begin{equation}
[x_i]=\prod_{j=1}^{l}d_j^{D_{ji}},
\end{equation}
where $d_j$ denotes the $j$-th fundamental dimension and $\mathbf{D} \in \mathbb{R}^{l \times p}$ is the \textit{dimensional matrix} whose column $i$ encodes the exponents of $x_i$. The dimensionless dependent variable $\Pi_y$ is excluded from $\mathbf{D}$, while a quantity $x_i$ that is already dimensionless corresponds to a zero column. The theorem reduces the physical relationship to
\begin{equation}
\Pi_y=\Phi\!\left(\Pi^{(1)},\ldots,\Pi^{(n)}\right),
\quad n=p-\operatorname{rank}(\mathbf{D}),
\label{eq:buckingham}
\end{equation}
where the $n$ basis dimensionless parameters are power-law products,
\begin{equation}
\Pi^{(q)}=\prod_{i=1}^{p}x_i^{w_i^{(q)}},\quad
\mathbf{D}\mathbf w^{(q)}=\mathbf0,\quad q=1,\ldots,n.
\label{eq:basisgroup}
\end{equation}

The exponent vectors $\{\mathbf{w}^{(1)},\ldots,\mathbf{w}^{(n)}\}$ therefore form a basis for the null space of $\mathbf{D}$. In AI Buckingham, this basis is constructed symbolically from the dimensional matrix using the null-space routine in SymPy. Arranging the transposes of the basis vectors as the rows of $\mathbf{B}\in\mathbb{R}^{n\times p}$ gives
\begin{equation}
\mathbf{w}(\boldsymbol{\gamma})
=
\mathbf{B}^{T}\boldsymbol{\gamma}
=
\sum_{q=1}^{n}\gamma_q\mathbf{w}^{(q)},
\label{eq:basis_param}
\end{equation}
where $\boldsymbol{\gamma}=(\gamma_1,\ldots,\gamma_n)^T\in\mathbb{R}^{n}$ contains the free coefficients to be optimized. The corresponding dimensionless group can equivalently be written as
\begin{equation}
\Pi(\boldsymbol{\gamma})
=
\prod_{i=1}^{p}x_i^{w_i(\boldsymbol{\gamma})}
=
\prod_{q=1}^{n}
\left(\Pi^{(q)}\right)^{\gamma_q}.
\label{eq:group_param}
\end{equation}
Every coefficient vector $\boldsymbol{\gamma}$ therefore defines a dimensionally admissible group in the Buckingham--$\Pi$ space. Dimensional homogeneity alone does not require the exponents of an admissible group to be integer or rational.

\textbf{The non-uniqueness problem.}
Equation~\eqref{eq:basis_param} reveals a fundamental limitation of classical dimensional analysis: the choice of null-space basis $\{\mathbf{w}^{(j)}\}_{j=1}^{n}$ is non-unique (any invertible linear transformation of a basis yields another valid basis), and the coefficients $\boldsymbol{\gamma}$ are entirely unconstrained by dimensional considerations. This generates an infinite family of permissible dimensionless representations, each formally equivalent under the Buckingham $\Pi$ theorem but potentially differing substantially in physical interpretability and relevance to the dependent variable $\Pi_y$. The classical analyst resolves this ambiguity through physical intuition, a process that is neither systematic nor reproducible. This non-uniqueness, together with the absence of a quantitative measure of a group's relevance, is the central problem that data-driven dimensional analysis~\cite{constantine2017data,bakarji2022dimensionally,xie2022data,yuan2025dimensionless,zhang2025mutual} aims to resolve.

\textbf{Type I and Type II Similarity.}

Barenblatt~\cite{barenblatt1996scaling,barenblatt2003scaling,barenblatt1972self} formalized a crucial distinction between two classes of physical similarity that profoundly affects how dimensionless groups govern a system's behavior.

\textbf{Type I: Complete similarity.} For the classical similarity formulation, let $\Pi_1,\ldots,\Pi_n$ denote the dimensionless parameters on which $\Pi_y$ depends:
\begin{equation}
\Pi_y = \Phi(\Pi_1, \ldots, \Pi_k, \Pi_{k+1}, \ldots, \Pi_n).
\end{equation}
Complete similarity (Type I) holds when, as the parameters $\Pi_{k+1}, \ldots, \Pi_n$ approach zero or infinity, the function $\Phi$ converges to a finite, non-zero limit:
\begin{equation}
\lim_{\Pi_{k+1},\ldots,\Pi_n \to 0\;\text{or}\;\infty}
\Phi(\Pi_1, \ldots, \Pi_n)
=
\Phi^{(0)}(\Pi_1, \ldots, \Pi_k).
\label{eq:type1_limit}
\end{equation}
Physically, the parameters $\Pi_{k+1}, \ldots, \Pi_n$ become irrelevant in the corresponding asymptotic regime, and the system reduces to a lower-dimensional similarity description involving only $\Pi_1,\ldots,\Pi_k$.

\textbf{Type II: Incomplete similarity.} When the limit in Eq.~\eqref{eq:type1_limit} does not exist (i.e., $\Phi$ diverges or vanishes as its arguments approach extreme values), the system exhibits \textit{incomplete similarity} (Type II). In this case, the dependent variable obeys a generalized homogeneity relation with \textit{anomalous exponents} $\xi_j$ and $\xi^{(i)}_j$~\cite{barenblatt1972self}:
\begin{equation}
\Pi_y \cdot \Pi_{k+1}^{\xi_{k+1}} \cdots \Pi_n^{\xi_n}
=
\Phi^{(1)}\!\left(
\Pi_1 \Pi_{k+1}^{\xi^{(1)}_{k+1}} \cdots \Pi_n^{\xi^{(1)}_n},
\ldots,
\Pi_k \Pi_{k+1}^{\xi^{(k)}_{k+1}} \cdots \Pi_n^{\xi^{(k)}_n}
\right).
\label{eq:type2_homogeneity}
\end{equation}
These additional exponents cannot be determined from dimensional analysis alone and must instead be obtained from additional physical analysis or inferred from data. In classical Type II problems, such exponents may be determined through nonlinear eigenvalue problems or renormalization-group analysis~\cite{goldenfeld1992lectures}. Examples include the Guderley converging shock wave~\cite{guderley1942starke} and the Goldenfeld scaling of turbulent friction in rough pipes~\cite{goldenfeld2006roughness,gioia2006turbulent}.

\textbf{Implications for data-driven discovery.}
The Type I/II distinction is not merely taxonomic; it determines \textit{what} must be discovered from data. For Type I problems, the task is to identify which combinations within the admissible Buckingham--$\Pi$ space capture the dominant dependence of $\Pi_y$. For Type II problems, the task additionally requires identifying the scaling transformation of the dimensionless dependent variable that achieves data collapse. In the computational formulation adopted here, these two cases are written as
\begin{align}
\text{Type I:}\qquad
\Pi_y
&\simeq
\Phi^{(0)}
\left(
\Pi_1^{*},\ldots,\Pi_k^{*}
\right),
\\
\text{Type II:}\qquad
\widetilde{\Pi}_y
=
\Pi_y\Pi(\boldsymbol{\gamma}_o)
&\simeq
\Phi^{(1)}
\left(
\Pi_1^{*},\ldots,\Pi_k^{*}
\right),
\end{align}
where $k\leq n$ is the number of retained dominant groups, $\Pi_j^{*}=\Pi(\boldsymbol{\gamma}_j^{*})$ denotes a dominant dimensionless group, and $\Pi(\boldsymbol{\gamma}_o)$ denotes the learned multiplicative rescaling factor for the dimensionless dependent variable in Type II similarity. AI Buckingham handles both Type~I and Type~II similarity problems within a unified agentic workflow.

\subsection{Gaussian-process regression surrogate and analytical gradients}\label{subsec:gpr_gradients}

For gradient-based identification of multiple physical regimes, AI Buckingham employs a Gaussian-process regression (GPR) surrogate to evaluate local gradients in the admissible Buckingham--$\Pi$ space. The dominant dimensionless groups within each identified regime are subsequently determined by mutual-information-based dimensionless learning.

\textbf{GPR formulation.}
Applying Gram--Schmidt orthonormalization to the rows of $\mathbf{B}$ gives the orthonormalized basis matrix $\mathbf{B}^{(S)}$, with corresponding dimensionless parameters $\Pi_1^{(S)},\ldots,\Pi_n^{(S)}$. The logarithmic dimensionless coordinates are
\begin{equation}
\mathbf{z}=\left[\log\Pi_1^{(S)},\ldots,\log\Pi_n^{(S)}\right]^T=\mathbf{B}^{(S)}\log\mathbf{x}.
\label{eq:gpr_coordinates}
\end{equation}
For the multi-regime cases considered here, the GPR surrogate approximates $g(\mathbf{z})\simeq \log\Pi_y$. The response function is modeled as a Gaussian process,
\begin{equation}
g(\mathbf{z})\sim\mathcal{GP}\left(m(\mathbf{z}),k(\mathbf{z},\mathbf{z}')\right),
\label{eq:gpr_prior}
\end{equation}
where $m(\mathbf{z})$ is the mean function and $k(\mathbf{z},\mathbf{z}')$ is the covariance function. The covariance function is taken as the isotropic squared-exponential form
\begin{equation}
k(\mathbf{z},\mathbf{z}') = \sigma_0^2 \exp \left[-\sigma_1^2\left\| \mathbf{z}-\mathbf{z}' \right\|_2^2 \right].
\label{eq:gpr_kernel}
\end{equation}
The observations are modeled as $\log\Pi_y=g(\mathbf{z})+\epsilon$, with independent Gaussian noise $\epsilon\sim\mathcal{N}(0,\sigma_u^2)$.

For $N_{\mathrm{tr}}$ training samples, let $\mathbf{y}=\left[\log\Pi_y^{(1)},\ldots,\log\Pi_y^{(N_{\mathrm{tr}})}\right]^T$ denote the transformed dependent-variable values and $\mathbf{Z}=[\mathbf{z}^{(1)},\ldots,\mathbf{z}^{(N_{\mathrm{tr}})}]^T$ the corresponding logarithmic dimensionless coordinates. The covariance matrix $\mathbf{K}$ has entries $K_{ij}=k(\mathbf{z}^{(i)},\mathbf{z}^{(j)})$, and the training covariance is $\mathbf{K}_y=\mathbf{K}+\sigma_u^2\mathbf{I}$. A small diagonal jitter $\delta\mathbf{I}$, with $\delta=10^{-10}$, is added to $\mathbf{K}_y$ in the numerical linear solves and is omitted from the notation below. For an evaluation point $\mathbf{z}$, the joint Gaussian distribution is
\begin{equation}
\begin{bmatrix}
\mathbf{y}\\
g(\mathbf{z})
\end{bmatrix}
\sim
\mathcal{N}
\left(
\begin{bmatrix}
m(\mathbf{Z})\\
m(\mathbf{z})
\end{bmatrix},
\begin{bmatrix}
\mathbf{K}_y & \mathbf{k}(\mathbf{Z},\mathbf{z})\\
\mathbf{k}(\mathbf{z},\mathbf{Z}) & k(\mathbf{z},\mathbf{z})
\end{bmatrix}
\right).
\end{equation}
Here, $\mathbf{k}(\mathbf{z},\mathbf{Z})$ contains the covariances between $\mathbf{z}$ and all training samples. Conditioning on the training data gives the posterior mean,
\begin{equation}
g_{\mathrm{pred}}(\mathbf{z})=m(\mathbf{z})+\mathbf{k}(\mathbf{z},\mathbf{Z})\mathbf{K}_y^{-1}\left[\mathbf{y}-m(\mathbf{Z})\right],
\label{eq:gpr_prediction}
\end{equation}
which is taken as the GPR prediction.

\textbf{Parameter optimization.}
The covariance hyperparameters and mean-function parameters are determined by minimizing the negative log marginal likelihood
\begin{equation}
\mathcal{L}=\frac{1}{2}\mathbf{r}^{T}\mathbf{K}_y^{-1}\mathbf{r}+\frac{1}{2}\log|\mathbf{K}_y|+\frac{N_{\mathrm{tr}}}{2}\log(2\pi),
\label{eq:gpr_nll}
\end{equation}
where $\mathbf{r}=\mathbf{y}-m(\mathbf{Z})$. Cholesky factorization is used for the covariance-system solves. In the implementation, the mean-function parameters and the logarithms of $\sigma_0^2$, $\sigma_1^2$, and $\sigma_u^2$ are optimized using bounded L-BFGS-B with multiple initializations.

With $\boldsymbol{\alpha}=\mathbf{K}_y^{-1}\mathbf{r}$ and $\mathbf{Q}=\mathbf{K}_y^{-1}-\boldsymbol{\alpha}\boldsymbol{\alpha}^{T}$, the analytical derivative with respect to a covariance hyperparameter $\theta$ is
\begin{equation}
\frac{\partial\mathcal{L}}{\partial\theta}=\frac{1}{2}\operatorname{tr}\left(\mathbf{Q}\frac{\partial\mathbf{K}_y}{\partial\theta}\right),
\label{eq:gpr_nll_cov_gradient}
\end{equation}
whereas for a mean-function parameter $\theta_m$,
\begin{equation}
\frac{\partial\mathcal{L}}{\partial\theta_m}=-\left(\frac{\partial m(\mathbf{Z})}{\partial\theta_m}\right)^T\boldsymbol{\alpha}.
\label{eq:gpr_nll_mean_gradient}
\end{equation}
These analytical derivatives are supplied directly to the optimizer.

\textbf{Mean functions.}
The default mean function is constant, $m(\mathbf{z})=\beta$, for which $\nabla_{\mathbf{z}}m(\mathbf{z})=\mathbf{0}$. For the rough-pipe case, the non-zero mean function used by Zhang et al.~\cite{zhang2024clustering} for sharp-gradient behavior is adopted in the affine--sigmoid form $m(\mathbf{z})=\beta+\left(\mathbf{w}_1^T\mathbf{z}+b_1\right)s\!\left(\mathbf{w}_2^T\mathbf{z}+b_2\right)$, where $s(a)=[1+\exp(-a)]^{-1}$ is the sigmoid function. Its gradient is $\nabla_{\mathbf{z}}m(\mathbf{z})=s\,\mathbf{w}_1+\left(\mathbf{w}_1^T\mathbf{z}+b_1\right)s(1-s)\mathbf{w}_2$, where $s=s(\mathbf{w}_2^T\mathbf{z}+b_2)$.

\textbf{Analytical posterior-mean gradients.}
Rather than using finite-difference approximations, the posterior-mean gradients are obtained analytically from the GPR prediction~\cite{zhang2024clustering}. Differentiating Eq.~\eqref{eq:gpr_prediction} with respect to $\mathbf{z}$, the covariance function in Eq.~\eqref{eq:gpr_kernel} gives
\begin{equation}
\nabla_{\mathbf{z}}k(\mathbf{z},\mathbf{z}^{(i)})=2\sigma_1^2k(\mathbf{z},\mathbf{z}^{(i)})\left(\mathbf{z}^{(i)}-\mathbf{z}\right),
\end{equation}
and therefore
\begin{equation}
\nabla_{\mathbf{z}}
g_{\mathrm{pred}}(\mathbf{z})
=
\nabla_{\mathbf{z}}m(\mathbf{z})
+
2\sigma_1^2
\sum_{i=1}^{N_{\mathrm{tr}}}
\alpha_i
k(\mathbf{z},\mathbf{z}^{(i)})
\left(
\mathbf{z}^{(i)}-\mathbf{z}
\right).
\label{eq:gpr_gradient}
\end{equation}
The resulting posterior-mean gradients provide the local directional information used for gradient-based regime identification.

\subsection{Mutual Information as an Objective Relevance Criterion for Dimensionless Learning}

Mutual information (MI) quantifies the statistical dependence between variables. For continuous random variables $X$ and $Y$ with joint probability density $p(x,y)$ and marginal densities $p_X(x)$ and $p_Y(y)$,
\begin{equation}
I(X; Y) = \iint p(x, y) \log \frac{p(x, y)}{p_X(x)\,p_Y(y)} \, \mathrm{d}x\,\mathrm{d}y.
\label{eq:mi_definition}
\end{equation}
MI is non-negative, symmetric, and vanishes if and only if the variables are statistically independent. Here, $X$ is a candidate dimensionless group, or a vector of such groups, and $Y$ is $\Pi_y$ for Type I similarity or $\widetilde\Pi_y$ for Type II similarity.

\textbf{Why MI provides an objective relevance criterion for dimensionless learning.} Three properties motivate its use for discovering dimensionless groups from data:

\begin{enumerate}
    \item \textbf{Model-free.} MI makes no assumptions about the functional form relating $\Pi$ to $\Pi_y$. Unlike regression-based methods that fit a specific parametric model $g_\theta(\Pi) \approx \Pi_y$, MI directly estimates the information content of $\Pi$ about $\Pi_y$. This avoids coupling the relevance criterion to a particular regression architecture or set of basis functions that could obscure a genuinely predictive dimensionless group.

    \item \textbf{Reparameterization-invariant.} For any invertible transformation $f$, $I(f(X); Y) = I(X; Y)$. Equivalent dimensionless representations may be related by invertible transformations, so an objective criterion for comparing groups should be invariant under such transformations. MI satisfies this requirement exactly~\cite{yuan2025dimensionless}. In contrast, gradient-based sensitivity measures~\cite{constantine2017data} depend on derivative magnitudes in specified coordinates, while neural-network-based approaches~\cite{bakarji2022dimensionally,xie2022data} inherit the inductive biases of their chosen architectures.

    \item \textbf{Multivariate generalization.} MI naturally extends to sets of variables: $I([\Pi_1,\ldots,\Pi_k];\Pi_y)$ measures the joint predictive information of $k$ dimensionless groups about the dependent variable. This enables AI Buckingham to assess whether an additional group provides genuinely new information or merely recapitulates what is already captured by existing groups. The corresponding numerical rule for adaptive dimensionality selection is given in Section~\ref{sec:si_mi_estimation}.
\end{enumerate}

\textbf{Comparison with alternative approaches.} Constantine et al.~\cite{constantine2017data} use the gradient of $\Pi_y$ with respect to input variables to construct an ``active subspace'' whose basis vectors define a unique set of dimensionless groups. While model-free, this approach relies on gradient information that is sensitive to variable scaling: rescaling an input variable changes the gradient covariance eigenvalues and can alter the identified active directions. Bakarji et al.~\cite{bakarji2022dimensionally} embed dimensional constraints into a neural network (BuckiNet), enforcing that the network's first-layer weights satisfy $\mathbf{D}\mathbf{W} = \mathbf{0}$. This provides a dimensional consistency guarantee, but the recovered dimensionless groups depend on the network architecture, initialization, and training procedure. Xie et al.~\cite{xie2022data} learn a linear projection in log-space whose rows span the dimensionless subspace; the projection is optimized to minimize regression error in a subsequent SINDy-style~\cite{brunton2016discovering} sparse regression stage, coupling the dimensionless discovery to the regression model's performance. Yuan \& Lozano-Dur\'an~\cite{yuan2025dimensionless} independently proposed MI as a criterion for dimensionless learning (IT-$\pi$), establishing an irreducibility theorem that provides a theoretical lower bound on prediction error given a set of dimensionless groups. AI Buckingham places information-based optimization within an autonomous agent orchestration that additionally performs variable identification, dimensional auditing, regime analysis, similarity-type selection, and automated parsimony enforcement.

\textbf{Reparameterization Invariance: A Proof Sketch.}

A fundamental requirement for any objective function in dimensionless learning is \textit{reparameterization invariance}: 
the relevance assigned to a candidate dimensionless group should not depend on an invertible re-expression of that group. Mutual information satisfies this requirement exactly.

\textbf{Theorem (informal).} Let $X$ and $Y$ be continuous random variables, and let $f$ be a differentiable one-to-one transformation with nonzero derivative on the support of $X$. Then $I(f(X); Y) = I(X; Y)$.

\textbf{Proof sketch.} The MI can be expressed in terms of differential entropies:
\begin{equation}
I(X; Y) = H(X) + H(Y) - H(X, Y).
\end{equation}
Under an invertible transformation $X' = f(X)$, the differential entropy transforms as
\begin{equation}
H(X') = H(X) + \mathbb{E}_X\!\left[\log\left|\frac{\mathrm{d}f}{\mathrm{d}x}\right|\right].
\end{equation}
For the joint entropy $H(X', Y)$, the same Jacobian term appears, and for $H(Y)$, no Jacobian appears. Since the Jacobian term $\mathbb{E}[\log|\mathrm{d}f/\mathrm{d}x|]$ appears in both $H(X')$ and $H(X', Y)$, it cancels in their difference:
\begin{equation}
I(X'; Y) = [H(X) + \mathbb{E}\log|f'|] + H(Y) - [H(X,Y) + \mathbb{E}\log|f'|] = I(X; Y).
\end{equation}

\textbf{Consequence for dimensional analysis.} Dimensionless groups are power-law products $\Pi = \prod x_i^{w_i}$. The space of all such products forms a multiplicative group, and any invertible transformation of a dimensionless group (e.g., $\Pi \mapsto \Pi^2$, $\Pi \mapsto 1/\Pi$, $\Pi \mapsto \ln\Pi$) yields another valid dimensionless combination. MI assigns the same score to all such representations, ensuring that the agent's ranking reflects \textit{physical information content} rather than the analyst's arbitrary choice of how to express a given group. This property, termed \textit{objectivity} in the dimensional-learning literature~\cite{zhang2025mutual}, distinguishes MI from gradient-based methods~\cite{constantine2017data} (where rescaling variables changes the gradient covariance) and regression-based methods (where the choice of basis functions influences the selection).

\subsection{Mutual-information estimation and adaptive count selection}\label{sec:si_mi_estimation}

In practice, the probability densities entering the MI definition are generally unknown, and MI must therefore be estimated from finite samples $\{(x_i,y_i)\}_{i=1}^{N}$. AI Buckingham employs either a Kraskov--St\"ogbauer--Grassberger (KSG)-type nearest-neighbor estimator~\cite{kraskov2004estimating} or a histogram-based estimator.

\textbf{Nearest-neighbor estimation.} Let $\rho_i$ denote the Chebyshev distance from sample $i$ to its $\omega$-th other sample in the joint space $(X,Y)$, and define $r_i=\max(\rho_i,10^{-10})$. Let $n_{x,i}$ and $n_{y,i}$ denote the numbers of samples within the corresponding closed marginal balls of radius $r_i$, including sample $i$ itself. The estimator implemented in AI Buckingham is
\begin{equation}
\hat{I}_{\mathrm{KSG}}(X; Y) = \psi(\omega) + \psi(N) - \frac{1}{N}\sum_{i=1}^{N}\bigl[\psi(n_{x,i}) + \psi(n_{y,i})\bigr],
\label{eq:ksg_estimator}
\end{equation}
where $\psi(x) = \Gamma'(x)/\Gamma(x)$ is the digamma function, $N$ is the sample size, and $\omega$ is the number of neighbors.

\textbf{Histogram-based estimation.} Each coordinate is divided into $b$ equal-width bins over its observed range. The empirical probabilities are used to evaluate
\begin{equation}
\hat{I}_{\mathrm{hist}}(X;Y)=\widehat{H}_b(X)+\widehat{H}_b(Y)-\widehat{H}_b(X,Y),\quad \widehat{H}_b=-\sum_{j:p_j>0}p_j\log p_j
\label{eq:hist_estimator}
\end{equation}
where $p_j$ denotes the empirical probability associated with an occupied bin.

\textbf{Key properties.}
\begin{itemize}
    \item \textbf{Adaptive resolution.} For KSG estimation, the nearest-neighbor distance automatically adapts to the local sample density: dense regions are resolved over smaller neighborhoods, whereas sparse regions are evaluated over larger neighborhoods. This avoids the bandwidth-selection problem of kernel-density methods.

    \item \textbf{Chebyshev metric and inclusive counting.}
    The $L_\infty$ metric allows a common neighborhood radius to be used consistently in the joint and marginal spaces. The implementation uses closed marginal balls and includes sample $i$ itself in the counts $n_{x,i}$ and $n_{y,i}$.

    \item \textbf{Automatic estimator selection.}
    Unless an estimator is explicitly prescribed, AI Buckingham uses the KSG estimator for $N<800$ and the histogram-based estimator for $N\geq800$, with default settings $\omega=5$ and $b=40$, respectively. When a histogram-based calculation is applied to a local subset containing fewer than 800 samples, the implementation switches to KSG for that subset.

    \item \textbf{Consistent candidate comparison.}
    Within each comparison of candidate dimensionalities, the MI estimator and its numerical parameter are held fixed so that changes in MI reflect changes in the candidate dimensionless representation rather than changes in the estimator configuration.
\end{itemize}

In dimensionless learning, MI is evaluated between candidate dimensionless groups and the dimensionless dependent variable. For Type I similarity, $Y=\Pi_y$; for Type II similarity, $Y=\widetilde\Pi_y$. For a candidate representation containing $k$ dimensionless groups, define
$\Pi_k(\boldsymbol{\Gamma}_k)=[\Pi(\boldsymbol{\gamma}_1),\ldots,\Pi(\boldsymbol{\gamma}_k)]$, where
$\boldsymbol{\Gamma}_k=[\boldsymbol{\gamma}_1,\ldots,\boldsymbol{\gamma}_k] \in\mathbb{R}^{n\times k}$. The selected MI estimator is then applied in the corresponding multivariate joint space.

\textbf{Adaptive count selection.}

A critical question in dimensional analysis is: \textit{how many dimensionless groups are needed?} Classical Buckingham analysis provides an upper bound $n = p - \mathrm{rank}(\mathbf{D})$ but gives no guidance on how many of these $n$ groups are \textit{relevant} for predicting the dependent variable. 
AI Buckingham therefore evaluates candidate representations sequentially from $k=1$ up to the admissible dimension $n$. The optimized mutual information for candidate count $k$ is denoted by
\begin{equation}
\hat{I}_k
=
\max
\hat{I}\left(
\Pi_k(\boldsymbol{\Gamma}_k);Y
\right).
\label{eq:id_max}
\end{equation}
where the maximization is performed over all free coefficients of the corresponding Type I or Type II representation. Building on the exponential dependence on R\'enyi mutual information in information-theoretic irreducible-error bounds~\cite{yuan2025dimensionless}, we map the optimized MI to a bounded, monotonically decreasing \textit{normalized information score},
\begin{equation}
\varepsilon_k = \exp\left[-\max(\hat{I}_k,0) \right],
\label{eq:normalized_information_score}
\end{equation}
Smaller values of $\varepsilon_k$ indicate stronger statistical dependence between the candidate dimensionless representation and the dependent variable. The clipping at zero ensures that $\varepsilon_k$ remains within $(0,1]$ when finite-sample MI estimates become slightly negative.

For $k\geq2$, the contribution of adding the $k$-th dimensionless group is quantified by the reduction in the normalized information score, $\Delta\varepsilon_k=\varepsilon_{k-1}-\varepsilon_k$. The search proceeds to the next candidate count when $\Delta\varepsilon_k \geq \Delta\varepsilon_{\min}$, with $\Delta\varepsilon_{\min}=0.08$ by default. If this threshold is not reached, the preceding representation is retained, giving $k^*=k-1$. This sequential criterion selects the most parsimonious dimensionless representation that captures the dominant predictive information.

\section{Semantic planning, workflow orchestration, and retrieval grounding}\label{sec:architecture}

\begin{figure}[!ht]
\centering
\begin{tikzpicture}[
>=Stealth,
font=\small,
mainbox/.style={
draw,
rounded corners,
fill=white,
align=center,
text width=66mm,
minimum height=8mm,
inner sep=3.5pt
},
modulebox/.style={
draw,
rounded corners,
fill=white,
align=center,
text width=31mm,
minimum height=13mm,
inner sep=3pt,
font=\footnotesize
},
outputbox/.style={
draw,
rounded corners,
fill=white,
align=center,
minimum height=10mm,
inner sep=3pt,
font=\footnotesize
},
kbbox/.style={
draw,
rounded corners,
dashed,
fill=white,
align=center,
text width=35mm,
minimum height=10mm,
inner sep=3pt,
font=\footnotesize
}
]


\node[align=center] (data) at (-1,0)
{Natural-language problem and dataset};


\node[mainbox] (problem) at (-1,-1.20)
{Problem formulation};

\node[mainbox] (check) at (-1,-2.30)
{Physical-consistency check};

\node[mainbox] (basis) at (-1,-3.40)
{Dimensionless-basis construction};

\node[mainbox] (plan) at (-1,-4.50)
{Analysis-protocol planning};

\draw[->] (data) -- (problem);
\draw[->] (problem) -- (check);
\draw[->] (check) -- (basis);
\draw[->] (basis) -- (plan);

\draw[->]
(check.west) -- ++(-0.85,0) |- (problem.west);

\node[
anchor=east,
align=center,
font=\footnotesize
] at ($(check.west)+(-0.9,0.35)$)
{structured\\correction};

\node[kbbox] (kbq) at (5,-2.30)
{Physical-quantity\\knowledge base};

\draw[dashed,->]
(kbq.west) -- (check.east);


\node[mainbox] (infer) at (-1,-5.70)
{Numerical inference};

\draw[->] (plan) -- (infer);

\node[modulebox] (regime) at (-4.8,-7.25)
{Physical-regime\\identification};

\node[modulebox] (similarity) at (-0.95,-7.25)
{Similarity-type\\screening};

\node[modulebox,text width=50mm] (learning) at (3.8,-7.25)
{MI-based dimensionless learning\\\& representation simplification};

\draw[->] (regime.east) -- (similarity.west);
\draw[->] (similarity.east) -- (learning.west);


\node[mainbox] (report) at (-1,-8.95)
{Synthesis and reporting};

\draw[->]
(infer.west) -- ++(-2.65,0) |- (report.west);

\node[outputbox,text width=55mm] (laws) at (-2.8,-10.15)
{Discovered dimensionless scaling laws};

\node[outputbox,text width=42mm] (visual) at (3.3,-10.15)
{Data-collapse visualization};

\node[kbbox] (kbr) at (5,-8.95)
{Dimensionless-relation\\knowledge base};

\draw[dashed,->]
(kbr.west) -- (report.east);


\begin{pgfonlayer}{background}

\coordinate (semNW) at (-7,-0.62);
\coordinate (semSE) at ( 7,-4.92);

\coordinate (numNW) at (-7,-5.08);
\coordinate (numSE) at ( 7,-8.20);

\coordinate (repNW) at (-7,-8.38);
\coordinate (repSE) at ( 7,-10.95);

\path[
draw=blue!45,
fill=blue!5,
rounded corners=3pt
] (semNW) rectangle (semSE);

\path[
draw=orange!55,
fill=orange!5,
rounded corners=3pt
] (numNW) rectangle (numSE);

\coordinate (subNW) at (-6.6,-6.40);
\coordinate (subSE) at ( 6.55,-8.00);

\path[
draw=orange!65,
fill=orange!13,
rounded corners=3pt
] (subNW) rectangle (subSE);

\path[
draw=green!45!black,
fill=green!5,
rounded corners=3pt
] (repNW) rectangle (repSE);

\coordinate (outNW) at (-6.6,-9.55);
\coordinate (outSE) at ( 6.55,-10.75);

\path[
draw=green!55!black,
fill=green!13,
rounded corners=3pt
] (outNW) rectangle (outSE);

\end{pgfonlayer}


\node[
anchor=west,
font=\scriptsize\bfseries,
text=blue!55!black,
fill=blue!5,
inner xsep=2pt,
inner ysep=1pt
] at (-7,-0.62)
{semantic planning and orchestration};

\node[
anchor=west,
font=\scriptsize\bfseries,
text=orange!70!black,
fill=orange!5,
inner xsep=2pt,
inner ysep=1pt
] at (-7,-5.08)
{numerical analysis};

\node[
anchor=west,
font=\scriptsize\bfseries,
text=orange!75!black,
fill=orange!13,
inner xsep=2pt,
inner ysep=1pt
] at (-5.5,-6.40)
{numerical modules};

\node[
anchor=west,
font=\scriptsize\bfseries,
text=green!45!black,
fill=green!5,
inner xsep=2pt,
inner ysep=1pt
] at (-7,-8.38)
{reporting and interpretation};

\end{tikzpicture}

\caption{
Functional architecture and information flow of AI Buckingham.
The three colored regions denote semantic planning and orchestration, numerical analysis, and reporting and interpretation, respectively; the nested orange region contains the principal numerical modules executed within numerical inference.
Dashed arrows denote retrieval from the physical-quantity and dimensionless-relation knowledge bases for pre-inference physical-consistency checking and post-inference physical interpretation, respectively.
}
\label{fig:macro}
\end{figure}
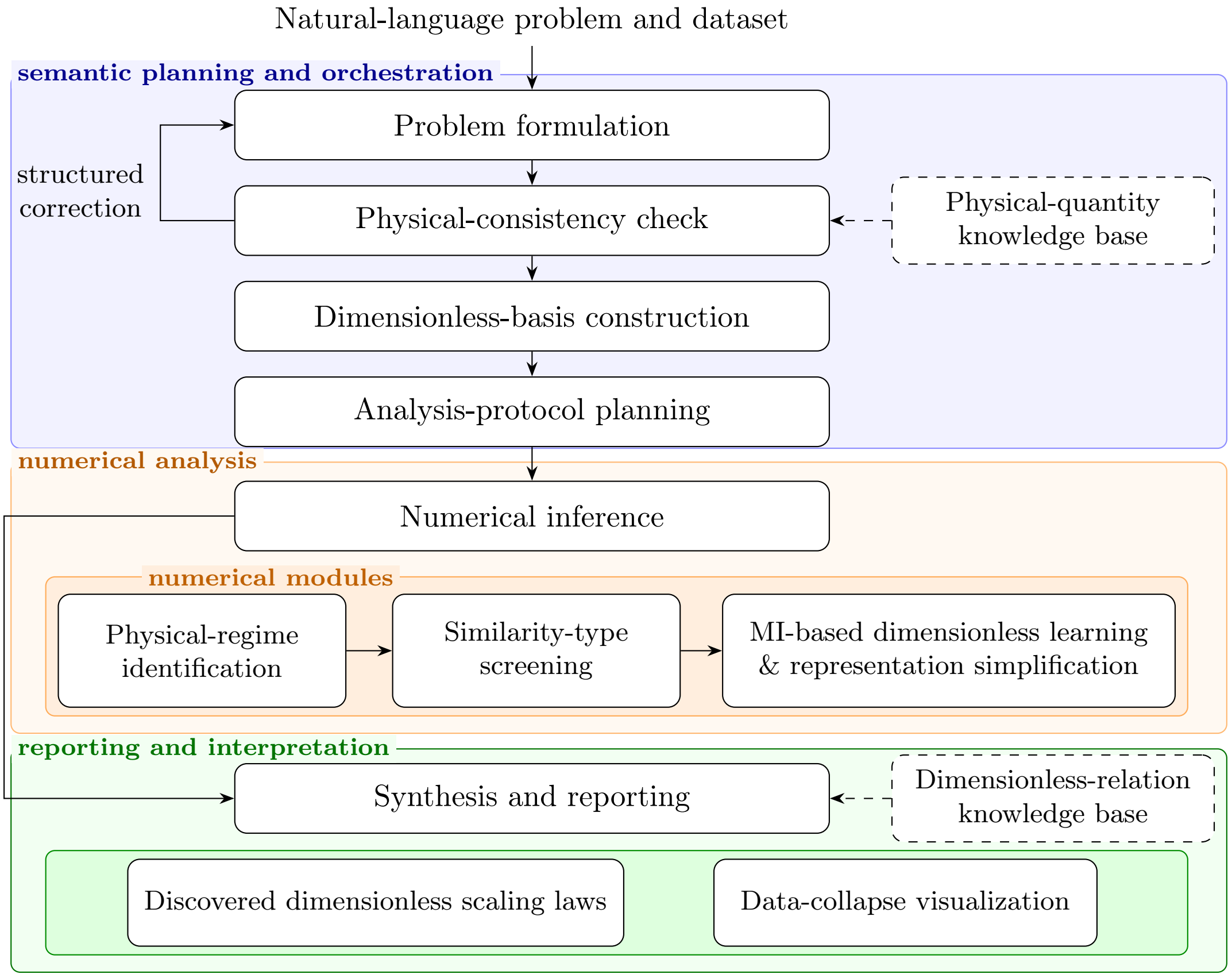

AI Buckingham combines LLM-based reasoning with predefined symbolic and numerical tools. The LLM handles tasks that require contextual reasoning, including problem formulation, workflow selection, and physical interpretation of the discovered relations, whereas dimensionless-basis construction, regime identification, mutual-information optimization, representation simplification, and validation are carried out by fixed computational routines. Supplementary Fig.~\ref{fig:macro} summarizes this division of responsibility and the overall information flow. All LLM interactions in the reported workflow used \texttt{gpt-5-mini}~\cite{gpt5_mini_250910} with temperature set to zero.

\subsection{Workflow stages and structured interfaces}

\begin{table}[!b]
\centering
\small
\setlength{\tabcolsep}{4pt}
\renewcommand{\arraystretch}{1.15}
\caption{Principal LLM interactions and tool interfaces in AI Buckingham.}
\label{tab:llm_interfaces}
\begin{tabular}{@{}
>{\raggedright\arraybackslash}p{34mm}
>{\raggedright\arraybackslash}p{55mm}
>{\raggedright\arraybackslash}p{65mm}
@{}}
\toprule
Interaction or step & Inputs & Outputs \\
\midrule

Problem formulation
& Problem description; dataset columns and preview; user constraints; revision feedback
& Dependent-variable selection with confidence and rationale; ordered governing quantities with physical meanings and candidate SI exponents \\
\cmidrule(lr){1-3}

Physical-consistency check
& Problem description; current quantity records; retrieved physical-quantity evidence
& Per-quantity decisions and reasons; candidate SI exponents under review and proposed corrections; support modes and cited record identifiers \\
\cmidrule(lr){1-3}

Basis construction (symbolic tool)
& Dimensional matrix $D$ and governing-variable order from accepted assignments
& Buckingham--$\Pi$ basis exponent vectors and corresponding dimensionless expressions \\
\cmidrule(lr){1-3}

Workflow planning
& Problem description; dataset summary; basis size and expressions; user constraints
& Regime-analysis and similarity-type policies; numerical settings; optional basis constraints \\
\cmidrule(lr){1-3}

Result interpretation
& Problem description; compiled numerical summary; retrieved dimensionless relations
& Physical interpretation of the discovered scalings and their connection to documented relations or physical analogues \\

\bottomrule
\end{tabular}
\end{table}

\begin{itemize}
\item \textbf{Problem formulation.} The LLM uses the natural-language problem description and a preview of the dataset to identify the governing physical quantities and the dimensionless dependent variable. It infers the physical meaning and candidate SI exponents of each governing quantity.

\item \textbf{Physical-consistency check.} Physical-quantity records are retrieved separately for each governing quantity through hybrid lexical--semantic matching and employed to check the proposed physical dimensions. Accepted assignments allow the workflow to proceed, while conflicting or unresolved assignments are returned to semantic parsing for revision.

\item \textbf{Dimensionless-basis construction.} The dimensional matrix $D$ is assembled from the accepted dimensional assignments. Symbolic null-space construction of $D$ yields the Buckingham--$\Pi$ basis defining the admissible search space.

\item \textbf{Analysis-protocol planning.} The structured plan specifies the regime-analysis and similarity-type policies, together with the numerical settings required by the downstream modules.

\item \textbf{Numerical inference.} The validated protocol is compiled into calls to fixed numerical routines that perform regime analysis, dimensionless learning and representation simplification according to the selected route and corresponding numerical criteria.

\item \textbf{Synthesis and reporting.} The completed numerical results are summarized in a final report, with retrieved dimensionless relations supporting the physical interpretation of the discovered scalings.
\end{itemize}

Supplementary Table~\ref{tab:llm_interfaces} summarizes the predefined inputs and returned information for the four types of LLM interaction and the symbolic basis tool. The dimensional assessment contains one record per governing quantity, with a decision of \texttt{PASSED}, \texttt{CONFLICT} or \texttt{UNCERTAIN}. The support mode is \texttt{KB\_RECORD}, \texttt{INTERNAL\_REASONING} or \texttt{UNCERTAIN}, and any cited record identifiers refer to the evidence supplied for that quantity. Subsequent checks determine the workflow outcomes \texttt{PASSED}, \texttt{RETRY} or \texttt{BLOCKED} shown in Supplementary Fig.~\ref{fig:rag_workflows}.

\subsection{Retrieval grounding and evidence flow}
\label{sec:retrieval_grounding}

\begin{table}[!b]
\centering
\small
\renewcommand{\arraystretch}{1.10}
\caption{The two knowledge bases used for retrieval grounding in AI Buckingham.}
\label{tab:rag_knowledge_bases}
\begin{tabular}{
>{\raggedright\arraybackslash}p{31mm}
>{\raggedright\arraybackslash}p{57mm}
>{\raggedright\arraybackslash}p{57mm}
}
\toprule
& Physical quantities & Dimensionless relations \\
\midrule

Coverage
& 247 records
& 251 records \\
\cmidrule(lr){1-3}

Record contents
& Quantity names, symbols and aliases; definitions; SI units and dimensional vectors; domain and ambiguity notes; sources
& Relation names and notation; defining expressions; physical interpretations; applicable domains; sources \\
\cmidrule(lr){1-3}

Lexical / semantic weights
& $0.65\,/\,0.35$
& $0.20\,/\,0.80$ \\
\cmidrule(lr){1-3}

Minimum hybrid score
& $0.22$
& No minimum-score cutoff \\
\cmidrule(lr){1-3}

Records retained
& Up to 3 per governing variable
& 5 highest-ranked records \\
\cmidrule(lr){1-3}

Evidence supplied to the LLM
& Up to 30 records per audit round, in governing-variable order
& All retained records \\

\bottomrule
\end{tabular}
\end{table}

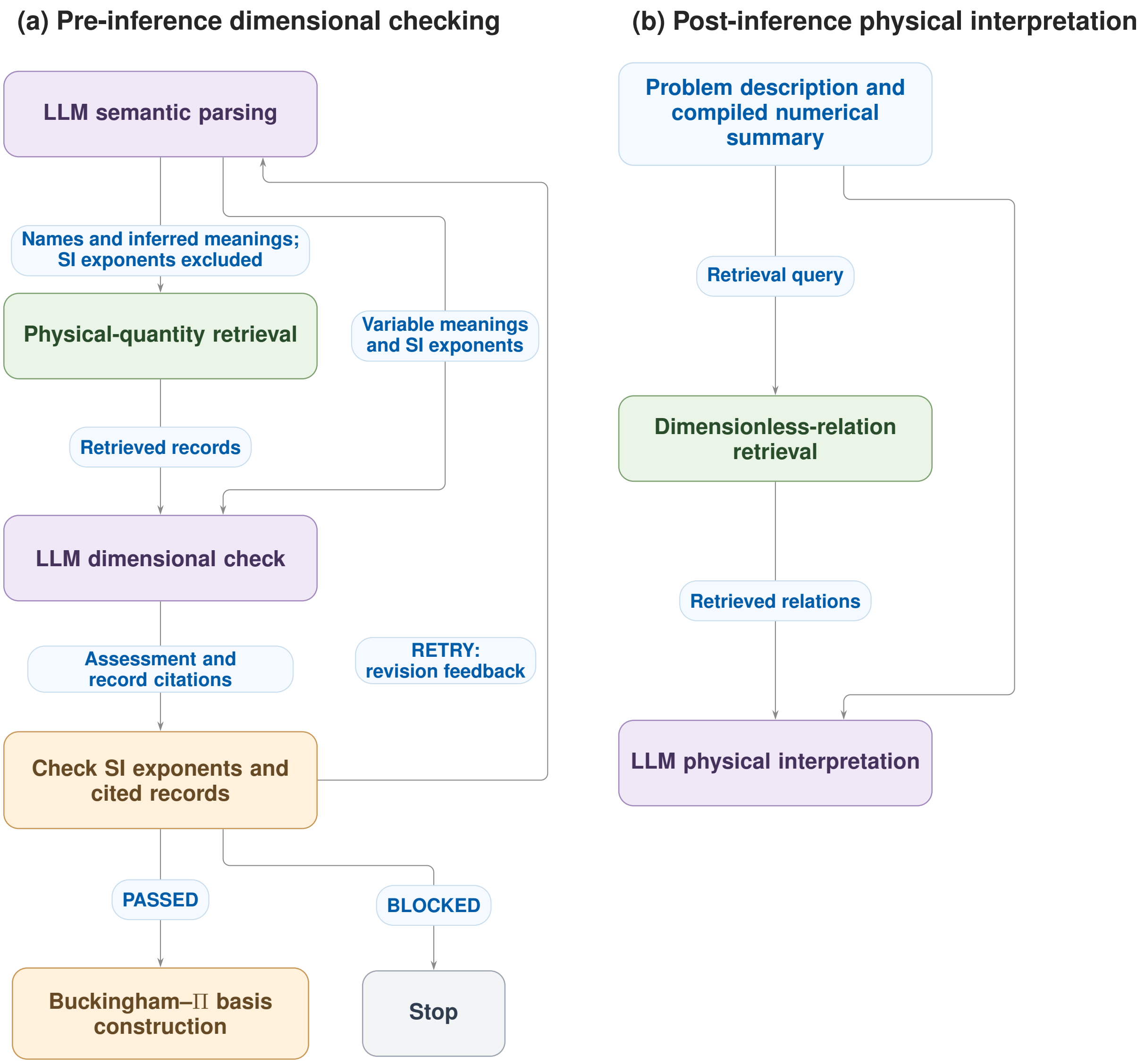
\begin{figure}[!t]
\centering
\resizebox{\linewidth}{!}{%
\definecolor{ragLLMfill}{HTML}{F0E8F7}%
\definecolor{ragLLMline}{HTML}{A38ABD}%
\definecolor{ragLLMtext}{HTML}{45325C}%
\definecolor{ragRetrievalfill}{HTML}{EAF5E7}%
\definecolor{ragRetrievalline}{HTML}{7EA171}%
\definecolor{ragRetrievaltext}{HTML}{294C2A}%
\definecolor{ragCheckfill}{HTML}{FFF1DB}%
\definecolor{ragCheckline}{HTML}{C99758}%
\definecolor{ragChecktext}{HTML}{654A25}%
\definecolor{ragGrayfill}{HTML}{F3F4F6}%
\definecolor{ragGrayline}{HTML}{9DA5B2}%
\definecolor{ragGraytext}{HTML}{2F3D4F}%
\definecolor{ragPillfill}{HTML}{F4FAFF}%
\definecolor{ragPillline}{HTML}{C9DDED}%
\definecolor{ragPilltext}{HTML}{005CA7}%
\begin{tikzpicture}[
    font=\fontfamily{phv}\selectfont\bfseries\fontsize{11}{12.5}\selectfont,
    box/.style={draw,rounded corners=2.6mm,align=center,minimum height=15mm,inner xsep=2mm,inner ysep=1.8mm,text width=51mm,line width=0.6pt},
    llm/.style={box,fill=ragLLMfill,draw=ragLLMline,text=ragLLMtext},
    retrieval/.style={box,fill=ragRetrievalfill,draw=ragRetrievalline,text=ragRetrievaltext},
    check/.style={box,fill=ragCheckfill,draw=ragCheckline,text=ragChecktext},
    neutral/.style={box,fill=ragGrayfill,draw=ragGrayline,text=ragGraytext},
    pill/.style={font=\fontfamily{phv}\selectfont\bfseries\fontsize{9.5}{10.5}\selectfont,draw=ragPillline,fill=ragPillfill,text=ragPilltext,rounded corners=3mm,minimum height=7mm,align=center,inner xsep=1.8mm,inner ysep=1mm,line width=0.5pt},
    arrow/.style={-{Stealth[length=1.7mm,width=1.1mm]},draw=black!46,line width=0.5pt,rounded corners=1.2mm},
    panel title/.style={align=left,anchor=west,font=\fontfamily{phv}\selectfont\bfseries\fontsize{13}{15}\selectfont,text=black!85}
]

\node[panel title] at (3.5mm,16mm) {(a) Pre-inference dimensional checking};

\node[llm] (parsing) at (30mm,0mm) {LLM semantic parsing};
\node[retrieval] (quantityretrieval) at (30mm,-39mm) {Physical-quantity retrieval};
\node[llm] (assessment) at (30mm,-78mm) {LLM dimensional check};
\node[check,minimum height=17mm] (validation) at (30mm,-117mm) {Check SI exponents and\\cited records};
\node[check,text width=48mm,minimum height=16mm] (basis) at (30mm,-158mm) {Buckingham--\scalebox{1.1}{\fontsize{10}{12}\selectfont$\Pi$} basis\\construction};
\node[neutral,text width=21mm,minimum height=15mm] (stop) at (78mm,-158mm) {Stop};

\draw[arrow] (parsing.south) -- (quantityretrieval.north);
\node[pill] at (30mm,-24mm) {Names and inferred meanings;\\SI exponents excluded};

\draw[arrow] (quantityretrieval.south) -- (assessment.north);
\node[pill] at (30mm,-58.5mm) {Retrieved records};

\draw[arrow] (assessment.south) -- (validation.north);
\node[pill,text width=43mm] at (30mm,-97.5mm) {Assessment and\\record citations};

\draw[arrow] ([xshift=11mm]parsing.south) -- (41mm,-18mm) -- (80mm,-18mm) -- (80mm,-66mm) -- (41mm,-66mm) -- ([xshift=11mm]assessment.north);
\node[pill] at (80mm,-39mm) {Variable meanings\\and SI exponents};

\draw[arrow] (validation.south) -- (basis.north);
\node[pill] at (30mm,-138mm) {PASSED};
\draw[arrow] ([xshift=11mm]validation.south) -- (41mm,-132mm) -- (78mm,-132mm) -- (stop.north);
\node[pill] at (78mm,-139mm) {BLOCKED};

\draw[arrow] (validation.east) -- (98mm,-117mm) -- (98mm,-12mm) -- (48mm,-12mm) -- ([xshift=18mm]parsing.south);
\node[pill,anchor=east] at (96mm,-96mm) {RETRY:\\ revision feedback};

\node[panel title] at (111.5mm,16mm) {(b) Post-inference physical interpretation};

\node[box,fill=ragPillfill,draw=ragPillline,text=ragPilltext,minimum height=18mm] (summary) at (138mm,0mm) {Problem description and\\compiled numerical\\summary};
\node[retrieval] (relationretrieval) at (138mm,-57mm) {Dimensionless-relation\\retrieval};
\node[llm] (interpretation) at (138mm,-114mm) {LLM physical interpretation};

\draw[arrow] (summary.south) -- (relationretrieval.north);
\node[pill] at (138mm,-28.5mm) {Retrieval query};
\draw[arrow] (relationretrieval.south) -- (interpretation.north);
\node[pill] at (138mm,-85.5mm) {Retrieved relations};

\draw[arrow] ([xshift=12mm]summary.south) -- (150mm,-15mm) -- (180mm,-15mm) -- (180mm,-102mm) -- (150mm,-102mm) -- ([xshift=12mm]interpretation.north);

\end{tikzpicture}%
}
\caption{Retrieval workflows in AI Buckingham: (a) pre-inference dimensional checking and (b) post-inference physical interpretation. Purple, green and orange boxes denote LLM operations, retrieval and computational operations, respectively. Blue text indicates the information passed between stages and routing outcomes.}
\label{fig:rag_workflows}
\end{figure}

Retrieval augmentation provides external physical knowledge, thereby grounding LLM reasoning in documented definitions and relations. AI Buckingham uses a physical-quantity knowledge base for dimensional checking and a dimensionless-relation knowledge base for interpreting discovered scalings. Their contents and retrieval settings are summarized in Supplementary Table~\ref{tab:rag_knowledge_bases}.

\begin{itemize}

\item \textbf{Hybrid lexical--semantic retrieval.} Each knowledge-base table row is treated as a complete retrieval record. Query and record descriptions are embedded using GLM-Embedding-3 and normalized to unit Euclidean norm. For a query $q$ and record $r$, the semantic score is
\begin{equation}
\label{eq:rag_semantic_score}
s_{\mathrm{sem}}(q,r)=\max\!\left(\widehat{\boldsymbol e}_q^{\,T}\widehat{\boldsymbol e}_r,0\right),
\end{equation}
where $\widehat{\boldsymbol e}_q$ and $\widehat{\boldsymbol e}_r$ are the normalized embeddings. The lexical score $s_{\mathrm{lex}}$ uses explicit name, symbol and alias matches together with contextual token overlap. For token comparison, symbols and separators are standardized, text is converted to lower case, and common stop words are removed. The resulting token sets $T_q$ and $T_r$ give the Jaccard overlap
\begin{equation}
\label{eq:rag_jaccard}
J(q,r)=\frac{|T_q\cap T_r|}{|T_q\cup T_r|}.
\end{equation}
The combined score is $s(q,r)=w_{\mathrm{lex}}s_{\mathrm{lex}}(q,r)+w_{\mathrm{sem}}s_{\mathrm{sem}}(q,r)$. Lexical matching receives greater weight for physical quantities because dimensional checking requires accurate identification of names and symbols. Semantic similarity receives greater weight for dimensionless relations because equivalent or related scalings may use different terminology.

\item \textbf{Pre-inference dimensional checking.} As shown in Supplementary Fig.~\ref{fig:rag_workflows}, the retrieval query for each governing physical quantity combines its exact dataset column name, the physical meaning inferred during semantic parsing and relevant excerpts from the problem description, excluding the inferred candidate SI exponents. The retrieved records are grouped by quantity to support the LLM's assessment of the assigned dimensions in the context of the physical problem, with the confirmed SI exponents or proposed corrections compared component by component with the dimensional vectors of the supporting records cited by the LLM. Up to two revisions are allowed by default after the initial formulation.

\item \textbf{Post-inference physical interpretation.} After numerical inference, the problem description and a compiled summary of the numerical results are used to retrieve documented dimensionless relations. The LLM uses the retrieved evidence to relate the discovered scalings to classical relations or relevant physical analogues.
\end{itemize}

\subsection{Tool interfaces and execution configuration}
\label{sec:tool_configuration}

\begin{itemize}
    \item \textbf{Dimensional-matrix assembly and basis construction.} The dimensional matrix $D$ is assembled computationally from the accepted SI exponents in governing-variable order. The nullity of $D$ is $n=p-\operatorname{rank}(D)$, where $p$ is the number of governing quantities. The workflow then supplies $D$ and the ordered variable names to a symbolic tool, which uses SymPy to compute the null-space exponent vectors and corresponding Buckingham--$\Pi$ basis parameters.
    \item \textbf{Protocol validation and configuration precedence.} The structured plan is validated against the permitted schema and compiled into arguments for the numerical modules. Optional fixed basis powers are checked against the available Buckingham--$\Pi$ coordinates, with output-side constraints applied only when Type II similarity is selected. Explicit user settings take precedence over planner-selected values, which in turn override predefined defaults.
    \item \textbf{Analysis policies and numerical modules.} Regime-analysis and similarity-type policies are specified independently. Automatic selections are determined from the data according to the numerical criteria in Section~\ref{sec:decisions}. Similarity-type screening is performed separately within each accepted regime, allowing different regimes to yield different similarity types. Supplementary Table~\ref{tab:planner_dispatch} summarizes the mapping between configuration groups and numerical modules.
\end{itemize}

\begin{table}[!htbp]
\centering
\small
\setlength{\tabcolsep}{4pt}
\renewcommand{\arraystretch}{1.15}
\caption{Configuration groups and corresponding numerical modules.}
\label{tab:planner_dispatch}
\begin{tabular}{@{}
>{\raggedright\arraybackslash}p{41mm}
>{\raggedright\arraybackslash}p{48mm}
>{\raggedright\arraybackslash}p{65mm}
@{}}
\toprule
Setting group & Numerical module & Configured behaviour \\
\midrule

Regime-analysis policy
& Workflow routing
& Prescribed single-regime or multi-regime analysis, or automatic regime assessment \\
\cmidrule(lr){1-3}

Partition settings
& Gradient-based regime identification
& Candidate cluster counts and regime-selection criteria \\
\cmidrule(lr){1-3}

Similarity-type policy
& Local similarity screening and inference
& Prescribed Type I or Type II, or automatic selection \\
\cmidrule(lr){1-3}

MI and optimization
& Dimensionless learning
& MI estimator, search settings, optimization restarts and sample requirements \\
\cmidrule(lr){1-3}

Basis constraints
& Constrained optimization
& Fixed powers in the Buckingham--$\Pi$ basis \\
\cmidrule(lr){1-3}

Simplification
& Representation simplification
& Sparsification, rationalization and their acceptance criteria \\

\bottomrule
\end{tabular}
\end{table}

\section{Numerical inference and implementation}
\label{sec:decisions}

This section details the numerical criteria governing physical-regime identification, similarity-type selection, dimensionless learning, simplification and validation. The principal decision thresholds are summarized in Supplementary Table~\ref{tab:numerical_criteria}.

\begin{table}[!t]
\centering
\small
\renewcommand{\arraystretch}{1.10}
\caption{Principal numerical criteria used in the data-dependent inference workflow.}
\label{tab:numerical_criteria}
\begin{tabular}{
>{\raggedright\arraybackslash}p{42mm}
>{\raggedright\arraybackslash}p{49mm}
>{\raggedright\arraybackslash}p{50mm}
}
\toprule
Numerical decision
& Criterion or diagnostic
& Role in the workflow \\
\midrule

Global gradient-alignment diagnostic
& $E_G=\lambda_1^{(G)}/\sum_j\lambda_j^{(G)}$
& Informs recommendations for single-regime treatment without determining partition selection \\
\cmidrule(lr){1-3}

Partition admissibility
& Eigenvalue ratio $\geq50$, fraction $\geq0.05$, and sample count $\geq60$ in every cluster
& Restricts the candidate partitions eligible for selection \\
\cmidrule(lr){1-3}

Directional separation
& $s_{\mathrm{dir}}<0.90$ for the candidate $K=2$ partition & Supports automatic selection of multiple-regime treatment when an admissible partition exists \\
\cmidrule(lr){1-3}

Automatic similarity-type screening
& $L_{\mathrm{out}}\leq0.15$
& Selects Type I; otherwise selects Type II \\
\cmidrule(lr){1-3}

Collinearity penalty
& Absolute cosine similarity $>0.70$
& Penalizes nearly redundant Buckingham--$\Pi$ coefficient directions \\
\cmidrule(lr){1-3}

Adaptive group-count selection
& $\Delta\varepsilon_k\geq0.08$
& Determines whether an additional dominant dimensionless group is retained \\
\cmidrule(lr){1-3}

Sparsification
& Physical-exponent threshold $0.05$ (max-normalized for input groups); information-score increase $\leq0.05$
& Identifies candidate weakly contributing physical variables for removal \\
\cmidrule(lr){1-3}

Rationalization
& Denominator $\leq5$; coefficient error $\leq0.05$; relative MI loss $\leq0.10$
& Accepts a simpler numerical representation when MI is sufficiently preserved \\
\cmidrule(lr){1-3}

Collapse validation
& Mean held-out $\overline{R^2}\geq0.85$
& Evaluates predictive collapse for the selected representation \\

\bottomrule
\end{tabular}
\end{table}

\subsection{Physical-regime identification and partition selection}
\label{sec:regimerules}

\begin{figure}[!t]
\centering
\resizebox{\linewidth}{!}{%
\definecolor{logicPolicyFill}{HTML}{F3F4F6}%
\definecolor{logicPolicyLine}{HTML}{9DA5B2}%
\definecolor{logicPolicyText}{HTML}{2F3D4F}%
\definecolor{logicCheckFill}{HTML}{FFF1DB}%
\definecolor{logicCheckLine}{HTML}{C99758}%
\definecolor{logicCheckText}{HTML}{654A25}%
\definecolor{logicInfoFill}{HTML}{F4FAFF}%
\definecolor{logicInfoLine}{HTML}{C9DDED}%
\definecolor{logicInfoText}{HTML}{005CA7}%
\begin{tikzpicture}[
    font=\fontfamily{phv}\selectfont\bfseries\fontsize{11}{12.5}\selectfont,
    base/.style={
        draw,
        rounded corners=2.3mm,
        align=center,
        minimum height=14mm,
        inner xsep=2mm,
        inner ysep=1.7mm,
        line width=0.6pt
    },
    policy/.style={
        base,
        fill=logicPolicyFill,
        draw=logicPolicyLine,
        text=logicPolicyText
    },
    check/.style={
        base,
        fill=logicCheckFill,
        draw=logicCheckLine,
        text=logicCheckText
    },
    result/.style={
        base,
        fill=white,
        draw=logicPolicyLine,
        text=logicPolicyText,
        minimum height=12mm
    },
    infolabel/.style={
        draw=logicInfoLine,
        fill=logicInfoFill,
        text=logicInfoText,
        rounded corners=2mm,
        align=center,
        minimum height=6mm,
        text height=1.5ex,
        text depth=0.3ex,
        inner xsep=1.5mm,
        inner ysep=0.7mm,
        line width=0.5pt,
        font=\fontfamily{phv}\selectfont\bfseries\fontsize{9.5}{10.5}\selectfont
    },
    mytitle/.style={
        font=\fontfamily{phv}\selectfont\bfseries\fontsize{13}{15}\selectfont
    },
    line/.style={
        draw=black!48,
        line width=0.55pt
    },
    arrow/.style={
        line,
        -{Stealth[length=1.8mm,width=1.15mm]}
    }
]

\node[mytitle] (root) at (100mm,11mm)
    {Regime-analysis policy};

\node[policy,text width=52mm] (single) at (30mm,-17mm)
    {Prescribed\\single regime};

\node[policy,text width=44mm] (multi) at (100mm,-17mm)
    {Prescribed\\multiple regimes};

\node[policy,text width=54mm] (auto) at (170mm,-17mm)
    {Automatic\\\mbox{regime assessment}};

\draw[line] (root.south) -- (100mm,0);
\draw[line] (30mm,0) -- (170mm,0);
\draw[arrow] (30mm,0) -- (single.north);
\draw[arrow] (100mm,0) -- (multi.north);
\draw[arrow] (170mm,0) -- (auto.north);

\node[result,text width=56mm] (whole) at (30mm,-48mm)
    {\mbox{Dimensionless learning}\\
     \mbox{on the complete dataset}};

\draw[arrow] (single.south) -- (whole.north);

\node[check,text width=68mm] (evaluate) at (126mm,-48mm)
    {\mbox{Evaluate candidate $K$;}\\
     check admissibility and record $E_G$};

\draw[line] (multi.south) -- (100mm,-33mm);
\draw[line] (auto.south) -- (170mm,-33mm);
\draw[line] (100mm,-33mm) -- (170mm,-33mm);
\draw[arrow] (126mm,-33mm) -- (evaluate.north);

\node[check,text width=52mm] (admissible) at (126mm,-77mm)
    {Any admissible partition?};

\draw[arrow] (evaluate.south) -- (admissible.north);

\draw[line] (admissible.south) -- (126mm,-92mm);
\draw[line] (80mm,-92mm) -- (172mm,-92mm);
\draw[line] (80mm,-92mm) -- (80mm,-108mm);
\draw[line] (172mm,-92mm) -- (172mm,-108mm);

\node[infolabel] at (80mm,-98mm) {yes};
\node[infolabel] at (172mm,-98mm) {no};

\node[check,text width=56mm] (direction) at (88mm,-133mm)
    {Directional separation\\
     \mbox{sufficient for $K=2$?}};

\node[result,text width=38mm] (unresolved) at (147mm,-133mm)
    {Unresolved\\assessment};

\node[result,text width=54mm] (multiple) at (30mm,-174mm)
    {Multiple regimes;\\
     \mbox{smallest admissible $K$}};

\node[result,text width=38mm] (global) at (178mm,-174mm)
    {Single-regime\\treatment};

\draw[line] (80mm,-108mm) -- (88mm,-108mm);
\draw[line] (80mm,-108mm) -- (30mm,-108mm);
\draw[line] (30mm,-108mm) -- (30mm,-161mm);
\draw[arrow] (88mm,-108mm) -- (direction.north);

\node[infolabel] at (30mm,-116mm) {Prescribed};
\node[infolabel] at (88mm,-116mm) {Automatic};

\draw[line] (172mm,-108mm) -- (147mm,-108mm);
\draw[arrow] (147mm,-108mm) -- (unresolved.north);

\draw[line] (172mm,-108mm) -- (178mm,-108mm);
\draw[line] (178mm,-108mm) -- (178mm,-161mm);

\node[infolabel] at (147mm,-116mm) {Prescribed};
\node[infolabel] at (178mm,-116mm) {Automatic};

\draw[line] (direction.south) -- (88mm,-146mm);
\draw[line] (64mm,-146mm) -- (112mm,-146mm);

\draw[line] (64mm,-146mm) -- (64mm,-161mm) -- (30mm,-161mm);
\draw[line] (112mm,-146mm) -- (112mm,-161mm) -- (178mm,-161mm);

\node[infolabel] at (64mm,-153mm) {yes};
\node[infolabel] at (112mm,-153mm) {no};

\draw[arrow] (30mm,-161mm) -- (multiple.north);
\draw[arrow] (178mm,-161mm) -- (global.north);

\pgfresetboundingbox
\path[use as bounding box]
    (-1mm,-182mm) rectangle (203mm,18mm);

\end{tikzpicture}%
}
\caption{Policy-dependent decision logic for physical-regime identification.}
\label{fig:regimetree}
\end{figure}
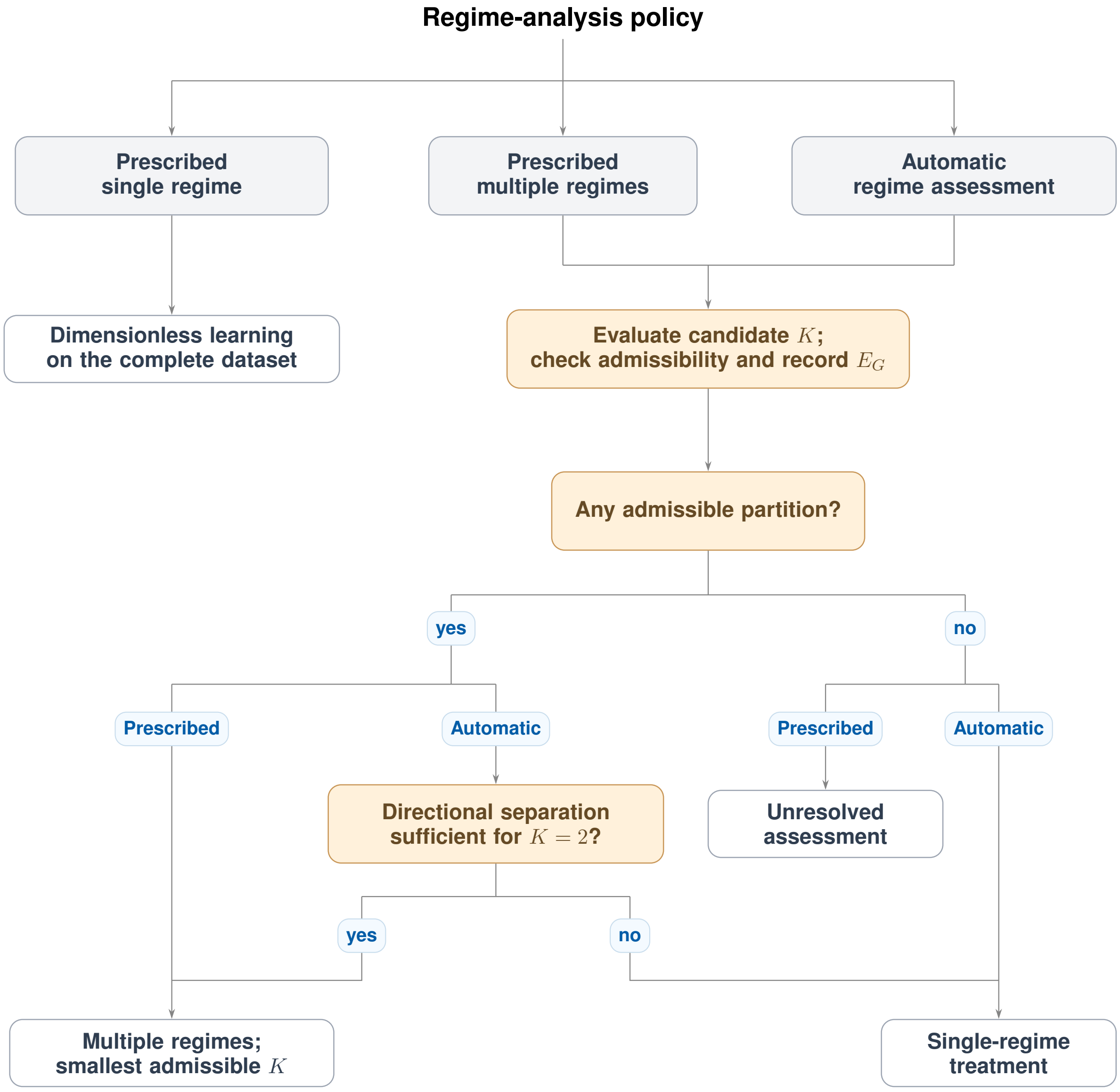

Physical-regime identification uses the analytical GPR posterior-mean gradients from Section~\ref{subsec:gpr_gradients}. For each candidate $K$, clustering alternates between assigning samples according to eigenvalue-weighted gradient similarity and updating the cluster covariance matrices. Convergence is reached when the sample assignments remain unchanged. Each cluster covariance matrix is computed as the average of the outer products of the unit-normalized gradients within that cluster. Multiple initializations are compared at fixed $K$, and the resulting partitions are assessed using the criteria below. Supplementary Fig.~\ref{fig:regimetree} shows the policy-dependent decision logic.

\begin{itemize}
\item \textbf{Candidate multi-regime partitions.}
When multi-regime analysis is prescribed or automatic regime assessment is requested, all configured candidate partitions ($K=2,3,4$ by default) are evaluated. The global leading-eigenvalue fraction $E_G$ is recorded as a diagnostic for single-regime recommendations, not as a partition-selection rule. A converged partition is admissible only when every cluster $\Omega^{(I)}$ satisfies an eigenvalue-ratio criterion,
\begin{equation}
\frac{\lambda_1^{(I)}}{\lambda_n^{(I)}}\geq E,
\quad E=50,
\label{eq:regime_eigen_ratio}
\end{equation}
a minimum cluster-fraction criterion,
\begin{equation}
\frac{N_I}{N}\geq p_{\min},
\quad p_{\min}=0.05,
\label{eq:regime_cluster_fraction}
\end{equation}
and a minimum sample-count criterion,
\begin{equation}
N_I\geq N_{\min},
\quad N_{\min}=60,
\label{eq:regime_sample_count}
\end{equation}
where $\lambda_1^{(I)}\geq\cdots\geq\lambda_n^{(I)}$ are the eigenvalues of the normalized-gradient covariance matrix within cluster $I$, $N_I$ is the number of samples assigned to that cluster, and $N$ is the total number of samples included in the partition.

The eigenvalue-ratio criterion supplies a lower bound on the useful number of clusters by requiring the existence of dominant dimensionless quantities within each cluster. The minimum cluster-fraction criterion prevents partitions from being refined into clusters containing too few samples, while the minimum sample-count criterion further limits excessive subdivision of small datasets and reduces the risk of unstable subsequent MI estimation. Together, the latter two criteria constrain the upper bound of the admissible cluster count. These conditions therefore identify admissible values of $K$ rather than a unique physical regime count. Among admissible candidates, smaller $K$ identifies the principal physical regimes, whereas larger $K$ provides progressively more detailed subdivisions. For an accepted multiple-regime treatment, AI Buckingham selects the smallest admissible $K$ by default, thereby retaining the coarsest partition that satisfies all three criteria.

\item \textbf{Directional separation.}
Partition admissibility checks for dominant gradient directions within each cluster but does not establish whether the associated physical dependences differ across clusters. To distinguish single- from multiple-regime treatments, automatic regime assessment therefore compares the leading covariance eigenvectors $\mathbf{v}_1^{(A)}$ and $\mathbf{v}_1^{(B)}$ of the two clusters in the candidate $K=2$ partition. Their alignment in the orthonormalized logarithmic Buckingham--$\Pi$ coordinates $\mathbf{z}$ is quantified by $s_{\mathrm{dir}}=\left|(\mathbf{v}_1^{(A)})^{T}\mathbf{v}_1^{(B)}\right|$, with a default threshold $\tau_{\mathrm{dir}}=0.90$. This directional criterion is evaluated only for $K=2$, independently of partition admissibility.

\item \textbf{Policy-dependent execution.}
A prescribed single-regime policy bypasses clustering and uses the complete dataset. A prescribed multi-regime policy selects the smallest admissible $K$ or returns an unresolved assessment if none exists; the directional cosine, when available, is diagnostic only. 
Automatic assessment selects multiple regimes only when an admissible partition exists and $s_{\mathrm{dir}}<\tau_{\mathrm{dir}}$. If no candidate is admissible or $s_{\mathrm{dir}}\geq\tau_{\mathrm{dir}}$, single-regime learning is applied to the complete dataset. The assessment remains unresolved when the numerical evidence needed for selection is unavailable.
\end{itemize}

\subsection{Similarity-type screening}
\label{sec:similarityrules}

Similarity-type selection is performed separately within each accepted regime. The similarity type can be prescribed through natural-language instructions in the problem prompt, specifying complete (incomplete) similarity or excluding (requiring) dependent-variable rescaling for Type I (Type II). An output-side basis-power constraint also implies Type II unless automatic screening is requested. Type I is used by default when no similarity type, output-rescaling requirement or automatic screening is specified.

\begin{itemize}
\item \textbf{Automatic similarity-type screening.}
When automatic screening is explicitly requested, two preliminary $k=1$ calculations are performed within each accepted regime using the same local samples, the same Buckingham--$\Pi$ basis, and the same MI estimator. In the Type I calculation, the dimensionless dependent variable is left unchanged. In the Type II calculation, one input dimensionless group and a multiplicative rescaling of the dependent variable are optimized jointly.

Let $\widehat I_{\mathrm{I}}$ and $\widehat I_{\mathrm{II}}$ denote the optimized MI values from the two preliminary calculations. The relative MI loss caused by omitting dependent-variable rescaling is defined as
\begin{equation}
L_{\mathrm{out}}=
\frac{\max\!\left(\widehat I_{\mathrm{II}}-\widehat I_{\mathrm{I}},0\right)}
{\max\!\left(|\widehat I_{\mathrm{II}}|,10^{-12}\right)}.
\label{eq:similarity_screen}
\end{equation}
Type I is selected when $L_{\mathrm{out}}\leq0.15$; otherwise Type II is selected.

\item \textbf{Post-screening dimensionless learning.}
The preliminary comparison uses screening-specific settings without fixed input- or output-side basis-power constraints. After similarity-type selection, local dimensionless learning is restarted using the analysis settings and any applicable constraints. The screening coefficients are not retained as discovered dimensionless groups.
\end{itemize}

\subsection{Numerical optimization and collinearity penalty}
\label{sec:optimization}

For each candidate representation, numerical optimization acts on the free Buckingham--$\Pi$ coefficients, with any prescribed coefficients held fixed. A candidate with no free coefficient is evaluated directly; one free coefficient is handled by bounded scalar minimization, whereas two or more are optimized using the covariance matrix adaptation evolution strategy (CMA-ES)~\cite{hansen2003reducing}.

\begin{figure}[!t]
\centering
\begingroup
\resizebox{\linewidth}{!}{%
\def\cmaBackgroundShift{-12mm}%
\def\cmaFinalX{118}%
\def\cmaFinalY{31}%
\def\cmaEllipseOpacity{0.07}%

\definecolor{cmaText}{HTML}{2D3E4F}%
\definecolor{cmaAxis}{HTML}{3B4A57}%
\definecolor{cmaContourLine}{HTML}{C8D1DC}%
\definecolor{cmaShadeA}{HTML}{F4F6F9}%
\definecolor{cmaShadeB}{HTML}{EDF1F5}%
\definecolor{cmaShadeC}{HTML}{E6ECF2}%
\definecolor{cmaShadeD}{HTML}{DDE5EE}%
\definecolor{cmaShadeE}{HTML}{D2DCE8}%
\definecolor{cmaShadeF}{HTML}{C4D1E0}%
\definecolor{cmaShadeG}{HTML}{B8C7D8}%
\definecolor{cmaTeal}{HTML}{3A8E9D}%
\definecolor{cmaTealDark}{HTML}{2B6F7A}%
\definecolor{cmaMean}{HTML}{C76D49}%

\begin{tikzpicture}[
    x=1mm,
    y=1mm,
    font=\fontfamily{phv}\fontsize{9}{10.5}\selectfont,
    contour/.style={
        draw=cmaContourLine,
        line width=0.45pt
    },
    dist/.style={
        draw=cmaTeal,
        fill=cmaTeal,
        fill opacity=\cmaEllipseOpacity,
        draw opacity=1,
        line width=0.85pt,
        dash pattern=on 2.3pt off 1.6pt
    },
    cand/.style={
        draw=cmaContourLine,
        fill=white,
        line width=0.55pt
    },
    good/.style={
        draw=cmaTealDark,
        fill=cmaTealDark,
        line width=0.55pt
    },
    patharrow/.style={
        draw=cmaMean,
        line width=1.0pt,
        shorten <=1.4mm,
        shorten >=1.6mm,
        -{Stealth[length=1.7mm,width=1.2mm]}
    },
    meanmark/.style={
        draw=cmaMean,
        fill=cmaMean,
        line width=0.5pt
    },
    axis/.style={
        draw=cmaAxis,
        line width=0.6pt,
        -{Stealth[length=1.7mm,width=1.1mm]}
    },
    meanlabel/.style={
        text=cmaText,
        fill=none,
        font=\fontfamily{phv}\bfseries
             \fontsize{9}{10}\selectfont
    },
    smallnote/.style={
        text=cmaText,
        font=\fontfamily{phv}
             \fontsize{8}{9.2}\selectfont
    }
]

\begin{scope}
\clip (0.5,2) rectangle (150,50);

\begin{scope}[xshift=\cmaBackgroundShift]

\foreach \cx/\cy/\sx/\sy/\aa/\bb/\cc/\dd/\ee/\ff/\cmaLayerColor in {
    92/28/96/18.5/0.24/0.11/0.05/0.15/0.08/0.04/cmaShadeA,
    101/28/79/15.0/0.23/0.10/0.05/0.14/0.08/0.04/cmaShadeB,
    110/28/63/12.0/0.22/0.10/0.05/0.14/0.07/0.04/cmaShadeC,
    117/28.5/47/9.0/0.21/0.09/0.04/0.13/0.07/0.03/cmaShadeD,
    121.5/30.0/32/6.2/0.20/0.08/0.04/0.12/0.06/0.03/cmaShadeE%
}{
    \filldraw[contour,fill=\cmaLayerColor]
        plot[
            smooth cycle,
            variable=\t,
            domain=0:358,
            samples=180
        ]
        ({\cx+\sx*(cos(\t)
                  +\aa*cos(2*\t+18)
                  -\bb*sin(3*\t-12)
                  +\cc*cos(4*\t+8))},
         {\cy+\sy*(sin(\t)
                  +\dd*sin(2*\t-24)
                  +\ee*cos(3*\t+6)
                  -\ff*sin(4*\t))})
        -- cycle;
}

\end{scope}

\foreach \rx/\ry/\amp/\cmaLayerColor in {
    16/3.7/0.06/cmaShadeF,
    8/2.0/0.03/cmaShadeG%
}{
    \filldraw[contour,fill=\cmaLayerColor]
        plot[
            smooth cycle,
            variable=\t,
            domain=0:358,
            samples=180
        ]
        ({\cmaFinalX
           +\rx*(1+\amp*cos(3*\t)-0.025*sin(2*\t))*cos(\t)},
         {\cmaFinalY
           +\ry*(1+\amp*cos(3*\t)-0.025*sin(2*\t))*sin(\t)
           +0.08*\rx*(1+\amp*cos(3*\t)-0.025*sin(2*\t))*cos(\t)})
        -- cycle;
}

\filldraw[dist,rotate around={-14:(23,18)}]
    (23,18) ellipse[x radius=14mm,y radius=9.7mm];

\filldraw[dist,rotate around={10:(71,25)}]
    (71,25) ellipse[x radius=18mm,y radius=5.8mm];

\filldraw[
    dist,
    rotate around={14:(\cmaFinalX,\cmaFinalY)}
]
    (\cmaFinalX,\cmaFinalY)
    ellipse[x radius=8.6mm,y radius=3.1mm];

\draw[patharrow] (23,18) -- (71,25);
\draw[patharrow] (71,25) -- (\cmaFinalX,\cmaFinalY);

\foreach \px/\py in {
    10.5/26.3,12.8/14.0,14.0/10.8,15.5/28.2,
    16.6/17.8,17.4/8.0,18.0/22.1,19.6/27.5,
    20.8/11.3,21.7/6.9,22.3/24.6,24.0/14.8,
    25.0/9.2,26.3/20.0,28.7/6.8,29.0/25.0,
    30.1/11.5,31.2/19.6,33.0/16.1,34.4/12.6
}{
    \draw[cand] (\px,\py) circle[radius=0.68mm];
}

\foreach \px/\py in {
    18.2/20.6,21.2/14.0,23.4/22.9,28.0/18.0,31.0/12.8
}{
    \filldraw[good] (\px,\py) circle[radius=0.75mm];
}

\foreach \px/\py in {
    56.0/26.8,57.4/20.9,59.0/28.1,60.8/23.5,
    62.5/18.9,64.0/29.6,66.0/21.0,67.7/25.6,
    69.0/31.2,71.4/19.2,73.0/24.5,74.8/29.3,
    76.2/21.8,77.6/27.1,79.3/32.0,80.6/23.8,
    82.4/28.8,84.0/26.2
}{
    \draw[cand] (\px,\py) circle[radius=0.68mm];
}

\foreach \px/\py in {
    68.2/23.1,71.7/27.2,74.5/25.0,77.8/30.0,81.0/26.0
}{
    \filldraw[good] (\px,\py) circle[radius=0.75mm];
}

\foreach \px/\py in {
    111.5/29.1,112.4/33.0,113.8/28.0,114.7/31.6,
    115.8/34.0,116.8/27.9,117.8/30.6,118.8/33.7,
    119.9/28.8,120.8/31.9,121.9/34.2,122.9/29.6,
    123.7/32.3,124.5/35.0,125.4/30.5,126.1/33.0,
    127.0/36.1,128.2/31.6
}{
    \draw[cand] (\px,\py) circle[radius=0.68mm];
}

\foreach \px/\py in {
    117.2/31.2,119.6/30.1,121.2/32.8,123.8/31.4,125.8/33.5
}{
    \filldraw[good] (\px,\py) circle[radius=0.75mm];
}

\foreach \mx/\my in {
    23/18,
    71/25,
    \cmaFinalX/\cmaFinalY
}{
    \draw[meanmark]
        (\mx,\my+1.2)
        -- (\mx+1.2,\my)
        -- (\mx,\my-1.2)
        -- (\mx-1.2,\my)
        -- cycle;
}

\end{scope}

\node[meanlabel] at (25,31.5)
    {$\mathbf{m}^{(g_1)}$};

\node[meanlabel] at (72,37.0)
    {$\mathbf{m}^{(g_2)}$};

\node[meanlabel] at (\cmaFinalX,{\cmaFinalY+8.5})
    {$\mathbf{m}^{(g_3)}$};

\node[smallnote,anchor=north east] at (148,49)
    {$g_1<g_2<g_3$};

\draw[axis] (0,0) -- (151,0)
    node[anchor=north east,yshift=-1mm] {$\theta_1$};

\draw[axis] (0,0) -- (0,52)
    node[anchor=south] {$\theta_2$};

\draw[cand] (18,-8) circle[radius=0.68mm];
\node[anchor=west,text=cmaText] at (21,-8)
    {Sampled candidate};

\filldraw[good] (70,-8) circle[radius=0.75mm];
\node[anchor=west,text=cmaText] at (73,-8)
    {Selected candidate};

\draw[meanmark]
    (122,-6.8)
    -- (123.2,-8)
    -- (122,-9.2)
    -- (120.8,-8)
    -- cycle;

\node[anchor=west,text=cmaText] at (125,-8)
    {Distribution mean};

\end{tikzpicture}%
}
\endgroup
\caption{Schematic CMA-ES search in two free-coefficient coordinates, $\theta_1$ and $\theta_2$. Darker shading denotes lower objective values; dashed teal ellipses show equal-density contours of the sampling distributions, and arrows connect their means at selected iterations.}
\label{fig:cmaes_schematic}
\end{figure}
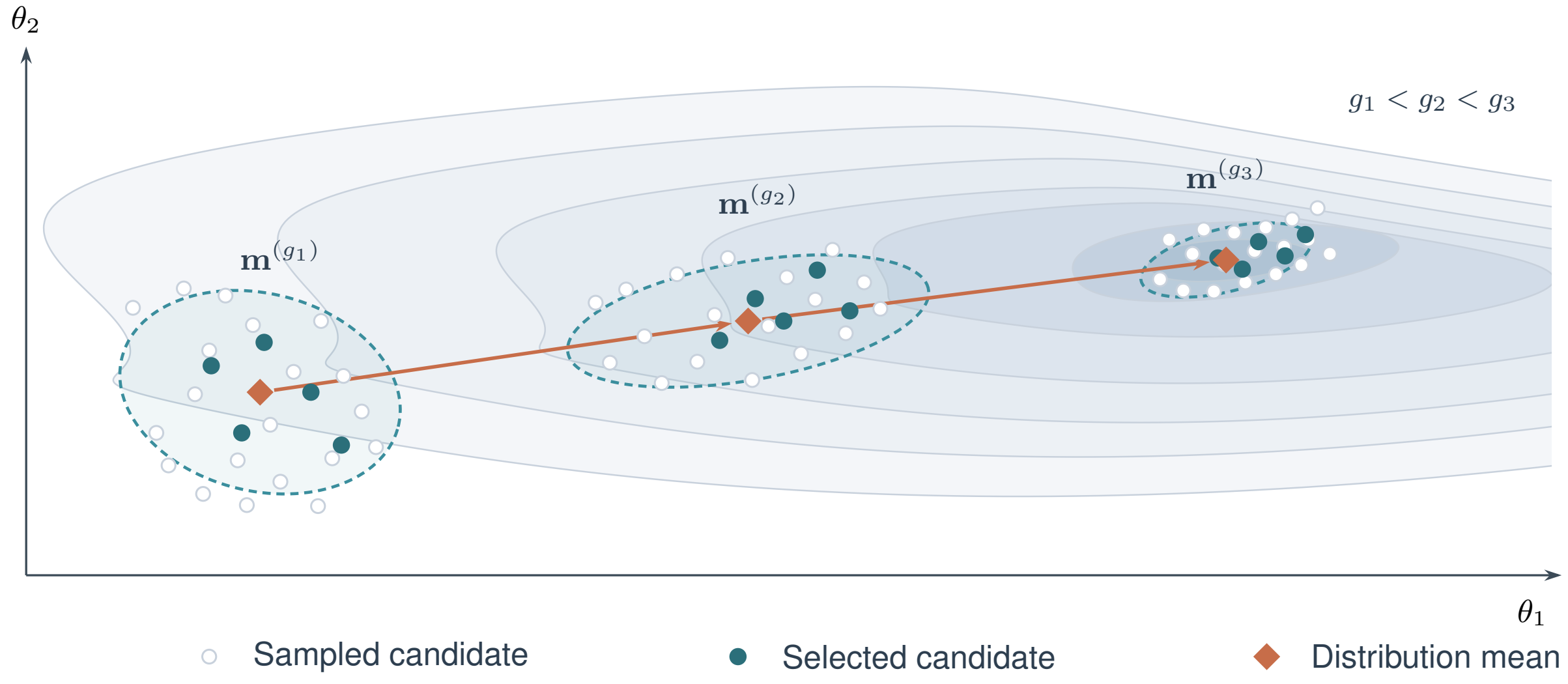

\begin{itemize}
\item \textbf{CMA-ES optimization.} The objective $f(\boldsymbol{\theta})=-\widehat I\!\left(\boldsymbol{\Pi}_k(\boldsymbol{\Gamma}_k);Y\right)$ is minimized to maximize the estimated statistical dependence between the candidate dimensionless groups and the dependent variable. Candidates that violate an enabled collinearity constraint or fail numerical evaluation receive large positive objective values instead. Each run terminates when an optimizer stopping criterion or the iteration limit is reached.

Let $\boldsymbol{\theta}\in\mathbb{R}^{d}$ denote the internal search coordinates associated with the free Buckingham--$\Pi$ coefficients. The following equations summarize the core CMA-ES updates before coefficient-bound handling. Supplementary Fig.~\ref{fig:cmaes_schematic} illustrates the evolution of the search distribution. At iteration $g$, the Gaussian sampling step generates
\begin{equation}
\boldsymbol{\theta}_r^{(g)}=\mathbf{m}^{(g)}+\sigma^{(g)}\mathbf{y}_r^{(g)},
\quad
\mathbf{y}_r^{(g)}\sim\mathcal{N}\!\left(\mathbf{0},\mathbf{C}^{(g)}\right),
\quad r=1,\ldots,\lambda,
\label{eq:cma_sampling}
\end{equation}
where $\mathbf{m}^{(g)}$ is the distribution mean, $\sigma^{(g)}$ is the global step size, $\mathbf{C}^{(g)}$ is the covariance matrix of the sampling steps, and $\lambda$ is the population size. The covariance of the trial vectors is $(\sigma^{(g)})^2\mathbf{C}^{(g)}$. Candidates are ranked in ascending order of objective value, and the best $\mu=\lfloor\lambda/2\rfloor$ candidates are assigned normalized logarithmic weights~\cite{hansen2016cma}:
\begin{equation}
w_i=
\frac{\ln\!\left((\lambda+1)/2\right)-\ln i}
{\sum_{j=1}^{\mu}\left[\ln\!\left((\lambda+1)/2\right)-\ln j\right]},
\quad i=1,\ldots,\mu.
\label{eq:cma_weights}
\end{equation}
The distribution mean is then updated as $\mathbf{m}^{(g+1)}=\sum_{i=1}^{\mu}w_i\boldsymbol{\theta}_{i:\lambda}^{(g)}$, where $\boldsymbol{\theta}_{i:\lambda}^{(g)}$ denotes the $i$th-ranked candidate. Define the step-size-normalized mean displacement and effective selection size as
\begin{equation}
\mathbf y_w^{(g)}
=
\frac{\mathbf m^{(g+1)}-\mathbf m^{(g)}}{\sigma^{(g)}}
=
\sum_{i=1}^{\mu}w_i\mathbf y_{i:\lambda}^{(g)},
\quad
\mu_{\mathrm{eff}}
=
\left(\sum_{i=1}^{\mu}w_i^2\right)^{-1}.
\label{eq:cma_mean_step}
\end{equation}

The covariance matrix $\mathbf C^{(g)}$ controls the orientation and relative spread of the search distribution. Successive mean displacements are accumulated in the covariance path,
\begin{equation}
\mathbf p_c^{(g+1)}
=
(1-c_c)\mathbf p_c^{(g)}
+
h_{\sigma}^{(g+1)}
\sqrt{c_c(2-c_c)\mu_{\mathrm{eff}}}\,
\mathbf y_w^{(g)},
\quad
\mathbf p_c^{(0)}=\mathbf 0,
\label{eq:cma_covariance_path}
\end{equation}
where $c_c$ controls the decay of earlier displacements. The indicator $h_{\sigma}^{(g+1)}$ is normally one but becomes zero when the normalized path $\mathbf p_{\sigma}$ defined below exceeds its prescribed length threshold. In compact form, the covariance update combines this history with the ranked sampling steps:
\begin{equation}
\mathbf C^{(g+1)}
=
a_g\mathbf C^{(g)}
+
c_1\mathbf p_c^{(g+1)}\bigl(\mathbf p_c^{(g+1)}\bigr)^T
+
c_{\mu}\sum_{i=1}^{\lambda}
\widetilde w_i^{(g)}
\mathbf y_{i:\lambda}^{(g)}
\bigl(\mathbf y_{i:\lambda}^{(g)}\bigr)^T.
\label{eq:cma_covariance}
\end{equation}
Here, $c_1$ and $c_{\mu}$ weight the history and current-population contributions, respectively. The coefficient $a_g$ collects the contributions multiplying the previous covariance, including compensation for the omitted path increment when $h_{\sigma}^{(g+1)}=0$. The effective covariance weights $\widetilde w_i^{(g)}$ assign positive contributions to better-ranked steps and negative contributions to poorly ranked steps.

The global step size $\sigma^{(g)}$ controls the overall sampling scale. Its adaptation uses a separate path that accumulates the same mean displacements after removing the directional scaling represented by $\mathbf C^{(g)}$:
\begin{equation}
\mathbf p_{\sigma}^{(g+1)}
=
(1-c_{\sigma})\mathbf p_{\sigma}^{(g)}
+
\sqrt{c_{\sigma}(2-c_{\sigma})\mu_{\mathrm{eff}}}\,
(\mathbf C^{(g)})^{-1/2}\mathbf y_w^{(g)},
\quad
\mathbf p_{\sigma}^{(0)}=\mathbf 0.
\label{eq:cma_stepsize_path}
\end{equation}
The core step-size update is
\begin{equation}
\sigma^{(g+1)}
=
\sigma^{(g)}
\exp\!\left[
\frac{c_{\sigma}}{d_{\sigma}}
\left(
\frac{\|\mathbf p_{\sigma}^{(g+1)}\|_2}{\chi_d}-1
\right)
\right],
\label{eq:cma_stepsize}
\end{equation}
where $\chi_d=\mathbb E_{\mathbf z\sim\mathcal N(\mathbf 0,\mathbf I_d)}[\|\mathbf z\|_2]$ is the expected Euclidean length of a $d$-dimensional standard normal vector, used as the random-selection reference. The step size increases when $\|\mathbf p_{\sigma}^{(g+1)}\|_2>\chi_d$ and decreases when it is smaller. The coefficient $c_{\sigma}$ controls the decay of earlier displacements, while $d_{\sigma}$ damps the step-size change. Detailed adaptation rules are given in~\cite{hansen2016cma}; adaptation parameters, numerical safeguards and coefficient-bound handling are supplied by the Python \texttt{cma} implementation.

\item \textbf{Collinearity penalty.} For candidate representations containing more than one input dimensionless group, nearly collinear Buckingham--$\Pi$ coefficient directions are penalized. For two coefficient vectors $\boldsymbol{\gamma}_i$ and $\boldsymbol{\gamma}_j$, their absolute cosine similarity is
\begin{equation}
c_{ij}=
\frac{|\boldsymbol{\gamma}_i^T\boldsymbol{\gamma}_j|}
{\|\boldsymbol{\gamma}_i\|_2\|\boldsymbol{\gamma}_j\|_2}.
\label{eq:direction_cosine}
\end{equation}
A penalty is applied when $c_{ij}>c_{\max}$. For Type II similarity, the same penalty can also be applied between an input-group direction and the dependent-variable rescaling direction. With prescribed basis coefficients, comparisons between input groups use only the free components. For comparisons with the dependent-variable rescaling vector, a configurable switch determines whether all basis coefficients are included or only those free in both vectors are used.
\end{itemize}

\subsection{Representation simplification}
\label{sec:simplification}

\begin{itemize}
\item \textbf{Sparsification.}
For each retained input dimensionless group, the physical-exponent vector $\mathbf w^{(j)}$ is normalized by its largest absolute component,
\begin{equation}
\widetilde{\mathbf w}^{(j)}
=
\frac{\mathbf w^{(j)}}{\max_r |w_r^{(j)}|},
\label{eq:sparsification_normalization}
\end{equation}
so that candidate screening is independent of the overall power used to represent the group. A physical variable is considered for removal only when its normalized exponent has magnitude below $0.05$ in every retained input group. For Type II similarity, its absolute exponent in the output-rescaling factor must also be below $0.05$, without normalization. Sparsification is disabled when fixed basis-power constraints are imposed.

Rather than directly setting optimized exponents to zero, sparsification restricts the search to a Buckingham--$\Pi$ subspace in which the candidate variables have zero physical exponents. Let $J$ index these variables and $\mathbf B$ denote the current basis matrix. A numerical null-space calculation constructs a matrix $\mathbf R$ whose rows span the coefficient space satisfying these constraints:
\begin{equation}
\mathbf R\mathbf B_{[:,J]}=\mathbf 0,
\qquad
\mathbf B_{\mathrm{restricted}}=\mathbf R\mathbf B.
\label{eq:restricted}
\end{equation}
The resulting basis $\mathbf B_{\mathrm{restricted}}$ represents dimensionless combinations that no longer involve these variables. The additional constraints may reduce the search-space dimension; if it falls below the retained group count $k^*$, the proposed removal is rejected. Otherwise, the representation is re-optimized at fixed $k^*$ and accepted only when
\begin{equation}
\varepsilon_{\mathrm{sparse}}-\varepsilon_{\mathrm{original}}\leq0.05.
\label{eq:sparsification_acceptance}
\end{equation}

\item \textbf{Rationalization.} Following the sparsification stage, the selected Buckingham--$\Pi$ basis coefficients are approximated using a bounded-denominator continued-fraction method, with a maximum denominator of $5$ by default and an absolute coefficient-error tolerance of $0.05$. Decimal rounding is employed when no rational approximation satisfies these conditions. The resulting representation is re-evaluated using the same MI estimator and accepted only when
\begin{equation}
\frac{
\max\!\left(\widehat I_{\mathrm{original}}-\widehat I_{\mathrm{rationalized}},0\right)
}{
\max\!\left(|\widehat I_{\mathrm{original}}|,10^{-12}\right)
}
\leq0.10,
\label{eq:rataccept}
\end{equation}
where $\widehat I_{\mathrm{original}}$ and $\widehat I_{\mathrm{rationalized}}$ denote the estimated MI before and after rationalization, respectively.
\end{itemize}

\subsection{Collapse validation and numerical uncertainty}
\label{sec:validation}
\begin{itemize}
\item \textbf{Collapse validation.} After group-count selection and any accepted simplification, the selected representation is evaluated using a separate GPR surrogate that is not involved in regime identification or dimensionless discovery. By default, the dimensionless groups and dependent variable, including the learned output rescaling for incomplete similarity, are logarithmically transformed after excluding samples with non-finite or non-positive values. The validation model uses a scaled anisotropic radial-basis-function kernel with an additive white-noise term. By default, three random 80/20 train--test splits are evaluated, and collapse validation is passed when the mean test-set score satisfies $\overline{R^2}\geq0.85$; under the default transformation, this score is computed in log-response space. Because the representation is selected before these splits, the reported score evaluates predictive collapse conditional on that representation, not out-of-sample performance of the complete discovery workflow.

\item \textbf{Numerical uncertainty.} Optimization variability is assessed through repeated runs with different random seeds. The best candidate from the designated primary-seed run is retained for subsequent analysis. Subsampling variability is assessed by re-evaluating MI on randomly selected data subsets while keeping the selected representation fixed. For both MI and the normalized information score, $\sigma_{\mathrm{restart}}$ and $\sigma_{\mathrm{subsample}}$ denote the respective standard deviations. They are retained separately, with the reported combined measure defined as
\begin{equation}
\sigma_{\mathrm{combined}}
=
\sqrt{\sigma_{\mathrm{restart}}^2+\sigma_{\mathrm{subsample}}^2}.
\label{eq:combined_uncertainty}
\end{equation}
These measures describe sensitivity to optimization randomness and data subsampling. For rationalized representations, no further coefficient optimization is performed: $\sigma_{\mathrm{restart}}$ is inherited from the preceding solution, whereas subsampling variability is recomputed.
\end{itemize}

\subsection{Numerical configuration and data sources}
\label{sec:implementation}

Supplementary Table~\ref{tab:defaults} summarizes the shared numerical settings; Table~\ref{tab:data} documents dataset sources and preparation; Table~\ref{tab:overrides} reports case-specific configurations and resolved policy outcomes. Variable definitions, dimensional assignments, and numerical ranges are provided in the individual case studies.

\begin{longtable}{
>{\raggedright\arraybackslash}p{38mm}
>{\raggedright\arraybackslash}p{72mm}
>{\raggedright\arraybackslash}p{40mm}
}
\caption{Shared numerical settings for the case studies. Regime-identification and collapse-validation GPR settings are listed separately because the two surrogates use different kernels and hyperparameter parameterizations.}
\label{tab:defaults}
\\

\toprule
Component & Parameter & Setting \\
\midrule
\endfirsthead

\caption[]{Shared numerical settings used in the case studies (continued).}\\
\toprule
Component & Parameter & Setting \\
\midrule
\endhead

\midrule
\multicolumn{3}{r}{\footnotesize Continued on next page}
\\
\endfoot

\bottomrule
\endlastfoot

Optimization
& Primary random seed; optimization runs per candidate
& 42; 3 \\

CMA-ES
& Initial coefficient value; initial step size
& 0.1; 0.5 \\

CMA-ES
& Default coefficient bounds
& $[-1,1]$ \\

CMA-ES
& Population size; maximum iterations
& 300; 30,000 \\

CMA-ES
& Position and objective convergence tolerances
& $10^{-8}$ \\

Scalar optimization
& Bounds for one free coefficient
& $[-2,2]$ \\

\cmidrule(lr){1-3}

MI estimation
& Automatic estimator policy
& KSG for $N<800$; histogram for $N\geq800$ \\

MI estimation
& Default KSG neighbor count; histogram bin count
& $\omega=5$; $b=40$ \\

Similarity-type screening
& Preliminary group count; maximum relative MI loss for Type I
& $k=1$; 0.15 \\

Adaptive group-count selection
& Minimum reduction in normalized information score
& $\Delta\varepsilon_k=0.08$ \\

Local inference
& Minimum local samples; nominal samples per MI dimension; minimum samples for the $k=2$ comparison
& 20; 20; 40 \\

Collinearity penalty
& Maximum absolute cosine similarity
& 0.70 \\

\cmidrule(lr){1-3}

Sparsification
& Physical-exponent threshold; maximum increase in normalized information score
& 0.05; 0.05 \\

Rationalization
& Maximum coefficient-approximation error
& 0.05 \\

Rationalization
& Maximum denominator $Q_{\max}$; maximum relative MI loss
& 5; 0.10 \\

\cmidrule(lr){1-3}

Automatic regime assessment
& Global leading-eigenvalue fraction
& Diagnostic only \\

Automatic regime assessment
& Directional-separation threshold
& $\tau_{\mathrm{dir}}=0.90$ \\

Regime partition
& Minimum cluster fraction; eigenvalue-ratio threshold; minimum samples per cluster
& 0.05; 50; 60 \\

Regime partition
& Candidate cluster counts; selection policy
& $\{2,3,4\}$; smallest admissible $K$ for an accepted multiple-regime treatment \\

Regime partition
& Random initializations; maximum clustering iterations
& 30; 200 \\

Regime partition
& Convergence requirement; gradient-norm cutoff
& required; $10^{-12}$ \\

\cmidrule(lr){1-3}

Regime GPR
& Kernel and default mean function
& isotropic squared-exponential; constant \\

Regime GPR
& Maximum training samples
& all available \\

Regime GPR
& Optimizer restarts; maximum optimizer iterations
& 3; 2000 \\

Regime GPR
& Bounds on log amplitude, log inverse length scale, and log noise variance
& $[-30,30]$; $[-30,30]$; $[-30,30]$ \\

Regime GPR
& Diagonal jitter
& $10^{-10}$ \\

Regime GPR
& Optimizer function and gradient tolerances
& $10^{-10}$; $10^{-6}$ \\

\cmidrule(lr){1-3}

Collapse-validation GPR
& Kernel
& Scaled anisotropic RBF $+$ white noise \\

Collapse-validation GPR
& Kernel-amplitude bounds; length-scale bounds; noise-level bounds
& $[10^{-3},10^{3}]$; $[10^{-3},10^{3}]$; $[10^{-8},10]$ \\

Collapse-validation GPR
& Diagonal regularization; optimizer restarts
& $10^{-10}$; 3 \\

Collapse validation
& Maximum validation samples; test fraction
& 800; 0.20 \\

Collapse validation
& Independent train--test splits
& 3 \\

Collapse validation
& Mean held-out $R^2$ threshold
& 0.85 \\

Numerical uncertainty
& Subsampling repetitions; nominal sample fraction
& 20; 0.50 \\

\end{longtable}

\begin{longtable}{
>{\raggedright\arraybackslash}p{28mm}
>{\centering\arraybackslash}p{10mm}
>{\raggedright\arraybackslash}p{41mm}
>{\raggedright\arraybackslash}p{62mm}
}
\caption{Data sources and preparation of the case-study datasets.}
\label{tab:data}
\\

\toprule
Case & $N$ & Source & Dataset preparation \\
\midrule
\endfirsthead

\caption[]{Data sources and preparation of the case-study datasets (continued).}\\
\toprule
Case & $N$ & Source & Dataset preparation \\
\midrule
\endhead

\midrule
\multicolumn{4}{r}{\footnotesize Continued on next page}\\
\endfoot

\bottomrule
\endlastfoot

MHD
& 483
& Glaws et al.; Yuan and Lozano-Dur\'an~\cite{glaws2016dimension,yuan2025dimensionless}
& Published numerical duct-flow benchmark; dimensionless mean velocity supplied. \\
\cmidrule(lr){1-4}

Pressure-gradient friction
& 144
& Dixit et al.; Yuan and Lozano-Dur\'an~\cite{dixit2024generalized,yuan2025dimensionless}
& Compiled experimental and numerical wall-friction observations with normalized flow variables. \\
\cmidrule(lr){1-4}

Rayleigh--B\'enard
& 182
& Xie et al.~\cite{xie2022data}
& Published convection measurements compiled for dimensional learning. \\
\cmidrule(lr){1-4}

Laser keyhole
& 90
& Xie et al.~\cite{xie2022data}
& Published high-speed X-ray measurements of laser-induced keyhole depth. \\
\cmidrule(lr){1-4}

Blasius
& 200
& Yuan and Lozano-Dur\'an~\cite{yuan2025dimensionless}
& Numerical integration of the Blasius equation. \\
\cmidrule(lr){1-4}

Sand-grain roughness & 104 & Forooghi et al.; Jouybari et al.; Flack and coauthors; Li et al. and Zhou et al.~\cite{forooghi2017universal,aghaeijouybari2021data,flack2016skin,barros2018measurements,flack2020skin,li2023orientation,li2023topology,zhou2024data} & Compiled experimental and numerical rough-surface observations. \\
\cmidrule(lr){1-4}

Compressible velocity
& 1109
& Trettel and Larsson; Modesti and Pirozzoli; Yuan and Lozano-Dur\'an~\cite{trettel2016mean,modesti2019direct,yuan2025dimensionless}
& Mean profiles from four channel and two pipe DNS cases. \\
\cmidrule(lr){1-4}

Goldenfeld
& 414
& Nikuradse; Goldenfeld; Sanfins et al.~\cite{nikuradse1950laws,goldenfeld2006roughness,sanfins2025similarity}
& Supplied rough-pipe dimensionless observations converted to randomized dimensional realizations. \\
\cmidrule(lr){1-4}

Cess
& 1000
& Zhang et al.~\cite{zhang2024clustering}
& Cess eddy-viscosity relation evaluated with stratified wall-normal sampling; no added noise. \\
\cmidrule(lr){1-4}

Rough circular pipe
& 1200
& Zhang et al.~\cite{zhang2024clustering}
& Piecewise laminar/Colebrook relation evaluated by rejection sampling; no added noise. \\

\end{longtable}

\begin{landscape}
\begin{table}[p]
\centering
\fontsize{8.6}{10.3}\selectfont
\setlength{\tabcolsep}{1.2pt}
\renewcommand{\arraystretch}{1.48}

\caption{Case-specific configurations and resolved policy outcomes.
``Auto $\rightarrow$ outcome'' denotes an automatic policy followed by its resolved outcome; ``prescr.'' denotes a prescribed policy, and ``Hist.'' denotes histogram-based MI estimation.
For CMA-ES, $m_0$ is the common component of the initial search mean, $\mathbf m^{(0)}=m_0\mathbf 1$, and $\sigma^{(0)}$ is the initial step size.
``Cos limit'' denotes the absolute cosine-similarity threshold $c_{\max}$ for Buckingham--$\Pi$ coefficient vectors. 
``All'' in the GPR-training-sample column indicates that all available samples are used to fit the regime-identification surrogate; otherwise, the listed value is the maximum training-sample count. The GPR-bound column lists the bounds on log amplitude, log inverse length scale, and log noise variance, in that order; ``shared'' refers to the settings in Supplementary Table~\ref{tab:defaults}.}
\label{tab:overrides}

\begin{tabular}{
@{}
>{\raggedright\arraybackslash}p{27mm}
>{\raggedright\arraybackslash}p{24mm}
>{\raggedright\arraybackslash}p{33mm}
>{\raggedright\arraybackslash}p{24mm}
>{\centering\arraybackslash}p{14mm}
>{\centering\arraybackslash}p{16mm}
>{\centering\arraybackslash}p{13mm}
>{\centering\arraybackslash}p{9mm}
>{\raggedright\arraybackslash}p{18mm}
>{\centering\arraybackslash}p{18mm}
>{\centering\arraybackslash}p{37mm}
@{}
}
\toprule

Case
& Regime
& \shortstack[l]{Similarity type}
& MI setting
& \multicolumn{2}{c}{CMA-ES}
& \shortstack{Cos limit}
& $Q_{\max}$
& \multicolumn{3}{c}{Regime GPR} \\

\cmidrule(lr){5-6}
\cmidrule(lr){9-11}

& & & &
$m_0;\,\sigma^{(0)}$
& Bounds
& &
& Mean
& \shortstack{Training samples\\($N$)}
& \shortstack{Log bounds\\
$\sigma_0^2/\sigma_1^2/\sigma_u^2$} \\

\midrule

MHD
& auto $\rightarrow$ single
& auto $\rightarrow$ Type I
& KSG ($\omega=4$)
& 0.1; 0.5
& $[-1,1]$
& 0.70
& 5
& constant
& all (483)
& shared \\

Pressure-gradient friction
& auto $\rightarrow$ single
& prescr. Type I
& KSG ($\omega=3$)
& 0.1; 0.5
& $[-1,1]$
& 0.70
& 5
& constant
& all (144)
& shared \\

Rayleigh--B\'enard
& auto $\rightarrow$ single
& auto $\rightarrow$ Type I
& KSG ($\omega=6$)
& 0.3; 0.5
& $[-1,1]$
& 0.70
& 4
& constant
& all (182)
& shared \\

Laser keyhole
& auto $\rightarrow$ single
& auto $\rightarrow$ Type I
& KSG ($\omega=4$)
& 0.1; 0.5
& $[-1,1]$
& 0.70
& 5
& constant
& all (90)
& shared \\

Blasius
& auto $\rightarrow$ single
& auto $\rightarrow$ Type I
& KSG ($\omega=5$)
& 0.6; 0.3
& $[-1,1]$
& 0.70
& 8
& constant
& 196 (200)
& shared \\

Sand-grain roughness 
& auto $\rightarrow$ single 
& prescr. Type I 
& KSG ($\omega=5$) 
& 0.1; 0.5 
& $[-1,1]$ 
& 0.70
& 5
& constant
& all (104)
& shared \\

Compressible velocity
& auto $\rightarrow$ single
& auto $\rightarrow$ Type II
& Hist. ($b=50$)
& 0.1; 0.2
& $[-1,1]$
& 0.80
& 5
& constant
& 1103 (1109)
& shared \\

Goldenfeld
& auto $\rightarrow$ single
& auto $\rightarrow$ Type II
& KSG ($\omega=5$)
& 0.1; 0.1
& $[-0.1,0.5]$
& 0.70
& 5
& constant
& all (414)
& shared \\

Cess
& auto $\rightarrow$ $K=3$
& auto $\rightarrow$ Type I (all)
& KSG ($\omega=5$)
& 0.1; 0.5
& $[-1,1]$
& 0.70
& 5
& constant
& 800 (1000)
& shared \\

Rough circular pipe
& auto $\rightarrow$ $K=2$
& auto $\rightarrow$ Type I (both)
& KSG ($\omega=5$) Hist. ($b=20$)
& 0.1; 0.5
& $[-1,1]$
& 0.70
& 5
& affine--sigmoid
& 1100 (1200)
& \shortstack{
$[-40,40]$ / $[-30,30]$ /\\
$[-30,5]$
} \\

\bottomrule
\end{tabular}

\end{table}
\end{landscape}

\section{Validation Case Details}\label{sec:cases}

This section provides the data preparation, dimensional formulations, and numerical evidence supporting the ten case studies in the main text.

\textbf{Notation and reporting conventions.}
Unless explicitly disabled, maximum-magnitude normalization is applied for presentation to the physical-variable exponents of input groups whose overall power scale is not fixed by prescribed powers. Type II dependent-variable rescaling exponents are not normalized, because the exponent of $\Pi_y$ is fixed to unity.

\textbf{Block A: regime assessment.}
Eigenvectors are expressed in the orthonormalized logarithmic Buckingham--$\Pi$ coordinates $\mathbf z$. The symbols $\checkmark$ and $\times$ mark satisfied and failed criteria, respectively: ratio, $\lambda_1^{(I)}/\lambda_n^{(I)}\geq50$; fraction, $N_I/N\geq0.05$; sample count, $N_I\geq60$.

\textbf{Block B: similarity-type screening.}
$\Pi_{1,\mathrm I}$ and $\Pi_{1,\mathrm{II}}$ denote the optimized $k=1$ input groups for the Type I and Type II screening comparison, respectively; $\widetilde\Pi_{y,\mathrm{II}}$ denotes the Type II rescaled dependent variable.

\textbf{Block C: dimensionless learning.}
$\Pi^{\mathrm{cand}}_{j|k}$ denotes the $j$th optimized input group for candidate count $k$, expressed in the case-specific basis $\Pi^{(q)}$. For Type II similarity, $\widetilde\Pi_{y|k}$ denotes the accompanying rescaled dependent variable. Here, $k^*$ denotes the retained group count, with $k^*=k-1$ when adding group $k\geq2$ yields a score reduction $\Delta\varepsilon_k=\varepsilon_{k-1}-\varepsilon_k<0.08$.

MI and the normalized information score $\varepsilon$ are reported as $a_{\pm\sigma_{\mathrm{combined}}}$, where $\sigma_{\mathrm{combined}}$ is defined in Eq.~\eqref{eq:combined_uncertainty}. The reported $\overline{R^2}$ is the mean held-out collapse-validation score; a dash denotes an unreported value.

\subsection{Case 1: Magnetohydrodynamic Power Generator}
\label{case:MHD}

In pressure-driven magnetohydrodynamic (MHD) duct flow, induced electric currents interact with a transverse magnetic field to produce a Lorentz force, so that the mean velocity reflects the balance among pressure forcing, viscous resistance, and electromagnetic effects~\cite{glaws2016dimension}. This case examines whether the mean-flow dependence can be reduced to a single dominant dimensionless group.

\noindent\textbf{Prompt.}
\begin{quote}\small\itshape
Perform dimensional analysis for a magnetohydrodynamics (MHD) generator to derive the dimensionless form of the average flow velocity (denoted as u*). 
This analysis uses ONLY the following 6 physical quantities as input parameters: 
the side length of the square duct (h), the dynamic viscosity ($\mu$), the fluid density ($\rho$), the electrical resistivity of the fluid ($\eta$), the applied pressure gradient (dp/dx), and the magnetic field magnitude (B).
\end{quote}

\noindent\textbf{Variable definitions.}

\begin{table}[!htb]
\centering
\setlength{\tabcolsep}{4pt}
\small
\caption{Physical descriptions, dimensional assignments, and numerical ranges for MHD duct flow.}
\label{tab:vars-MHD}
\begin{tabular}{@{}
>{\raggedright\arraybackslash}p{16mm}
>{\raggedright\arraybackslash}p{44mm}
>{\raggedright\arraybackslash}p{25mm}
>{\centering\arraybackslash}p{31mm}
>{\raggedright\arraybackslash}p{32mm}
@{}}
\toprule
Symbol & Description & SI Unit &
\shortstack{Dimensional vector\\$(M,L,T,I)$} & Range\\
\midrule
$h$ & Square-duct side length & $\mathrm m$ & $(0,1,0,0)$ & $1$\\
$\mu$ & Dynamic viscosity & $\mathrm{kg\,m^{-1}s^{-1}}$ & $(1,-1,-1,0)$ & $0.001003$--\allowbreak $0.009987$\\
$\rho$ & Mass density & $\mathrm{kg\,m^{-3}}$ & $(1,-3,0,0)$ & $0.1002$--\allowbreak $9.899$\\
$\eta$ & Electrical resistivity & $\Omega\,\mathrm m$ & $(1,3,-3,-2)$ & $0.1005$--\allowbreak $9.934$\\
$\mathrm{d}p/\mathrm{d}x$ & Positive driving pressure-gradient magnitude & $\mathrm{Pa\,m^{-1}}$ & $(1,-2,-2,0)$ & $0.1011$--\allowbreak $0.4985$\\
$B$ & Applied magnetic-flux density & $\mathrm T$ & $(1,0,-2,-1)$ & $0.1008$--\allowbreak $0.9815$\\
\midrule
$u^*$ & Dimensionless mean duct velocity & $1$ & $(0,0,0,0)$ & $5.185$--\allowbreak $4.811\times10^{4}$\\
\bottomrule
\end{tabular}
\end{table}

\noindent\textbf{Dimensional matrix and basis dimensionless parameters.}

\begin{equation}
\mathbf{D}=
\begin{array}{c|rrrrrr}
\! & h & \mu & \rho & \eta & \mathrm{d}p/\mathrm{d}x & B\\
\hline
M & 0 & 1 & 1 & 1 & 1 & 1\\
L & 1 & -1 & -3 & 3 & -2 & 0\\
T & 0 & -1 & 0 & -3 & -2 & -2\\
I & 0 & 0 & 0 & -2 & 0 & -1
\end{array}.
\end{equation}

The rank is 4 and the nullity is $n=2$. A basis is
\begin{equation}
\begin{aligned}
\Pi^{(1)}&=h^3\rho(\mathrm{d}p/\mathrm{d}x)/\mu^2,
&
\qquad
\Pi^{(2)}&=hB/(\mu\eta)^{1/2}.
\end{aligned}
\end{equation}

\begingroup
\small
\setlength{\tabcolsep}{2pt}
\setlength{\LTpre}{5pt}
\setlength{\LTpost}{5pt}
\setlength{\LTcapwidth}{\textwidth}
\renewcommand{\arraystretch}{1.00}
\begin{longtable}{@{}
>{\raggedright\arraybackslash}p{30mm}
>{\raggedleft\arraybackslash}p{30mm}
>{\raggedleft\arraybackslash}p{25mm}
>{\raggedright\arraybackslash}p{70mm}
@{}}
\caption{Numerical evidence and intermediate results for MHD duct flow.}
\label{tab:case-MHD}
\label{tab:partition-MHD}
\label{tab:metrics-MHD}
\label{tab:screen-MHD}
\\
\toprule
\endfirsthead
\caption[]{Numerical evidence and intermediate results for MHD duct flow (continued).}\\
\toprule
\endhead
\midrule
\endfoot
\bottomrule
\endlastfoot
\multicolumn{4}{@{}l}{\textbf{A. Regime assessment}}\\*
\multicolumn{4}{@{}l}{$N=483$; $E_G=0.998988$.}\\*
$(K,I)$ & $N_I\ (N_I/N)$ & $\lambda_1^{(I)}/\lambda_n^{(I)}$ & $\mathbf{v}_1^{(I)}$\\*
\multicolumn{4}{@{}l}{$K=2$ (ratio $\checkmark$, fraction $\checkmark$, sample count $\times$)}\\*
$(2,A)$ & $29\ (0.060)$ & $819.909$ & $\begin{smallmatrix}(-0.99725,\ -0.07410)^T\end{smallmatrix}$\\*
$(2,B)$ & $454\ (0.940)$ & $1799.875$ & $\begin{smallmatrix}(-0.99993,\ 0.01194)^T\end{smallmatrix}$\\
\cmidrule(lr){1-4}
\multicolumn{4}{@{}l}{$K=3$ (ratio $\checkmark$, fraction $\checkmark$, sample count $\times$)}\\*
$(3,A)$ & $26\ (0.054)$ & $854.007$ & $\begin{smallmatrix}(-0.99691,\ -0.07852)^T\end{smallmatrix}$\\*
$(3,B)$ & $398\ (0.824)$ & $7663.617$ & $\begin{smallmatrix}(-0.99999,\ 0.00398)^T\end{smallmatrix}$\\*
$(3,C)$ & $59\ (0.122)$ & $2284.212$ & $\begin{smallmatrix}(-0.99799,\ 0.06336)^T\end{smallmatrix}$\\
\cmidrule(lr){1-4}
\multicolumn{4}{@{}l}{$K=4$ (--: no usable partition)}\\*
$(4,--)$ & -- & -- & --\\
\midrule
\multicolumn{4}{@{}l}{$K=2$ direction: $s_{\mathrm{dir}}=0.996295\geq\tau_{\mathrm{dir}}=0.90$ ($\times$); selected $K=1$.}\\
\midrule
\multicolumn{4}{@{}l}{\textbf{B. Similarity-type screening}}\\*
Type and $\widehat I$ & \multicolumn{3}{l}{Optimized representation ($k=1$)}\\*
Type~I\newline $\widehat I=3.555557$ & \multicolumn{3}{l}{$\begin{aligned}\Pi_{1,\mathrm{I}}&=\bigl(\Pi^{(1)}\bigr)^{0.70507}\bigl(\Pi^{(2)}\bigr)^{-0.0112}\\\Pi_{y,\mathrm{I}}&=\Pi_y\end{aligned}$}\\*
Type~II\newline $\widehat I=3.640408$ & \multicolumn{3}{l}{$\begin{aligned}\Pi_{1,\mathrm{II}}&=\bigl(\Pi^{(1)}\bigr)^{0.89283}\bigl(\Pi^{(2)}\bigr)^{0.54196}\\\widetilde\Pi_{y,\mathrm{II}}&=\Pi_y\bigl(\Pi^{(1)}\bigr)^{0.17097}\bigl(\Pi^{(2)}\bigr)^{0.7344}\end{aligned}$}\\*
\multicolumn{4}{@{}l}{$L_{\mathrm{out}}=0.023308\leq0.15$; Type~I.}\\
\midrule
\multicolumn{4}{@{}l}{\textbf{C. Dimensionless learning}}\\*
Stage & $\widehat I$ & $\varepsilon$ & \multicolumn{1}{c}{$\overline{R^2}$}\\*
$k=1$ & $3.556_{\pm 0.041}$ & $0.029_{\pm 0.002}$ & \multicolumn{1}{c}{$0.999$}\\*
\multicolumn{4}{@{}p{\textwidth}@{}}{\hspace{1em}
$\Pi^{\mathrm{cand}}_{1|1}
=\bigl(\Pi^{(1)}\bigr)^{0.70507}
\bigl(\Pi^{(2)}\bigr)^{-0.0112}$}\\
\addlinespace[1pt]
$k=2$ & $3.557_{\pm 0.041}$ & $0.029_{\pm 0.002}$ & \multicolumn{1}{c}{$1.000$}\\*
\multicolumn{4}{@{}p{\textwidth}@{}}{\hspace{1em}
$\Pi^{\mathrm{cand}}_{1|2}
=\bigl(\Pi^{(1)}\bigr)^{-0.33923}
\bigl(\Pi^{(2)}\bigr)^{0.82285}$}\\
\multicolumn{4}{@{}p{\textwidth}@{}}{\hspace{1em}
$\Pi^{\mathrm{cand}}_{2|2}
=\bigl(\Pi^{(1)}\bigr)^{0.70507}
\bigl(\Pi^{(2)}\bigr)^{-0.01121}$}\\
\addlinespace[1pt]
Sparsification & $3.483_{\pm 0.047}$ & $0.031_{\pm 0.002}$ & \multicolumn{1}{c}{--}\\*
Rationalization & $3.394_{\pm 0.047}$ & $0.034_{\pm 0.003}$ & \multicolumn{1}{c}{$0.999$}\\
\midrule
\multicolumn{4}{@{}l}{$k^*=1$; $\Delta\varepsilon_2=3.566\times10^{-5}$.}\\*
\multicolumn{4}{@{}l}{Relative MI loss after rationalization: $0.026$.}\\
\end{longtable}
\endgroup

\noindent\textbf{Dominant dimensionless group.}

\begin{equation}
\Pi_1^*
=(\Pi^{(1)})^{1/3}
=\frac{h\rho^{1/3}(\mathrm{d}p/\mathrm{d}x)^{1/3}}{\mu^{2/3}},
\qquad
\Pi_y=u^*.
\end{equation}

The retained group is an invertible cube-root representation of the pressure--viscous parameter, with $B$ and $\eta$ removed from the learned representation during sparsification.

\clearpage
\subsection{Case 2: Skin Friction Under Pressure Gradients}
\label{case:PRESGRAD}

Streamwise pressure gradients modify turbulent boundary-layer profiles and wall friction. This case examines whether integral-profile quantities can be combined into a single dominant dimensionless coordinate across different pressure-gradient conditions~\cite{dixit2024generalized}.

\noindent\textbf{Prompt.}
\begin{quote}\small\itshape
Perform dimensional analysis for turbulent boundary layers under pressure gradient effects to derive the dimensionless form of the friction velocity (denoted as U\_tau/U\_inf). 
This analysis uses ONLY the following 6 physical quantities as input parameters: 
the free-stream velocity (U\_inf), the kinematic viscosity (nu), the integrated mean-flow kinetic energy (MKE, defined by the integral: $\int$(0 to delta) U\textasciicircum{}2 dy), the boundary layer thickness at 99\% of free-stream (delta), the boundary layer displacement thickness (delta*), and the boundary layer momentum thickness (theta).
\end{quote}

\noindent\textbf{Variable definitions.}

\begin{table}[!h]
\centering
\setlength{\tabcolsep}{4pt}
\small
\caption{Physical descriptions, dimensional assignments, and numerical ranges for pressure-gradient wall friction.}
\label{tab:vars-PRESGRAD}
\begin{tabular}{@{}
>{\raggedright\arraybackslash}p{16mm}
>{\raggedright\arraybackslash}p{44mm}
>{\raggedright\arraybackslash}p{25mm}
>{\centering\arraybackslash}p{31mm}
>{\raggedright\arraybackslash}p{32mm}
@{}}
\toprule
Symbol & Description & SI Unit &
\shortstack{Dimensional vector\\$(M,L,T)$} & Range\\
\midrule
$U_\infty$ & Free-stream velocity & $\mathrm{m\,s^{-1}}$ & $(0,1,-1)$ & $1$\\
$\nu$ & Kinematic viscosity & $\mathrm{m^2s^{-1}}$ & $(0,2,-1)$ & $4.989\times10^{-8}$--\allowbreak $0.0005515$\\
$M$ & Integrated mean-flow kinetic energy, $M=\int_0^\delta U^2\,\mathrm{d}y$ & $\mathrm{m^3s^{-2}}$ & $(0,3,-2)$ & $0.5157$--\allowbreak $0.9188$\\
$\delta$ & Boundary-layer thickness & $\mathrm m$ & $(0,1,0)$ & $1$\\
$\delta^*$ & Displacement thickness & $\mathrm m$ & $(0,1,0)$ & $0.04579$--\allowbreak $0.3507$\\
$\theta_m$ & Momentum thickness & $\mathrm m$ & $(0,1,0)$ & $0.03541$--\allowbreak $0.2571$\\
\midrule
$u_\tau/U_\infty$ & Normalized friction velocity & $1$ & $(0,0,0)$ & $0.02034$--\allowbreak $0.0557$\\
\bottomrule
\end{tabular}
\end{table}

\noindent\textbf{Dimensional matrix and basis dimensionless parameters.}

\begin{equation}
\mathbf{D}=
\begin{array}{c|rrrrrr}
\! & U_\infty & \nu & M & \delta & \delta^* & \theta_m\\
\hline
L & 1 & 2 & 3 & 1 & 1 & 1\\
T & -1 & -1 & -2 & 0 & 0 & 0
\end{array}.
\end{equation}

The rank is 2 and the nullity is $n=4$. A basis is
\begin{equation}
\begin{aligned}
\Pi^{(1)}&=M/(U_\infty\nu),
&
\Pi^{(2)}&=U_\infty\delta/\nu,
&
\Pi^{(3)}&=U_\infty\delta^*/\nu,
&
\Pi^{(4)}&=U_\infty\theta_m/\nu.
\end{aligned}
\end{equation}

\begingroup
\small
\setlength{\tabcolsep}{2pt}
\setlength{\LTpre}{5pt}
\setlength{\LTpost}{5pt}
\setlength{\LTcapwidth}{\textwidth}
\renewcommand{\arraystretch}{1.00}
\begin{longtable}{@{}
>{\raggedright\arraybackslash}p{30mm}
>{\raggedleft\arraybackslash}p{30mm}
>{\raggedleft\arraybackslash}p{25mm}
>{\raggedright\arraybackslash}p{70mm}
@{}}
\caption{Numerical evidence and intermediate results for pressure-gradient wall friction.}
\label{tab:case-PRESGRAD}
\label{tab:partition-PRESGRAD}
\label{tab:metrics-PRESGRAD}
\\
\toprule
\endfirsthead
\caption[]{Numerical evidence and intermediate results for pressure-gradient wall friction (continued).}\\
\toprule
\endhead
\midrule
\endfoot
\bottomrule
\endlastfoot
\multicolumn{4}{@{}l}{\textbf{A. Regime assessment}}\\*
\multicolumn{4}{@{}l}{$N=144$; $E_G=0.984287$.}\\*
$(K,I)$ & $N_I\ (N_I/N)$ & $\lambda_1^{(I)}/\lambda_n^{(I)}$ & $\mathbf{v}_1^{(I)}$\\*
\multicolumn{4}{@{}l}{$K=2$ (ratio $\checkmark$, fraction $\checkmark$, sample count $\checkmark$)}\\*
$(2,A)$ & $81\ (0.562)$ & $42065.973$ & $\begin{smallmatrix}(-0.19012,\ 0.36106,\ 0.53082,\ -0.74278)^T\end{smallmatrix}$\\*
$(2,B)$ & $63\ (0.438)$ & $57408.470$ & $\begin{smallmatrix}(-0.11392,\ 0.20705,\ 0.58906,\ -0.77277)^T\end{smallmatrix}$\\
\cmidrule(lr){1-4}
\multicolumn{4}{@{}l}{$K=3$ (ratio $\checkmark$, fraction $\checkmark$, sample count $\times$)}\\*
$(3,A)$ & $42\ (0.292)$ & $75226.986$ & $\begin{smallmatrix}(-0.20871,\ 0.40364,\ 0.52375,\ -0.72056)^T\end{smallmatrix}$\\*
$(3,B)$ & $16\ (0.111)$ & $1.72\!\times\!10^{5}$ & $\begin{smallmatrix}(-0.06753,\ 0.09487,\ 0.67033,\ -0.73287)^T\end{smallmatrix}$\\*
$(3,C)$ & $86\ (0.597)$ & $2.01\!\times\!10^{5}$ & $\begin{smallmatrix}(-0.14744,\ 0.27586,\ 0.54943,\ -0.77478)^T\end{smallmatrix}$\\
\cmidrule(lr){1-4}
\multicolumn{4}{@{}l}{$K=4$ (ratio $\checkmark$, fraction $\checkmark$, sample count $\times$)}\\*
$(4,A)$ & $50\ (0.347)$ & $2.38\!\times\!10^{5}$ & $\begin{smallmatrix}(-0.12756,\ 0.23919,\ 0.56250,\ -0.78110)^T\end{smallmatrix}$\\*
$(4,B)$ & $26\ (0.181)$ & $2.44\!\times\!10^{5}$ & $\begin{smallmatrix}(-0.21591,\ 0.43389,\ 0.51562,\ -0.70658)^T\end{smallmatrix}$\\*
$(4,C)$ & $13\ (0.090)$ & $2.86\!\times\!10^{5}$ & $\begin{smallmatrix}(-0.05801,\ 0.07669,\ 0.68209,\ -0.72492)^T\end{smallmatrix}$\\*
$(4,D)$ & $55\ (0.382)$ & $4.66\!\times\!10^{5}$ & $\begin{smallmatrix}(0.17748,\ -0.32588,\ -0.53659,\ 0.75788)^T\end{smallmatrix}$\\
\midrule
\multicolumn{4}{@{}l}{$K=2$ direction: $s_{\mathrm{dir}}=0.983092\geq\tau_{\mathrm{dir}}=0.90$ ($\times$); selected $K=1$.}\\
\midrule
\multicolumn{4}{@{}l}{\textbf{B. Similarity-type screening}}\\*
\multicolumn{4}{@{}l}{Prescribed Type I; no automatic comparison.}\\
\midrule
\multicolumn{4}{@{}l}{\textbf{C. Dimensionless learning}}\\*
Stage & $\widehat I$ & $\varepsilon$ & \multicolumn{1}{c}{$\overline{R^2}$}\\*
$k=1$ & $1.869_{\pm 0.102}$ & $0.154_{\pm 0.023}$ & \multicolumn{1}{c}{$0.913$}\\*
\multicolumn{4}{@{}p{\textwidth}@{}}{\hspace{1em}
$\Pi^{\mathrm{cand}}_{1|1}
=\bigl(\Pi^{(1)}\bigr)^{0.65145}
\bigl(\Pi^{(2)}\bigr)^{-0.65959}
\bigl(\Pi^{(3)}\bigr)^{-0.99629}
\bigl(\Pi^{(4)}\bigr)^{0.6182}$}\\
\addlinespace[1pt]
$k=2$ & $1.871_{\pm 0.120}$ & $0.154_{\pm 0.025}$ & \multicolumn{1}{c}{$0.981$}\\*
\multicolumn{4}{@{}p{\textwidth}@{}}{\hspace{1em}
$\Pi^{\mathrm{cand}}_{1|2}
=\bigl(\Pi^{(1)}\bigr)^{0.65202}
\bigl(\Pi^{(2)}\bigr)^{-0.66061}
\bigl(\Pi^{(3)}\bigr)^{-0.99778}
\bigl(\Pi^{(4)}\bigr)^{0.62029}$}\\
\multicolumn{4}{@{}p{\textwidth}@{}}{\hspace{1em}
$\Pi^{\mathrm{cand}}_{2|2}
=\bigl(\Pi^{(1)}\bigr)^{-0.99678}
\bigl(\Pi^{(2)}\bigr)^{0.17751}
\bigl(\Pi^{(3)}\bigr)^{-0.24509}
\bigl(\Pi^{(4)}\bigr)^{0.29879}$}\\
\addlinespace[1pt]
Rationalization & $1.859_{\pm 0.102}$ & $0.156_{\pm 0.023}$ & \multicolumn{1}{c}{$0.912$}\\
\midrule
\multicolumn{4}{@{}l}{$k^*=1$; $\Delta\varepsilon_2=0.0002646$.}\\*
\multicolumn{4}{@{}l}{Relative MI loss after rationalization: $0.006$.}\\
\end{longtable}
\endgroup

\noindent\textbf{Dominant dimensionless group.}

\begin{equation}
\begin{aligned}
\Pi_1^*
&=(\Pi^{(1)})^{2/3}
(\Pi^{(2)})^{-2/3}
(\Pi^{(3)})^{-1}
(\Pi^{(4)})^{3/5}
&=
\frac{\nu^{2/5}M^{2/3}\theta_m^{3/5}}
{U_\infty^{26/15}\delta^{2/3}\delta^*},
\qquad
\Pi_y=\frac{u_\tau}{U_\infty}.
\end{aligned}
\end{equation}

\begin{figure}[!hb]
\centering
\includegraphics[width=0.8\textwidth]{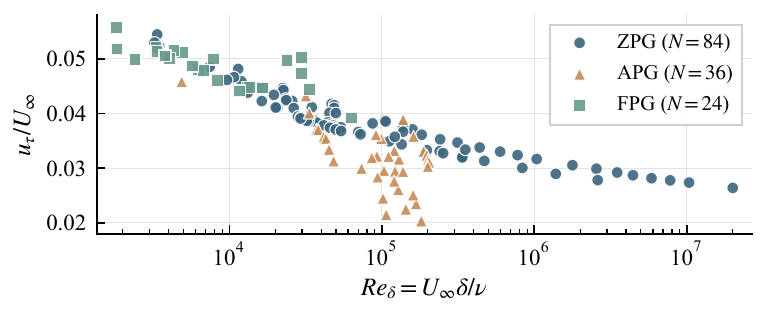}
\caption{Normalized friction velocity $u_{\tau}/U_{\infty}$ versus $Re_{\delta}=U_{\infty}\delta/\nu$ for the pressure-gradient wall-friction observations. Symbols distinguish ZPG, APG, and FPG cases.}
\label{fig:presgrad-redelta}
\end{figure}

The retained group combines viscous, integral-profile, and geometric contributions into a single coordinate for wall friction. For comparison, Supplementary Fig.~\ref{fig:presgrad-redelta} shows that the ZPG, APG, and FPG cases do not collapse onto a common relation when represented by $Re_{\delta}$ alone.

\clearpage
\subsection{Case 3: Rayleigh--B\'{e}nard Convection}
\label{case:RAYLEIGH}

Rayleigh--B\'enard convection is driven by buoyancy in a fluid layer heated from below and cooled from above, with viscous dissipation and thermal diffusion opposing the motion. This case tests whether dimensionless learning recovers the dominant Rayleigh-number dependence of heat transport.

\noindent\textbf{Prompt.}
\begin{quote}\small\itshape
Perform dimensional analysis for Rayleigh-Bénard convection in a planar horizontal layer of fluid heated from below, to derive the dimensionless form of the Nusselt number (denoted as Nu). 
This analysis uses ONLY the following 7 physical quantities as input parameters: 
the height of the container (h), the thermal conductivity (lambda), the temperature difference between the top and bottom surfaces (delta\_T), the gravitational acceleration (g), the thermal expansion coefficient (alpha), the kinematic viscosity (nu), and the thermal diffusivity (kappa).
\end{quote}

\noindent\textbf{Variable definitions.}

\begin{table}[!htbp]
\centering
\setlength{\tabcolsep}{4pt}
\small
\caption{Physical descriptions, dimensional assignments, and numerical ranges for Rayleigh--B\'enard convection.}
\label{tab:vars-RAYLEIGH}
\begin{tabular}{@{}
>{\raggedright\arraybackslash}p{16mm}
>{\raggedright\arraybackslash}p{44mm}
>{\raggedright\arraybackslash}p{25mm}
>{\centering\arraybackslash}p{31mm}
>{\raggedright\arraybackslash}p{32mm}
@{}}
\toprule
Symbol & Description & SI Unit &
\shortstack{Dimensional vector\\$(M,L,T,\Theta)$} & Range\\
\midrule
$h$ & Fluid-layer height & $\mathrm m$ & $(0,1,0,0)$ & $0.125$--\allowbreak $0.2$\\
$\lambda$ & Thermal conductivity & $\mathrm{W\,m^{-1}K^{-1}}$ & $(1,1,-3,-1)$ & $0.001426$--\allowbreak $21.55$\\
$\Delta T$ & Temperature difference & $\mathrm K$ & $(0,0,0,1)$ & $0.0017$--\allowbreak $0.92$\\
$g$ & Gravitational acceleration & $\mathrm{m\,s^{-2}}$ & $(0,1,-2,0)$ & $9.81$\\
$\alpha$ & Volumetric thermal-expansion coefficient & $\mathrm{K^{-1}}$ & $(0,0,0,-1)$ & $0.07242$--\allowbreak $2759$\\
$\nu$ & Kinematic viscosity & $\mathrm{m^2s^{-1}}$ & $(0,2,-1,0)$ & $3.505\times10^{-8}$--\allowbreak $0.001528$\\
$\kappa$ & Thermal diffusivity & $\mathrm{m^2s^{-1}}$ & $(0,2,-1,0)$ & $2.381\times10^{-9}$--\allowbreak $0.00228$\\
\midrule
$Nu$ & Nusselt number & $1$ & $(0,0,0,0)$ & $1.01$--\allowbreak $6890$\\
\bottomrule
\end{tabular}
\end{table}

\noindent\textbf{Dimensional matrix and basis dimensionless parameters.}

\begin{equation}
\mathbf{D}=
\begin{array}{c|rrrrrrr}
\! & h & \lambda & \Delta T & g & \alpha & \nu & \kappa\\
\hline
M & 0 & 1 & 0 & 0 & 0 & 0 & 0\\
L & 1 & 1 & 0 & 1 & 0 & 2 & 2\\
T & 0 & -3 & 0 & -2 & 0 & -1 & -1\\
\Theta & 0 & -1 & 1 & 0 & -1 & 0 & 0
\end{array}.
\end{equation}

The rank is 4 and the nullity is $n=3$. A basis is
\begin{equation}
\begin{aligned}
\Pi^{(1)}&=\Delta T\alpha,
&
\Pi^{(2)}&=\nu/(h^{3/2}g^{1/2}),
&
\Pi^{(3)}&=\kappa/(h^{3/2}g^{1/2}).
\end{aligned}
\end{equation}

\begingroup
\small
\setlength{\tabcolsep}{2pt}
\setlength{\LTpre}{5pt}
\setlength{\LTpost}{5pt}
\setlength{\LTcapwidth}{\textwidth}
\renewcommand{\arraystretch}{1.00}
\begin{longtable}{@{}
>{\raggedright\arraybackslash}p{30mm}
>{\raggedleft\arraybackslash}p{30mm}
>{\raggedleft\arraybackslash}p{25mm}
>{\raggedright\arraybackslash}p{70mm}
@{}}
\caption{Numerical evidence and intermediate results for Rayleigh--B\'enard convection.}
\label{tab:case-RAYLEIGH}
\label{tab:partition-RAYLEIGH}
\label{tab:metrics-RAYLEIGH}
\label{tab:screen-RAYLEIGH}
\\
\toprule
\endfirsthead
\caption[]{Numerical evidence and intermediate results for Rayleigh--B\'enard convection (continued).}\\
\toprule
\endhead
\midrule
\endfoot
\bottomrule
\endlastfoot
\multicolumn{4}{@{}l}{\textbf{A. Regime assessment}}\\*
\multicolumn{4}{@{}l}{$N=182$; $E_G=0.917603$.}\\*
$(K,I)$ & $N_I\ (N_I/N)$ & $\lambda_1^{(I)}/\lambda_n^{(I)}$ & $\mathbf{v}_1^{(I)}$\\*
\multicolumn{4}{@{}l}{$K=2$ (ratio $\checkmark$, fraction $\checkmark$, sample count $\times$)}\\*
$(2,A)$ & $54\ (0.297)$ & $102.621$ & $\begin{smallmatrix}(-0.35933,\ 0.53657,\ 0.76352)^T\end{smallmatrix}$\\*
$(2,B)$ & $128\ (0.703)$ & $125.090$ & $\begin{smallmatrix}(-0.44221,\ 0.81713,\ 0.36980)^T\end{smallmatrix}$\\
\cmidrule(lr){1-4}
\multicolumn{4}{@{}l}{$K=3$ (ratio $\checkmark$, fraction $\times$, sample count $\times$)}\\*
$(3,A)$ & $49\ (0.269)$ & $932.870$ & $\begin{smallmatrix}(-0.39757,\ 0.59324,\ 0.70000)^T\end{smallmatrix}$\\*
$(3,B)$ & $9\ (0.049)$ & $568.825$ & $\begin{smallmatrix}(-0.03890,\ 0.18844,\ 0.98131)^T\end{smallmatrix}$\\*
$(3,C)$ & $124\ (0.681)$ & $145.697$ & $\begin{smallmatrix}(-0.44371,\ 0.81940,\ 0.36290)^T\end{smallmatrix}$\\
\cmidrule(lr){1-4}
\multicolumn{4}{@{}l}{$K=4$ (ratio $\checkmark$, fraction $\times$, sample count $\times$)}\\*
$(4,A)$ & $52\ (0.286)$ & $2270.572$ & $\begin{smallmatrix}(-0.38656,\ 0.87988,\ 0.27638)^T\end{smallmatrix}$\\*
$(4,B)$ & $75\ (0.412)$ & $396.490$ & $\begin{smallmatrix}(-0.47831,\ 0.76659,\ 0.42845)^T\end{smallmatrix}$\\*
$(4,C)$ & $9\ (0.049)$ & $568.825$ & $\begin{smallmatrix}(-0.03890,\ 0.18844,\ 0.98131)^T\end{smallmatrix}$\\*
$(4,D)$ & $46\ (0.253)$ & $899.707$ & $\begin{smallmatrix}(-0.39837,\ 0.58346,\ 0.70773)^T\end{smallmatrix}$\\
\midrule
\multicolumn{4}{@{}l}{$K=2$ direction: $s_{\mathrm{dir}}=0.879701<\tau_{\mathrm{dir}}=0.90$ ($\checkmark$); selected $K=1$.}\\
\midrule
\multicolumn{4}{@{}l}{\textbf{B. Similarity-type screening}}\\*
Type and $\widehat I$ & \multicolumn{3}{l}{Optimized representation ($k=1$)}\\*
Type~I\newline $\widehat I=2.600264$ & \multicolumn{3}{l}{$\begin{aligned}\Pi_{1,\mathrm{I}}&=\bigl(\Pi^{(1)}\bigr)^{0.25052}\bigl(\Pi^{(2)}\bigr)^{-0.20077}\bigl(\Pi^{(3)}\bigr)^{-0.27908}\\\Pi_{y,\mathrm{I}}&=\Pi_y\end{aligned}$}\\*
Type~II\newline $\widehat I=2.694659$ & \multicolumn{3}{l}{$\begin{aligned}\Pi_{1,\mathrm{II}}&=\bigl(\Pi^{(1)}\bigr)^{0.36103}\bigl(\Pi^{(2)}\bigr)^{-0.41002}\bigl(\Pi^{(3)}\bigr)^{-0.20404}\\\widetilde\Pi_{y,\mathrm{II}}&=\Pi_y\bigl(\Pi^{(1)}\bigr)^{0.12901}\bigl(\Pi^{(2)}\bigr)^{-0.26234}\bigl(\Pi^{(3)}\bigr)^{0.1239}\end{aligned}$}\\*
\multicolumn{4}{@{}l}{$L_{\mathrm{out}}=0.035030\leq0.15$; Type~I.}\\
\midrule
\multicolumn{4}{@{}l}{\textbf{C. Dimensionless learning}}\\*
Stage & $\widehat I$ & $\varepsilon$ & \multicolumn{1}{c}{$\overline{R^2}$}\\*
$k=1$ & $2.600_{\pm 0.031}$ & $0.074_{\pm 0.004}$ & \multicolumn{1}{c}{$0.999$}\\*
\multicolumn{4}{@{}p{\textwidth}@{}}{\hspace{1em}
$\Pi^{\mathrm{cand}}_{1|1}
=\bigl(\Pi^{(1)}\bigr)^{0.25052}
\bigl(\Pi^{(2)}\bigr)^{-0.20077}
\bigl(\Pi^{(3)}\bigr)^{-0.27908}$}\\
\addlinespace[1pt]
$k=2$ & $2.612_{\pm 0.031}$ & $0.073_{\pm 0.004}$ & \multicolumn{1}{c}{$1.000$}\\*
\multicolumn{4}{@{}p{\textwidth}@{}}{\hspace{1em}
$\Pi^{\mathrm{cand}}_{1|2}
=\bigl(\Pi^{(1)}\bigr)^{-0.00706}
\bigl(\Pi^{(2)}\bigr)^{0.79078}
\bigl(\Pi^{(3)}\bigr)^{-0.96802}$}\\
\multicolumn{4}{@{}p{\textwidth}@{}}{\hspace{1em}
$\Pi^{\mathrm{cand}}_{2|2}
=\bigl(\Pi^{(1)}\bigr)^{0.25043}
\bigl(\Pi^{(2)}\bigr)^{-0.20095}
\bigl(\Pi^{(3)}\bigr)^{-0.27894}$}\\
\addlinespace[1pt]
Sparsification & $2.600_{\pm 0.031}$ & $0.074_{\pm 0.004}$ & \multicolumn{1}{c}{--}\\*
Rationalization & $2.531_{\pm 0.035}$ & $0.080_{\pm 0.005}$ & \multicolumn{1}{c}{$0.999$}\\
\midrule
\multicolumn{4}{@{}l}{$k^*=1$; $\Delta\varepsilon_2=0.0008662$.}\\*
\multicolumn{4}{@{}l}{Relative MI loss after rationalization: $0.026$.}\\
\end{longtable}
\endgroup

\noindent\textbf{Dominant dimensionless group.}

\begin{equation}
\Pi_1^*
=
(\Pi^{(1)})^{1/3}
(\Pi^{(2)}\Pi^{(3)})^{-1/3}
=
\left(
\frac{g\alpha\Delta T h^3}{\nu\kappa}
\right)^{1/3}
=
Ra^{1/3},
\qquad
\Pi_y=Nu.
\end{equation}

The retained group is an invertible reparameterization of the Rayleigh number; its exponent $1/3$ does not imply a heat-transfer law $Nu\propto Ra^{1/3}$.

\clearpage
\subsection{Case 4: Laser--Metal Interaction in Additive Manufacturing}

In laser powder-bed fusion, sufficiently high energy input causes metal vaporization, whose recoil pressure depresses the melt-pool surface and forms a keyhole. This case tests whether dimensionless learning recovers the Keyhole-number scaling that combines laser parameters and material properties~\cite{xie2022data}.

\noindent\textbf{Prompt.}
\begin{quote}\small\itshape
Perform dimensional analysis for laser-melted liquid metal to derive the dimensionless form of the keyhole depth (denoted as e*). 
This analysis uses ONLY the following 7 physical quantities as input parameters: 
the material density ($\rho$), the temperature difference between melting temperature and ambient temperature (Tl - T0), the laser beam radius (r0), and the thermal diffusivity ($\alpha$), the effective laser power ($\eta$P), the heat capacity (cp), the laser scan speed (Vs).
\end{quote}

\noindent\textbf{Variable definitions.}

\begin{table}[!htbp]
\centering
\setlength{\tabcolsep}{4pt}
\small
\caption{Physical descriptions, dimensional assignments, and numerical ranges for laser keyhole depth.}
\label{tab:vars-KEYHOLE}
\begin{tabular}{@{}
>{\raggedright\arraybackslash}p{16mm}
>{\raggedright\arraybackslash}p{44mm}
>{\raggedright\arraybackslash}p{25mm}
>{\centering\arraybackslash}p{31mm}
>{\raggedright\arraybackslash}p{32mm}
@{}}
\toprule
Symbol & Description & SI Unit &
\shortstack{Dimensional vector\\$(M,L,T,\Theta)$} & Range\\
\midrule
$\rho$ & Material density & $\mathrm{kg\,m^{-3}}$ & $(1,-3,0,0)$ & $2415$--\allowbreak $6881$\\
$T_l-T_0$ & Melting-to-ambient temperature difference $T_l-T_0$ & $\mathrm K$ & $(0,0,0,1)$ & $622$--\allowbreak $1630$\\
$r_0$ & Laser-beam radius & $\mathrm m$ & $(0,1,0,0)$ & $4.4\times10^{-5}$--\allowbreak $7\times10^{-5}$\\
$\alpha$ & Thermal diffusivity & $\mathrm{m^2s^{-1}}$ & $(0,2,-1,0)$ & $4.95\times10^{-6}$--\allowbreak $3.19\times10^{-5}$\\
$\eta P$ & Effective absorbed laser power $\eta P$ & $\mathrm W$ & $(1,2,-3,0)$ & $31.4$--\allowbreak $342.6$\\
$C_p$ & Specific heat capacity & $\mathrm{J\,kg^{-1}K^{-1}}$ & $(0,2,-2,-1)$ & $790$--\allowbreak $1170$\\
$V_s$ & Laser scan speed & $\mathrm{m\,s^{-1}}$ & $(0,1,-1,0)$ & $0.3$--\allowbreak $1.2$\\
\midrule
$e^*$ & Normalized keyhole depth & $1$ & $(0,0,0,0)$ & $0.2087$--\allowbreak $12.77$\\
\bottomrule
\end{tabular}
\end{table}

\noindent\textbf{Dimensional matrix and basis dimensionless parameters.}

\begin{equation}
\mathbf{D}=
\begin{array}{c|rrrrrrr}
\! & \rho & T_l-T_0 & r_0 & \alpha & \eta P & C_p & V_s\\
\hline
M & 1 & 0 & 0 & 0 & 1 & 0 & 0\\
L & -3 & 0 & 1 & 2 & 2 & 2 & 1\\
T & 0 & 0 & 0 & -1 & -3 & -2 & -1\\
\Theta & 0 & 1 & 0 & 0 & 0 & -1 & 0
\end{array}.
\end{equation}

The rank is 4 and the nullity is $n=3$. A basis is
\begin{equation}
\begin{aligned}
\Pi^{(1)}&=r_0(\eta P)/(\rho\alpha^3),
&
\Pi^{(2)}&=(T_l-T_0)r_0^2C_p/\alpha^2,
&
\Pi^{(3)}&=r_0V_s/\alpha.
\end{aligned}
\end{equation}

\begingroup
\small
\setlength{\tabcolsep}{2pt}
\setlength{\LTpre}{5pt}
\setlength{\LTpost}{5pt}
\setlength{\LTcapwidth}{\textwidth}
\renewcommand{\arraystretch}{1.00}
\begin{longtable}{@{}
>{\raggedright\arraybackslash}p{30mm}
>{\raggedleft\arraybackslash}p{30mm}
>{\raggedleft\arraybackslash}p{25mm}
>{\raggedright\arraybackslash}p{70mm}
@{}}
\caption{Numerical evidence and intermediate results for laser keyhole depth.}
\label{tab:case-KEYHOLE}
\label{tab:partition-KEYHOLE}
\label{tab:metrics-KEYHOLE}
\label{tab:screen-KEYHOLE}
\\
\toprule
\endfirsthead
\caption[]{Numerical evidence and intermediate results for laser keyhole depth (continued).}\\
\toprule
\endhead
\midrule
\endfoot
\bottomrule
\endlastfoot
\multicolumn{4}{@{}l}{\textbf{A. Regime assessment}}\\*
\multicolumn{4}{@{}l}{$N=90$; $E_G=0.810978$.}\\*
$(K,I)$ & $N_I\ (N_I/N)$ & $\lambda_1^{(I)}/\lambda_n^{(I)}$ & $\mathbf{v}_1^{(I)}$\\*
\multicolumn{4}{@{}l}{$K=2$ (ratio $\checkmark$, fraction $\checkmark$, sample count $\times$)}\\*
$(2,A)$ & $68\ (0.756)$ & $51.094$ & $\begin{smallmatrix}(-0.00810,\ 0.98370,\ 0.17962)^T\end{smallmatrix}$\\*
$(2,B)$ & $22\ (0.244)$ & $79.504$ & $\begin{smallmatrix}(-0.78415,\ 0.53309,\ 0.31769)^T\end{smallmatrix}$\\
\cmidrule(lr){1-4}
\multicolumn{4}{@{}l}{$K=3$ (ratio $\checkmark$, fraction $\checkmark$, sample count $\times$)}\\*
$(3,A)$ & $45\ (0.500)$ & $71.411$ & $\begin{smallmatrix}(-0.12142,\ -0.97750,\ -0.17249)^T\end{smallmatrix}$\\*
$(3,B)$ & $16\ (0.178)$ & $86.664$ & $\begin{smallmatrix}(-0.85220,\ 0.41391,\ 0.32007)^T\end{smallmatrix}$\\*
$(3,C)$ & $29\ (0.322)$ & $57.038$ & $\begin{smallmatrix}(-0.32210,\ 0.92365,\ 0.20764)^T\end{smallmatrix}$\\
\cmidrule(lr){1-4}
\multicolumn{4}{@{}l}{$K=4$ (ratio $\checkmark$, fraction $\checkmark$, sample count $\times$)}\\*
$(4,A)$ & $27\ (0.300)$ & $84.747$ & $\begin{smallmatrix}(-0.15857,\ 0.97890,\ 0.12885)^T\end{smallmatrix}$\\*
$(4,B)$ & $35\ (0.389)$ & $79.739$ & $\begin{smallmatrix}(-0.16700,\ -0.96734,\ -0.19067)^T\end{smallmatrix}$\\*
$(4,C)$ & $15\ (0.167)$ & $91.585$ & $\begin{smallmatrix}(-0.86283,\ 0.39241,\ 0.31865)^T\end{smallmatrix}$\\*
$(4,D)$ & $13\ (0.144)$ & $86.918$ & $\begin{smallmatrix}(-0.46504,\ 0.83333,\ 0.29883)^T\end{smallmatrix}$\\
\midrule
\multicolumn{4}{@{}l}{$K=2$ direction: $s_{\mathrm{dir}}=0.587812<\tau_{\mathrm{dir}}=0.90$ ($\checkmark$); selected $K=1$.}\\
\midrule
\multicolumn{4}{@{}l}{\textbf{B. Similarity-type screening}}\\*
Type and $\widehat I$ & \multicolumn{3}{l}{Optimized representation ($k=1$)}\\*
Type~I\newline $\widehat I=1.633773$ & \multicolumn{3}{l}{$\begin{aligned}\Pi_{1,\mathrm{I}}&=\bigl(\Pi^{(1)}\bigr)^{0.53907}\bigl(\Pi^{(2)}\bigr)^{-0.51167}\bigl(\Pi^{(3)}\bigr)^{-0.28925}\\\Pi_{y,\mathrm{I}}&=\Pi_y\end{aligned}$}\\*
Type~II\newline $\widehat I=1.881104$ & \multicolumn{3}{l}{$\begin{aligned}\Pi_{1,\mathrm{II}}&=\bigl(\Pi^{(1)}\bigr)^{0.92603}\bigl(\Pi^{(2)}\bigr)^{-0.96536}\bigl(\Pi^{(3)}\bigr)^{0.31473}\\\widetilde\Pi_{y,\mathrm{II}}&=\Pi_y\bigl(\Pi^{(1)}\bigr)^{0.39208}\bigl(\Pi^{(2)}\bigr)^{-0.45255}\bigl(\Pi^{(3)}\bigr)^{0.98716}\end{aligned}$}\\*
\multicolumn{4}{@{}l}{$L_{\mathrm{out}}=0.131481\leq0.15$; Type~I.}\\
\midrule
\multicolumn{4}{@{}l}{\textbf{C. Dimensionless learning}}\\*
Stage & $\widehat I$ & $\varepsilon$ & \multicolumn{1}{c}{$\overline{R^2}$}\\*
$k=1$ & $1.634_{\pm 0.063}$ & $0.195_{\pm 0.018}$ & \multicolumn{1}{c}{$0.928$}\\*
\multicolumn{4}{@{}p{\textwidth}@{}}{\hspace{1em}
$\Pi^{\mathrm{cand}}_{1|1}
=\bigl(\Pi^{(1)}\bigr)^{0.53907}
\bigl(\Pi^{(2)}\bigr)^{-0.51167}
\bigl(\Pi^{(3)}\bigr)^{-0.28925}$}\\
\addlinespace[1pt]
$k=2$ & $1.633_{\pm 0.062}$ & $0.195_{\pm 0.018}$ & \multicolumn{1}{c}{$0.959$}\\*
\multicolumn{4}{@{}p{\textwidth}@{}}{\hspace{1em}
$\Pi^{\mathrm{cand}}_{1|2}
=\bigl(\Pi^{(1)}\bigr)^{0.54029}
\bigl(\Pi^{(2)}\bigr)^{-0.51248}
\bigl(\Pi^{(3)}\bigr)^{-0.29115}$}\\
\multicolumn{4}{@{}p{\textwidth}@{}}{\hspace{1em}
$\Pi^{\mathrm{cand}}_{2|2}
=\bigl(\Pi^{(1)}\bigr)^{0.38569}
\bigl(\Pi^{(2)}\bigr)^{-0.58748}
\bigl(\Pi^{(3)}\bigr)^{0.76966}$}\\
\addlinespace[1pt]
Rationalization & $1.483_{\pm 0.068}$ & $0.227_{\pm 0.021}$ & \multicolumn{1}{c}{$0.927$}\\
\midrule
\multicolumn{4}{@{}l}{$k^*=1$; $\Delta\varepsilon_2=-0.0001063$.}\\*
\multicolumn{4}{@{}l}{Relative MI loss after rationalization: $0.092$.}\\
\end{longtable}
\endgroup

\noindent\textbf{Dominant dimensionless group.}

\begin{equation}
\begin{aligned}
\Pi_1^*
&=(\Pi^{(1)})^{2/3}
(\Pi^{(2)})^{-2/3}
(\Pi^{(3)})^{-1/3}
&=
\frac{(\eta P)^{2/3}}
{\rho^{2/3}(T_l-T_0)^{2/3}
r_0\alpha^{1/3}C_p^{2/3}V_s^{1/3}},
\qquad
\Pi_y=e/r_0.
\end{aligned}
\end{equation}

The retained group satisfies $\Pi_1^*=(\pi Ke)^{2/3}$, an invertible reparameterization of the Keyhole number defined in the main text.

\clearpage
\subsection{Case 5: Blasius laminar boundary layer}

The Blasius boundary layer describes steady, incompressible laminar flow over a flat plate under zero pressure gradient. It provides a controlled test of whether dimensionless learning recovers the known one-dimensional similarity coordinate.

\noindent\textbf{Prompt.}
\begin{quote}\small\itshape
Perform dimensional analysis for the Blasius laminar boundary layer over a flat plate to derive the dimensionless form of the streamwise velocity (denoted as u/U). 
This analysis uses ONLY the following 4 physical quantities as input parameters: 
the free-stream velocity (U), the kinematic viscosity (nu), the streamwise distance from the leading edge (x), and the wall-normal distance (y).
\end{quote}

\noindent\textbf{Variable definitions.}

\begin{table}[!htbp]
\centering
\setlength{\tabcolsep}{4pt}
\small
\caption{Physical descriptions, dimensional assignments, and numerical ranges for the Blasius boundary layer.}
\label{tab:vars-BLASIUS}
\begin{tabular}{@{}
>{\raggedright\arraybackslash}p{16mm}
>{\raggedright\arraybackslash}p{44mm}
>{\raggedright\arraybackslash}p{25mm}
>{\centering\arraybackslash}p{31mm}
>{\raggedright\arraybackslash}p{32mm}
@{}}
\toprule
Symbol & Description & SI Unit &
\shortstack{Dimensional vector\\$(M,L,T)$} & Range\\
\midrule
$U$ & Free-stream velocity & $\mathrm{m\,s^{-1}}$ & $(0,1,-1)$ & $0.01$\\
$\nu$ & Kinematic viscosity & $\mathrm{m^2s^{-1}}$ & $(0,2,-1)$ & $1\times10^{-6}$\\
$x$ & Distance from the plate leading edge & $\mathrm m$ & $(0,1,0)$ & $0.001$--\allowbreak $0.1$\\
$y$ & Wall-normal distance & $\mathrm m$ & $(0,1,0)$ & $0.0001$--\allowbreak $0.01088$\\
\midrule
$u/U$ & Normalized streamwise velocity & $1$ & $(0,0,0)$ & $0$--\allowbreak $1$\\
\bottomrule
\end{tabular}
\end{table}

\noindent\textbf{Dimensional matrix and basis dimensionless parameters.}

\begin{equation}
\mathbf{D}=
\begin{array}{c|rrrr}
\! & U & \nu & x & y\\
\hline
L & 1 & 2 & 1 & 1\\
T & -1 & -1 & 0 & 0
\end{array}.
\end{equation}

The rank is 2 and the nullity is $n=2$. A basis is
\begin{equation}
\begin{aligned}
\Pi^{(1)}&=Ux/\nu,
&
\qquad
\Pi^{(2)}&=Uy/\nu.
\end{aligned}
\end{equation}

\begingroup
\small
\setlength{\tabcolsep}{2pt}
\setlength{\LTpre}{5pt}
\setlength{\LTpost}{5pt}
\setlength{\LTcapwidth}{\textwidth}
\renewcommand{\arraystretch}{1.00}
\begin{longtable}{@{}
>{\raggedright\arraybackslash}p{30mm}
>{\raggedleft\arraybackslash}p{30mm}
>{\raggedleft\arraybackslash}p{25mm}
>{\raggedright\arraybackslash}p{70mm}
@{}}
\caption{Numerical evidence and intermediate results for the Blasius boundary layer.}
\label{tab:case-BLASIUS}
\label{tab:partition-BLASIUS}
\label{tab:metrics-BLASIUS}
\label{tab:screen-BLASIUS}
\\
\toprule
\endfirsthead
\caption[]{Numerical evidence and intermediate results for the Blasius boundary layer (continued).}\\
\toprule
\endhead
\midrule
\endfoot
\bottomrule
\endlastfoot
\multicolumn{4}{@{}l}{\textbf{A. Regime assessment}}\\*
\multicolumn{4}{@{}l}{$N=196$; $E_G=0.965588$.}\\*
$(K,I)$ & $N_I\ (N_I/N)$ & $\lambda_1^{(I)}/\lambda_n^{(I)}$ & $\mathbf{v}_1^{(I)}$\\*
\multicolumn{4}{@{}l}{$K=2$ (ratio $\times$, fraction $\times$, sample count $\times$)}\\*
$(2,A)$ & $189\ (0.964)$ & $127.832$ & $\begin{smallmatrix}(0.25409,\ 0.96718)^T\end{smallmatrix}$\\*
$(2,B)$ & $7\ (0.036)$ & $5.142$ & $\begin{smallmatrix}(-0.99522,\ -0.09767)^T\end{smallmatrix}$\\
\cmidrule(lr){1-4}
\multicolumn{4}{@{}l}{$K=3$ (ratio $\times$, fraction $\times$, sample count $\times$)}\\*
$(3,A)$ & $1\ (0.005)$ & $1.00\!\times\!10^{15}$ & $\begin{smallmatrix}(-0.16800,\ 0.98579)^T\end{smallmatrix}$\\*
$(3,B)$ & $188\ (0.959)$ & $144.093$ & $\begin{smallmatrix}(0.25605,\ 0.96666)^T\end{smallmatrix}$\\*
$(3,C)$ & $7\ (0.036)$ & $5.142$ & $\begin{smallmatrix}(-0.99522,\ -0.09767)^T\end{smallmatrix}$\\
\cmidrule(lr){1-4}
\multicolumn{4}{@{}l}{$K=4$ (ratio $\times$, fraction $\times$, sample count $\times$)}\\*
$(4,A)$ & $3\ (0.015)$ & $40.357$ & $\begin{smallmatrix}(-0.92456,\ 0.38104)^T\end{smallmatrix}$\\*
$(4,B)$ & $182\ (0.929)$ & $508.962$ & $\begin{smallmatrix}(0.24171,\ 0.97035)^T\end{smallmatrix}$\\*
$(4,C)$ & $6\ (0.031)$ & $33.282$ & $\begin{smallmatrix}(-0.86361,\ -0.50416)^T\end{smallmatrix}$\\*
$(4,D)$ & $5\ (0.026)$ & $97.249$ & $\begin{smallmatrix}(0.53333,\ 0.84591)^T\end{smallmatrix}$\\
\midrule
\multicolumn{4}{@{}l}{$K=2$ direction: $s_{\mathrm{dir}}=0.347337<\tau_{\mathrm{dir}}=0.90$ ($\checkmark$); selected $K=1$.}\\
\midrule
\multicolumn{4}{@{}l}{\textbf{B. Similarity-type screening}}\\*
Type and $\widehat I$ & \multicolumn{3}{l}{Optimized representation ($k=1$)}\\*
Type~I\newline $\widehat I=2.763663$ & \multicolumn{3}{l}{$\begin{aligned}\Pi_{1,\mathrm{I}}&=\bigl(\Pi^{(1)}\bigr)^{-0.14492}\bigl(\Pi^{(2)}\bigr)^{0.29519}\\\Pi_{y,\mathrm{I}}&=\Pi_y\end{aligned}$}\\*
Type~II\newline $\widehat I=2.850951$ & \multicolumn{3}{l}{$\begin{aligned}\Pi_{1,\mathrm{II}}&=\bigl(\Pi^{(1)}\bigr)^{-0.15774}\bigl(\Pi^{(2)}\bigr)^{0.3137}\\\widetilde\Pi_{y,\mathrm{II}}&=\Pi_y\bigl(\Pi^{(1)}\bigr)^{-0.02964}\bigl(\Pi^{(2)}\bigr)^{0.04991}\end{aligned}$}\\*
\multicolumn{4}{@{}l}{$L_{\mathrm{out}}=0.030617\leq0.15$; Type~I.}\\
\midrule
\multicolumn{4}{@{}l}{\textbf{C. Dimensionless learning}}\\*
Stage & $\widehat I$ & $\varepsilon$ & \multicolumn{1}{c}{$\overline{R^2}$}\\*
$k=1$ & $2.764_{\pm 0.038}$ & $0.063_{\pm 0.004}$ & \multicolumn{1}{c}{$0.982$}\\*
\multicolumn{4}{@{}p{\textwidth}@{}}{\hspace{1em}
$\Pi^{\mathrm{cand}}_{1|1}
=\bigl(\Pi^{(1)}\bigr)^{-0.14492}
\bigl(\Pi^{(2)}\bigr)^{0.29519}$}\\
\addlinespace[1pt]
$k=2$ & $2.773_{\pm 0.037}$ & $0.062_{\pm 0.004}$ & \multicolumn{1}{c}{$0.999$}\\*
\multicolumn{4}{@{}p{\textwidth}@{}}{\hspace{1em}
$\Pi^{\mathrm{cand}}_{1|2}
=\bigl(\Pi^{(1)}\bigr)^{-0.13772}
\bigl(\Pi^{(2)}\bigr)^{0.28126}$}\\
\multicolumn{4}{@{}p{\textwidth}@{}}{\hspace{1em}
$\Pi^{\mathrm{cand}}_{2|2}
=\bigl(\Pi^{(1)}\bigr)^{-0.73657}
\bigl(\Pi^{(2)}\bigr)^{-0.99955}$}\\
\addlinespace[1pt]
Rationalization & $2.699_{\pm 0.041}$ & $0.067_{\pm 0.005}$ & \multicolumn{1}{c}{$0.976$}\\
\midrule
\multicolumn{4}{@{}l}{$k^*=1$; $\Delta\varepsilon_2=0.0005692$.}\\*
\multicolumn{4}{@{}l}{Relative MI loss after rationalization: $0.023$.}\\
\end{longtable}
\endgroup

\noindent\textbf{Dominant dimensionless group.}

\begin{equation}
\Pi_1^*
=
(\Pi^{(1)})^{-1/2}\Pi^{(2)}
=
y\sqrt{\frac{U}{\nu x}},
\qquad
\Pi_y=u/U.
\end{equation}

The retained group is the classical Blasius similarity coordinate $\eta$.

\clearpage
\subsection{Case 6: Equivalent sand-grain roughness}

In rough-wall turbulence, equivalent sand-grain roughness $k_s$ relates roughness-induced drag to the downward shift of the logarithmic mean velocity profile in the fully rough regime~\cite{yang2023prediction}. This case tests whether AI Buckingham can improve predictions of $k_s$ by discovering informative dimensionless combinations of surface statistics.

\noindent\textbf{Prompt.}
\begin{quote}\small\itshape
Perform dimensional analysis for sand-grain roughness parameterization to derive the dimensionless relationship governing the normalized equivalent sand-grain roughness (denoted as $k_s/R_a$). This analysis uses ONLY the following 7 physical quantities as input parameters: the first-order absolute moment of height fluctuations, $R_a$; the peak-to-valley height, $k_t$; the root-mean-square height, $k_{rms}$; the skewness of roughness height fluctuations, $S_k$; the kurtosis of roughness height fluctuations, $K_u$; the effective streamwise slope of the roughness, $E_s$; the surface porosity, $P_o$.
\end{quote}

\noindent\textbf{Variable definitions.}

Roughness statistics related to the rough surface height distribution $k=k(x,z)$ are defined as follows~\cite{ma2025machine}:
\begin{equation}
\begin{gathered}
k_{\mathrm{avg}}=\frac{1}{A_t}\int_{x,z}k\,\mathrm dA, \quad
R_a=\frac{1}{A_t}\int_{x,z}|k-k_{\mathrm{avg}}|\,\mathrm dA,\quad
k_t=\operatorname{mean}\!\left\{\max_{\delta\times\delta}k-\min_{\delta\times\delta}k\right\},\\
k_{rms}=\sqrt{\frac{1}{A_t}\int_{x,z}(k-k_{\mathrm{avg}})^2\,\mathrm dA},\quad
S_k=\frac{1}{A_t k_{rms}^{3}}\int_{x,z}(k-k_{\mathrm{avg}})^3\,\mathrm dA,\\
K_u=\frac{1}{A_t k_{rms}^{4}}\int_{x,z}(k-k_{\mathrm{avg}})^4\,\mathrm dA,\quad
E_s=\frac{1}{A_t}\int_{x,z}\left|\frac{\partial k}{\partial x}\right|\,\mathrm dA,\quad
P_o=\frac{1}{A_t k_c}\int_0^{k_c}A_f(y)\,\mathrm dy.
\end{gathered}
\label{eq:sg-surface-statistics-v01}
\end{equation}
Here, $k_c=\max k-\min k$ is the global peak-to-valley height, $A_t$ is the total plan area, and $A_f(y)$ is the fluid area at height $y$. The coordinates $x$ and $z$ are streamwise and spanwise, respectively. The mean in $k_t$ is taken over $\delta\times\delta$ square surface tiles, where $\delta$ is the channel half-height.

\begin{table}[!htbp]
\centering
\setlength{\tabcolsep}{4pt}
\small
\caption{Physical descriptions, dimensional assignments, and numerical ranges for sand-grain roughness parameterization.}
\label{tab:sg-variables-v01}
\label{tab:vars-SANDGRAIN}
\begin{tabular}{@{}
>{\raggedright\arraybackslash}p{16mm}
>{\raggedright\arraybackslash}p{44mm}
>{\raggedright\arraybackslash}p{25mm}
>{\centering\arraybackslash}p{31mm}
>{\raggedright\arraybackslash}p{32mm}
@{}}
\toprule
Symbol & Description & SI Unit &
\shortstack{Dimensional vector\\$(M,L,T)$} & Range\\
\midrule
$R_a$ & First-order absolute moment of height fluctuations & $\mathrm m$ & $(0,1,0)$ & $0.0079$--\allowbreak $0.0715$\\
$k_t$ & Peak-to-valley height & $\mathrm m$ & $(0,1,0)$ & $0.0497$--\allowbreak $0.7167$\\
$k_{rms}$ & Root-mean-square height & $\mathrm m$ & $(0,1,0)$ & $0.0097$--\allowbreak $0.0895$\\
$S_k$ & Height-distribution skewness & $1$ & $(0,0,0)$ & $-1.4448$--\allowbreak $2.5190$\\
$K_u$ & Height-distribution kurtosis & $1$ & $(0,0,0)$ & $1.4355$--\allowbreak $8.7396$\\
$E_s$ & Effective streamwise slope & $1$ & $(0,0,0)$ & $0.1581$--\allowbreak $1.1769$\\
$P_o$ & Surface porosity & $1$ & $(0,0,0)$ & $0.2400$--\allowbreak $0.9292$\\
\midrule
$k_s/R_a$ & Normalized equivalent sand-grain roughness & $1$ & $(0,0,0)$ & $1.3386$--\allowbreak $24.3600$\\
\bottomrule
\end{tabular}
\end{table}

\noindent\textbf{Dimensional matrix and basis dimensionless parameters.}
\begin{equation}
\mathbf{D}=
\begin{array}{c|rrrrrrr}
\! & R_a & k_t & k_{rms} & S_k & K_u & E_s & P_o\\
\hline
L & 1 & 1 & 1 & 0 & 0 & 0 & 0
\end{array}.
\label{eq:sg-dimension-v01}
\end{equation}

The rank is 1 and the nullity is $n=6$. A basis is
\begin{equation}
\left(\Pi^{(1)},\ldots,\Pi^{(6)}\right)
=\left(k_t/R_a,k_{rms}/R_a,S_k+2,K_u,E_s,P_o\right).
\label{eq:sg-basis-v01}
\end{equation}

\begingroup
\small
\setlength{\tabcolsep}{2pt}
\setlength{\LTpre}{5pt}
\setlength{\LTpost}{5pt}
\setlength{\LTcapwidth}{\textwidth}
\renewcommand{\arraystretch}{1.00}
\begin{longtable}{@{}
>{\raggedright\arraybackslash}p{30mm}
>{\raggedleft\arraybackslash}p{30mm}
>{\raggedleft\arraybackslash}p{25mm}
>{\raggedright\arraybackslash}p{70mm}
@{}}
\caption{Numerical evidence and intermediate results for sand-grain roughness parameterization.}
\label{tab:sg-evidence-v01}
\label{tab:case-SANDGRAIN}
\label{tab:partition-SANDGRAIN}
\label{tab:metrics-SANDGRAIN}
\label{tab:screen-SANDGRAIN}
\\
\toprule
\endfirsthead
\caption[]{Numerical evidence and intermediate results for sand-grain roughness parameterization (continued).}\\
\toprule
\endhead
\midrule
\endfoot
\bottomrule
\endlastfoot
\multicolumn{4}{@{}l}{\textbf{A. Regime assessment}}\\*
\multicolumn{4}{@{}l}{$N=104$; $E_G=0.865561$.}\\*
$(K,I)$ & $N_I\ (N_I/N)$ & $\lambda_1^{(I)}/\lambda_n^{(I)}$ & $\mathbf v_1^{(I)}$\\*
\multicolumn{4}{@{}l}{$K=2$ (ratio $\checkmark$, fraction $\checkmark$, sample count $\times$)}\\*
$(2,A)$ & $55\ (0.529)$ & $4749.322$ & $\begin{smallmatrix}(-0.04591,-0.21568,-0.81523,\\\quad -0.31744,-0.14185,0.40730)^T\end{smallmatrix}$\\
$(2,B)$ & $49\ (0.471)$ & $857.593$ & $\begin{smallmatrix}(-0.18548,0.41734,0.66541,\\\quad 0.23826,0.37513,-0.38879)^T\end{smallmatrix}$\\
\cmidrule(lr){1-4}
\multicolumn{4}{@{}l}{$K=3$ (ratio $\checkmark$, fraction $\times$, sample count $\times$)}\\*
$(3,A)$ & $51\ (0.490)$ & $855.024$ & $\begin{smallmatrix}(-0.16536,0.40328,0.67771,\\\quad 0.22401,0.34119,-0.42912)^T\end{smallmatrix}$\\
$(3,B)$ & $5\ (0.048)$ & $8.718\times10^{14}$ & $\begin{smallmatrix}(-0.14554,-0.15428,-0.15315,\\\quad -0.61690,-0.69172,-0.26931)^T\end{smallmatrix}$\\
$(3,C)$ & $48\ (0.462)$ & $4262.991$ & $\begin{smallmatrix}(-0.05309,-0.20124,-0.82762,\\\quad -0.32549,-0.12345,0.38801)^T\end{smallmatrix}$\\
\cmidrule(lr){1-4}
\multicolumn{4}{@{}l}{$K=4$ (ratio $\checkmark$, fraction $\times$, sample count $\times$)}\\*
$(4,A)$ & $26\ (0.250)$ & $742.149$ & $\begin{smallmatrix}(0.27510,-0.41967,-0.71351,\\\quad -0.21333,-0.31090,0.31135)^T\end{smallmatrix}$\\
$(4,B)$ & $5\ (0.048)$ & $8.718\times10^{14}$ & $\begin{smallmatrix}(-0.14554,-0.15428,-0.15315,\\\quad -0.61690,-0.69172,-0.26931)^T\end{smallmatrix}$\\
$(4,C)$ & $36\ (0.346)$ & $3500.481$ & $\begin{smallmatrix}(-0.04686,-0.19347,-0.84393,\\\quad -0.36031,-0.09450,0.33076)^T\end{smallmatrix}$\\
$(4,D)$ & $37\ (0.356)$ & $34085.275$ & $\begin{smallmatrix}(-0.00913,0.32562,0.66927,\\\quad 0.22402,0.31404,-0.54512)^T\end{smallmatrix}$\\
\cmidrule(lr){1-4}
\multicolumn{4}{@{}l}{$K=2$ direction: $s_{\mathrm{dir}}=0.911157\geq\tau_{\mathrm{dir}}=0.90$ ($\times$); selected $K=1$.}\\
\midrule
\multicolumn{4}{@{}l}{\textbf{B. Similarity-type screening}}\\*
\multicolumn{4}{@{}l}{Prescribed Type~I; no automatic comparison.}\\
\midrule
\multicolumn{4}{@{}l}{\textbf{C. Dimensionless learning}}\\*
Stage & $\widehat I$ & $\varepsilon$ & \multicolumn{1}{c}{$\overline{R^2}$}\\*
$k=1$ & $0.650_{\pm 0.080}$ & $0.522_{\pm 0.048}$ & \multicolumn{1}{c}{$0.639$}\\*
\multicolumn{4}{@{}p{\textwidth}@{}}{\hspace{1em}
$\begin{aligned}\Pi^{\mathrm{cand}}_{1|1}&=\bigl(\Pi^{(1)}\bigr)^{-0.14736}\bigl(\Pi^{(2)}\bigr)^{-0.32847}\bigl(\Pi^{(3)}\bigr)^{0.99990}\bigl(\Pi^{(4)}\bigr)^{0.61518}\bigl(\Pi^{(5)}\bigr)^{0.36712}\bigl(\Pi^{(6)}\bigr)^{-0.13983}\end{aligned}$}\\
\addlinespace[1pt]
$k=2$ & $0.858_{\pm 0.064}$ & $0.424_{\pm 0.036}$ & \multicolumn{1}{c}{$0.882$}\\*
\multicolumn{4}{@{}p{\textwidth}@{}}{\hspace{1em}
$\begin{aligned}\Pi^{\mathrm{cand}}_{1|2}&=\bigl(\Pi^{(1)}\bigr)^{0.41242}\bigl(\Pi^{(2)}\bigr)^{-0.99365}\bigl(\Pi^{(3)}\bigr)^{0.99999}\bigl(\Pi^{(4)}\bigr)^{0.18151}\bigl(\Pi^{(5)}\bigr)^{0.11747}\bigl(\Pi^{(6)}\bigr)^{-0.32895}\end{aligned}$}\\
\multicolumn{4}{@{}p{\textwidth}@{}}{\hspace{1em}
$\begin{aligned}\Pi^{\mathrm{cand}}_{2|2}&=\bigl(\Pi^{(1)}\bigr)^{0.99773}\bigl(\Pi^{(2)}\bigr)^{-0.22445}\bigl(\Pi^{(3)}\bigr)^{-0.49907}\bigl(\Pi^{(4)}\bigr)^{-0.68817}\bigl(\Pi^{(5)}\bigr)^{-0.57652}\bigl(\Pi^{(6)}\bigr)^{0.00077}\end{aligned}$}\\
\addlinespace[1pt]
$k=3$ & $0.860_{\pm 0.075}$ & $0.423_{\pm 0.038}$ & \multicolumn{1}{c}{$0.902$}\\*
\multicolumn{4}{@{}p{\textwidth}@{}}{\hspace{1em}
$\begin{aligned}\Pi^{\mathrm{cand}}_{1|3}&=\bigl(\Pi^{(1)}\bigr)^{0.47569}\bigl(\Pi^{(2)}\bigr)^{-0.98141}\bigl(\Pi^{(3)}\bigr)^{0.99991} \bigl(\Pi^{(4)}\bigr)^{0.10061}\bigl(\Pi^{(5)}\bigr)^{0.07491}\bigl(\Pi^{(6)}\bigr)^{-0.37258}\end{aligned}$}\\
\multicolumn{4}{@{}p{\textwidth}@{}}{\hspace{1em}
$\begin{aligned}\Pi^{\mathrm{cand}}_{2|3}&=\bigl(\Pi^{(1)}\bigr)^{0.99999}\bigl(\Pi^{(2)}\bigr)^{-0.32470}\bigl(\Pi^{(3)}\bigr)^{-0.39039} \bigl(\Pi^{(4)}\bigr)^{-0.78639}\bigl(\Pi^{(5)}\bigr)^{-0.66465}\bigl(\Pi^{(6)}\bigr)^{-0.10572}\end{aligned}$}\\
\multicolumn{4}{@{}p{\textwidth}@{}}{\hspace{1em}
$\begin{aligned}\Pi^{\mathrm{cand}}_{3|3}&=\bigl(\Pi^{(1)}\bigr)^{-0.34444}\bigl(\Pi^{(2)}\bigr)^{0.23694}\bigl(\Pi^{(3)}\bigr)^{-0.19729} \bigl(\Pi^{(4)}\bigr)^{-0.98504}\bigl(\Pi^{(5)}\bigr)^{-0.04319}\bigl(\Pi^{(6)}\bigr)^{-0.14058}\end{aligned}$}\\
\addlinespace[1pt]
Rationalization & $0.816_{\pm0.060}$ & $0.442_{\pm0.034}$ & \multicolumn{1}{c}{$0.879$}\\
\midrule
\multicolumn{4}{@{}l}{$k^*=2$; $\Delta\varepsilon_2=0.098038$; $\Delta\varepsilon_3=0.000593$.}\\*
\multicolumn{4}{@{}l}{Relative MI loss after rationalization: $0.0493$.}\\
\midrule
\multicolumn{4}{@{}l}{\textbf{D. Predictive performance across input representations}}\\*
MARE (\%) / $R^2$ & Train (83) & Test (21) & \multicolumn{1}{c}{Overall (104)}\\*
1 group & $18.52\,/\,0.817$ & $31.24\,/\,0.639$ & \multicolumn{1}{c}{$21.09\,/\,0.784$}\\*
2 groups & $16.60\,/\,0.749$ & $18.36\,/\,0.882$ & \multicolumn{1}{c}{$16.96\,/\,0.773$}\\*
3 groups & $16.56\,/\,0.748$ & $16.66\,/\,0.902$ & \multicolumn{1}{c}{$16.58\,/\,0.777$}\\ \midrule
2 groups (\textbf{rat.}) & $16.52\,/\,0.749$ & $18.75\,/\,0.879$ & \multicolumn{1}{c}{$16.97\,/\,0.773$}\\*
Raw 7 & $22.56\,/\,0.711$ & $26.74\,/\,0.766$ & \multicolumn{1}{c}{$23.40\,/\,0.721$}\\
\end{longtable}
\endgroup

\noindent\textbf{Dominant dimensionless group.}
\begin{equation}
\begin{aligned}
\Pi_1^*=\frac{R_a^{3/5}k_t^{2/5}(S_k+2)K_u^{1/5}E_s^{0.12}}{k_{rms}P_o^{1/3}},\quad \Pi_2^*=\frac{k_t}{R_a^{4/5}k_{rms}^{1/5}(S_k+2)^{1/2}K_u^{2/3}E_s^{3/5}},\quad \Pi_y=k_s/R_a.
\end{aligned}
\label{eq:sg-final-v01}
\end{equation}

\clearpage
\subsection{Case 7: Velocity scaling in compressible wall-bounded turbulence}

In compressible wall-bounded turbulence, wall-normal density and viscosity variations disrupt the collapse of mean-velocity profiles under conventional wall-unit scaling~\cite{trettel2016mean,modesti2019direct}. This case tests whether Type~II similarity learning can recover property-based corrections to both the wall-normal coordinate and the mean velocity.

\noindent\textbf{Prompt.}
\begin{quote}\small\itshape
Perform dimensional analysis for compressible wall-bounded channel turbulence to derive the dimensionless form of the normalized mean velocity (denoted as u+). 
This analysis uses ONLY the following 6 physical quantities as input parameters: 
the wall-normal distance (y), the wall density ($\rho$\_w), the wall dynamic viscosity ($\mu$\_w), the fluid density ($\rho$), the dynamic viscosity ($\mu$), and the friction velocity (u\_tau).\\
To ensure comparability with the log-law and uniqueness of results, the power of wall units must be constrained to 1 for the INPUT and 0 for the OUTPUT formula.
\end{quote}

\noindent\textbf{Variable definitions.}

\begin{table}[!htbp]
\centering
\setlength{\tabcolsep}{4pt}
\small
\caption{Physical descriptions, dimensional assignments, and numerical ranges for compressible velocity scaling.}
\label{tab:vars-VELOTRANS}
\begin{tabular}{@{}
>{\raggedright\arraybackslash}p{16mm}
>{\raggedright\arraybackslash}p{44mm}
>{\raggedright\arraybackslash}p{25mm}
>{\centering\arraybackslash}p{31mm}
>{\raggedright\arraybackslash}p{32mm}
@{}}
\toprule
Symbol & Description & SI Unit &
\shortstack{Dimensional vector\\$(M,L,T)$} & Range\\
\midrule
$y$ & Wall-normal distance & $\mathrm m$ & $(0,1,0)$ & $0$--\allowbreak $1$\\
$\rho_w$ & Wall density & $\mathrm{kg\,m^{-3}}$ & $(1,-3,0)$ & $0.0004167$--\allowbreak $3.449$\\
$\mu_w$ & Wall dynamic viscosity & $\mathrm{kg\,m^{-1}s^{-1}}$ & $(1,-1,-1)$ & $1.704\times10^{-5}$--\allowbreak $0.0001$\\
$\rho$ & Local mean density & $\mathrm{kg\,m^{-3}}$ & $(1,-3,0)$ & $0.0002928$--\allowbreak $3.449$\\
$\mu$ & Local dynamic viscosity & $\mathrm{kg\,m^{-1}s^{-1}}$ & $(1,-1,-1)$ & $1.704\times10^{-5}$--\allowbreak $0.0002634$\\
$u_\tau$ & Friction velocity & $\mathrm{m\,s^{-1}}$ & $(0,1,-1)$ & $0.02928$--\allowbreak $21.28$\\
\midrule
$U^+$ & Velocity in wall units & $1$ & $(0,0,0)$ & $0$--\allowbreak $38.55$\\
\bottomrule
\end{tabular}
\end{table}

\noindent\textbf{Dimensional matrix and basis dimensionless parameters.}

\begin{equation}
\mathbf{D}=
\begin{array}{c|rrrrrr}
\! & y & \rho_w & \mu_w & \rho & \mu & u_\tau\\
\hline
M & 0 & 1 & 1 & 1 & 1 & 0\\
L & 1 & -3 & -1 & -3 & -1 & 1\\
T & 0 & 0 & -1 & 0 & -1 & -1
\end{array}.
\end{equation}

The rank is 3 and the nullity is $n=3$. A basis is
\begin{equation}
\begin{aligned}
\Pi^{(1)}&=\rho/\rho_w,
&
\Pi^{(2)}&=\mu/\mu_w,
&
\Pi^{(3)}&=y\rho_wu_\tau/\mu_w.
\end{aligned}
\end{equation}

The power of $\Pi^{(3)}=y^+$ is fixed to 1 in each dominant input group and to 0 in the dependent-variable rescaling factor, while the density- and viscosity-ratio powers remain free. These constraints are removed during similarity-type screening and restored for the final dimensionless learning.

\begingroup
\small
\setlength{\tabcolsep}{2pt}
\setlength{\LTpre}{5pt}
\setlength{\LTpost}{5pt}
\setlength{\LTcapwidth}{\textwidth}
\renewcommand{\arraystretch}{1.00}
\begin{longtable}{@{}
>{\raggedright\arraybackslash}p{30mm}
>{\raggedleft\arraybackslash}p{30mm}
>{\raggedleft\arraybackslash}p{25mm}
>{\raggedright\arraybackslash}p{70mm}
@{}}
\caption{Numerical evidence and intermediate results for compressible velocity scaling.}
\label{tab:case-VELOTRANS}
\label{tab:partition-VELOTRANS}
\label{tab:metrics-VELOTRANS}
\label{tab:screen-VELOTRANS}
\\
\toprule
\endfirsthead
\caption[]{Numerical evidence and intermediate results for compressible velocity scaling (continued).}\\
\toprule
\endhead
\midrule
\endfoot
\bottomrule
\endlastfoot
\multicolumn{4}{@{}l}{\textbf{A. Regime assessment}}\\*
\multicolumn{4}{@{}l}{$N=1103$; $E_G=0.811460$.}\\*
$(K,I)$ & $N_I\ (N_I/N)$ & $\lambda_1^{(I)}/\lambda_n^{(I)}$ & $\mathbf{v}_1^{(I)}$\\*
\multicolumn{4}{@{}l}{$K=2$ (ratio $\times$, fraction $\checkmark$, sample count $\checkmark$)}\\*
$(2,A)$ & $238\ (0.216)$ & $13.714$ & $\begin{smallmatrix}(-0.23012,\ 0.96946,\ -0.08484)^T\end{smallmatrix}$\\*
$(2,B)$ & $865\ (0.784)$ & $143.957$ & $\begin{smallmatrix}(-0.69289,\ -0.72103,\ 0.00281)^T\end{smallmatrix}$\\
\cmidrule(lr){1-4}
\multicolumn{4}{@{}l}{$K=3$ (ratio $\times$, fraction $\checkmark$, sample count $\checkmark$)}\\*
$(3,A)$ & $184\ (0.167)$ & $14.188$ & $\begin{smallmatrix}(-0.34869,\ 0.93565,\ -0.05449)^T\end{smallmatrix}$\\*
$(3,B)$ & $266\ (0.241)$ & $54.472$ & $\begin{smallmatrix}(-0.41520,\ -0.90960,\ 0.01541)^T\end{smallmatrix}$\\*
$(3,C)$ & $653\ (0.592)$ & $213.622$ & $\begin{smallmatrix}(-0.75390,\ -0.65693,\ 0.00861)^T\end{smallmatrix}$\\
\cmidrule(lr){1-4}
\multicolumn{4}{@{}l}{$K=4$ (ratio $\times$, fraction $\checkmark$, sample count $\checkmark$)}\\*
$(4,A)$ & $245\ (0.222)$ & $90.875$ & $\begin{smallmatrix}(-0.45947,\ -0.88799,\ -0.01881)^T\end{smallmatrix}$\\*
$(4,B)$ & $62\ (0.056)$ & $11.802$ & $\begin{smallmatrix}(-0.97293,\ 0.21912,\ 0.07350)^T\end{smallmatrix}$\\*
$(4,C)$ & $608\ (0.551)$ & $376.428$ & $\begin{smallmatrix}(-0.73957,\ -0.67303,\ 0.00812)^T\end{smallmatrix}$\\*
$(4,D)$ & $188\ (0.170)$ & $25.145$ & $\begin{smallmatrix}(-0.20432,\ 0.97193,\ -0.11661)^T\end{smallmatrix}$\\
\midrule
\multicolumn{4}{@{}l}{$K=2$ direction: $s_{\mathrm{dir}}=0.539801<\tau_{\mathrm{dir}}=0.90$ ($\checkmark$); selected $K=1$.}\\
\midrule
\multicolumn{4}{@{}l}{\textbf{B. Similarity-type screening}}\\*
Type and $\widehat I$ & \multicolumn{3}{l}{Optimized representation ($k=1$)}\\*
Type~I\newline $\widehat I=2.734221$ & \multicolumn{3}{l}{$\begin{aligned}\Pi_{1,\mathrm{I}}&=\bigl(\Pi^{(1)}\bigr)^{-0.92279}\bigl(\Pi^{(2)}\bigr)^{-0.84018}\bigl(\Pi^{(3)}\bigr)^{0.10507}\\\Pi_{y,\mathrm{I}}&=\Pi_y\end{aligned}$}\\*
Type~II\newline $\widehat I=3.232046$ & \multicolumn{3}{l}{$\begin{aligned}\Pi_{1,\mathrm{II}}&=\bigl(\Pi^{(1)}\bigr)^{0.11976}\bigl(\Pi^{(2)}\bigr)^{-0.35864}\bigl(\Pi^{(3)}\bigr)^{0.5044}\\\widetilde\Pi_{y,\mathrm{II}}&=\Pi_y\bigl(\Pi^{(1)}\bigr)^{0.81893}\bigl(\Pi^{(2)}\bigr)^{-0.06056}\bigl(\Pi^{(3)}\bigr)^{0.47342}\end{aligned}$}\\*
\multicolumn{4}{@{}l}{$L_{\mathrm{out}}=0.154028>0.15$; Type~II.}\\
\midrule
\multicolumn{4}{@{}l}{\textbf{C. Dimensionless learning}}\\*
Stage & $\widehat I$ & $\varepsilon$ & \multicolumn{1}{c}{$\overline{R^2}$}\\*
$k=1$ & $2.128_{\pm 0.024}$ & $0.119_{\pm 0.003}$ & \multicolumn{1}{c}{$0.998$}\\*
\multicolumn{4}{@{}p{\textwidth}@{}}{\hspace{1em}
$\Pi^{\mathrm{cand}}_{1|1}
=\bigl(\Pi^{(1)}\bigr)^{0.65658}
\bigl(\Pi^{(2)}\bigr)^{-0.32426}
\Pi^{(3)}$}\\
\multicolumn{4}{@{}p{\textwidth}@{}}{\hspace{1em}
$\widetilde\Pi_{y|1}
=\Pi_y\,
\bigl(\Pi^{(1)}\bigr)^{0.86012}
\bigl(\Pi^{(2)}\bigr)^{0.41658}$}\\
\addlinespace[1pt]
$k=2$ & $2.879_{\pm 0.039}$ & $0.056_{\pm 0.002}$ & \multicolumn{1}{c}{$0.999$}\\*
\multicolumn{4}{@{}p{\textwidth}@{}}{\hspace{1em}
$\Pi^{\mathrm{cand}}_{1|2}
=\Pi^{(1)}
\bigl(\Pi^{(2)}\bigr)^{-0.71366}
\Pi^{(3)}$}\\
\multicolumn{4}{@{}p{\textwidth}@{}}{\hspace{1em}
$\Pi^{\mathrm{cand}}_{2|2}
=\bigl(\Pi^{(1)}\bigr)^{-0.14399}
\bigl(\Pi^{(2)}\bigr)^{0.99958}
\Pi^{(3)}$}\\
\multicolumn{4}{@{}p{\textwidth}@{}}{\hspace{1em}
$\widetilde\Pi_{y|2}
=\Pi_y\,
\bigl(\Pi^{(1)}\bigr)^{-0.83617}
\bigl(\Pi^{(2)}\bigr)^{-0.61983}$}\\
\addlinespace[1pt]
Rationalization & $2.128_{\pm 0.026}$ & $0.119_{\pm 0.003}$ & \multicolumn{1}{c}{$0.998$}\\
\midrule
\multicolumn{4}{@{}l}{$k^*=1$; $\Delta\varepsilon_2=0.06288$.}\\*
\multicolumn{4}{@{}l}{Relative MI change after rationalization: $<0.1\%$.}\\
\end{longtable}
\endgroup

\noindent\textbf{Dominant dimensionless group and similarity representation.}

\begin{equation}
\begin{aligned}
\Pi_1^*
&=
y^+
\left(\frac{\rho}{\rho_w}\right)^{2/3}
\left(\frac{\mu}{\mu_w}\right)^{-1/3},
\\
\widetilde\Pi_y
&=
U^+
\left(\frac{\rho}{\rho_w}\right)^{0.86}
\left(\frac{\mu}{\mu_w}\right)^{0.40}
\simeq
\Phi^{(1)}(\Pi_1^*).
\end{aligned}
\end{equation}

The learned transformation incorporates local-to-wall density and viscosity ratios into both the similarity coordinate and velocity rescaling, while retaining the prescribed unit power of $y^+$.

\clearpage
\subsection{Case 8: Goldenfeld's rough-pipe similarity}

Goldenfeld's rough-pipe similarity connects the hydraulically smooth and fully rough limits through an incomplete-similarity representation~\cite{goldenfeld2006roughness}. This case tests the joint identification of the input similarity coordinate and the friction-factor rescaling.

\noindent\textbf{Prompt.}
\begin{quote}\small\itshape
Perform dimensional analysis for rough-wall turbulent pipe flow to derive the dimensionless form of the friction factor (denoted as cf). 
This analysis uses ONLY the following 5 physical quantities as input parameters: 
the fluid density ($\rho$), the dynamic viscosity ($\mu$), the pipe diameter (D), the flow velocity (U), and the pipe wall roughness height (r).\\ 
To ensure uniqueness of results, the power of relative roughness height must be constrained to 0 for the OUTPUT formula.
\end{quote}

\noindent\textbf{Variable definitions.}

\begin{table}[!htbp]
\centering
\setlength{\tabcolsep}{4pt}
\small
\caption{Physical descriptions, dimensional assignments, and numerical ranges for Goldenfeld scaling.}
\label{tab:vars-GOLDENFELD}
\begin{tabular}{@{}
>{\raggedright\arraybackslash}p{16mm}
>{\raggedright\arraybackslash}p{44mm}
>{\raggedright\arraybackslash}p{25mm}
>{\centering\arraybackslash}p{31mm}
>{\raggedright\arraybackslash}p{32mm}
@{}}
\toprule
Symbol & Description & SI Unit &
\shortstack{Dimensional vector\\$(M,L,T)$} & Range\\
\midrule
$\rho$ & Mass density & $\mathrm{kg\,m^{-3}}$ & $(1,-3,0)$ & $6.057$--\allowbreak $990.5$\\
$\mu$ & Dynamic viscosity & $\mathrm{kg\,m^{-1}s^{-1}}$ & $(1,-1,-1)$ & $5.627\times10^{-5}$--\allowbreak $0.009997$\\
$D$ & Pipe diameter & $\mathrm m$ & $(0,1,0)$ & $0.1313$--\allowbreak $4.99$\\
$U$ & Mean velocity & $\mathrm{m\,s^{-1}}$ & $(0,1,-1)$ & $2.998\times10^{-5}$--\allowbreak $334.1$\\
$r$ & Wall-roughness height & $\mathrm m$ & $(0,1,0)$ & $0.0001263$--\allowbreak $0.3317$\\
\midrule
$f$ & Friction factor & $1$ & $(0,0,0)$ & $1.28$--\allowbreak $11.63$\\
\bottomrule
\end{tabular}
\end{table}

\noindent\textbf{Dimensional matrix and basis dimensionless parameters.}

\begin{equation}
\mathbf{D}=
\begin{array}{c|rrrrr}
\! & \rho & \mu & D & U & r\\
\hline
M & 1 & 1 & 0 & 0 & 0\\
L & -3 & -1 & 1 & 1 & 1\\
T & 0 & -1 & 0 & -1 & 0
\end{array}.
\end{equation}

The rank is 3 and the nullity is $n=2$. A basis is
\begin{equation}
\begin{aligned}
\Pi^{(1)}&=\rho DU/\mu,
&
\qquad
\Pi^{(2)}&=r/D.
\end{aligned}
\end{equation}

\begingroup
\small
\setlength{\tabcolsep}{2pt}
\setlength{\LTpre}{5pt}
\setlength{\LTpost}{5pt}
\setlength{\LTcapwidth}{\textwidth}
\renewcommand{\arraystretch}{1.00}
\begin{longtable}{@{}
>{\raggedright\arraybackslash}p{30mm}
>{\raggedleft\arraybackslash}p{30mm}
>{\raggedleft\arraybackslash}p{25mm}
>{\raggedright\arraybackslash}p{70mm}
@{}}
\caption{Numerical evidence and intermediate results for Goldenfeld's rough-pipe similarity.}
\label{tab:case-GOLDENFELD}
\label{tab:partition-GOLDENFELD}
\label{tab:metrics-GOLDENFELD}
\label{tab:screen-GOLDENFELD}
\\
\toprule
\endfirsthead
\caption[]{Numerical evidence and intermediate results for Goldenfeld's rough-pipe similarity (continued).}\\
\toprule
\endhead
\midrule
\endfoot
\bottomrule
\endlastfoot
\multicolumn{4}{@{}l}{\textbf{A. Regime assessment}}\\*
\multicolumn{4}{@{}l}{$N=414$; $E_G=0.641328$.}\\*
$(K,I)$ & $N_I\ (N_I/N)$ & $\lambda_1^{(I)}/\lambda_n^{(I)}$ & $\mathbf{v}_1^{(I)}$\\*
\multicolumn{4}{@{}l}{$K=2$ (ratio $\times$, fraction $\checkmark$, sample count $\checkmark$)}\\*
$(2,A)$ & $269\ (0.650)$ & $12.717$ & $\begin{smallmatrix}(-0.20584,\ 0.97859)^T\end{smallmatrix}$\\*
$(2,B)$ & $145\ (0.350)$ & $10.718$ & $\begin{smallmatrix}(-0.99655,\ -0.08299)^T\end{smallmatrix}$\\
\cmidrule(lr){1-4}
\multicolumn{4}{@{}l}{$K=3$ (ratio $\times$, fraction $\checkmark$, sample count $\checkmark$)}\\*
$(3,A)$ & $77\ (0.186)$ & $18.487$ & $\begin{smallmatrix}(0.19895,\ 0.98001)^T\end{smallmatrix}$\\*
$(3,B)$ & $136\ (0.329)$ & $15.203$ & $\begin{smallmatrix}(-0.99751,\ -0.07053)^T\end{smallmatrix}$\\*
$(3,C)$ & $201\ (0.486)$ & $27.326$ & $\begin{smallmatrix}(-0.33003,\ 0.94397)^T\end{smallmatrix}$\\
\cmidrule(lr){1-4}
\multicolumn{4}{@{}l}{$K=4$ (ratio $\times$, fraction $\checkmark$, sample count $\times$)}\\*
$(4,A)$ & $245\ (0.592)$ & $17.271$ & $\begin{smallmatrix}(-0.16684,\ 0.98598)^T\end{smallmatrix}$\\*
$(4,B)$ & $69\ (0.167)$ & $43.071$ & $\begin{smallmatrix}(-0.99760,\ 0.06921)^T\end{smallmatrix}$\\*
$(4,C)$ & $67\ (0.162)$ & $24.688$ & $\begin{smallmatrix}(-0.94351,\ -0.33136)^T\end{smallmatrix}$\\*
$(4,D)$ & $33\ (0.080)$ & $24.828$ & $\begin{smallmatrix}(-0.71711,\ 0.69696)^T\end{smallmatrix}$\\
\midrule
\multicolumn{4}{@{}l}{$K=2$ direction: $s_{\mathrm{dir}}=0.123914<\tau_{\mathrm{dir}}=0.90$ ($\checkmark$); selected $K=1$.}\\
\midrule
\multicolumn{4}{@{}l}{\textbf{B. Similarity-type screening}}\\*
Type and $\widehat I$ & \multicolumn{3}{l}{Optimized representation ($k=1$)}\\*
Type~I\newline $\widehat I=1.178833$ & \multicolumn{3}{l}{$\begin{aligned}\Pi_{1,\mathrm{I}}&=\bigl(\Pi^{(1)}\bigr)^{0.32296}\bigl(\Pi^{(2)}\bigr)^{0.94641}\\\Pi_{y,\mathrm{I}}&=\Pi_y\end{aligned}$}\\*
Type~II\newline $\widehat I=1.994507$ & \multicolumn{3}{l}{$\begin{aligned}\Pi_{1,\mathrm{II}}&=\bigl(\Pi^{(1)}\bigr)^{0.57502}\bigl(\Pi^{(2)}\bigr)^{0.81814}\\\widetilde\Pi_{y,\mathrm{II}}&=\Pi_y\bigl(\Pi^{(1)}\bigr)^{0.25537}\bigl(\Pi^{(2)}\bigr)^{0.04193}\end{aligned}$}\\*
\multicolumn{4}{@{}l}{$L_{\mathrm{out}}=0.408960>0.15$; Type~II.}\\
\midrule
\multicolumn{4}{@{}l}{\textbf{C. Dimensionless learning}}\\*
Stage & $\widehat I$ & $\varepsilon$ & \multicolumn{1}{c}{$\overline{R^2}$}\\*
$k=1$ & $1.813_{\pm 0.067}$ & $0.163_{\pm 0.013}$ & \multicolumn{1}{c}{$0.930$}\\*
\multicolumn{4}{@{}p{\textwidth}@{}}{\hspace{1em}
$\Pi^{\mathrm{cand}}_{1|1}
=\bigl(\Pi^{(1)}\bigr)^{0.59409}
\bigl(\Pi^{(2)}\bigr)^{0.80439}$}\\
\multicolumn{4}{@{}p{\textwidth}@{}}{\hspace{1em}
$\widetilde\Pi_{y|1}
=\Pi_y\,
\bigl(\Pi^{(1)}\bigr)^{0.23764}$}\\
\addlinespace[1pt]
$k=2$ & $1.812_{\pm 0.067}$ & $0.163_{\pm 0.013}$ & \multicolumn{1}{c}{$0.966$}\\*
\multicolumn{4}{@{}p{\textwidth}@{}}{\hspace{1em}
$\Pi^{\mathrm{cand}}_{1|2}
=\bigl(\Pi^{(1)}\bigr)^{-0.42876}
\bigl(\Pi^{(2)}\bigr)^{0.90342}$}\\
\multicolumn{4}{@{}p{\textwidth}@{}}{\hspace{1em}
$\Pi^{\mathrm{cand}}_{2|2}
=\bigl(\Pi^{(1)}\bigr)^{0.59402}
\bigl(\Pi^{(2)}\bigr)^{0.80445}$}\\
\multicolumn{4}{@{}p{\textwidth}@{}}{\hspace{1em}
$\widetilde\Pi_{y|2}
=\Pi_y\,
\bigl(\Pi^{(1)}\bigr)^{0.23771}$}\\
\addlinespace[1pt]
Rationalization & $1.803_{\pm 0.070}$ & $0.165_{\pm 0.014}$ & \multicolumn{1}{c}{$0.929$}\\
\midrule
\multicolumn{4}{@{}l}{$k^*=1$; $\Delta\varepsilon_2=-0.0002484$.}\\*
\multicolumn{4}{@{}l}{Relative MI loss after rationalization: $0.005$.}\\
\end{longtable}
\endgroup

\noindent\textbf{Dominant dimensionless group and similarity representation.}

\begin{equation}
\begin{aligned}
\Pi_1^*
&=
(\Pi^{(1)})^{3/4}\Pi^{(2)}
=
Re^{3/4}\frac{r}{D},
\\
\widetilde\Pi_y
&=
f(\Pi^{(1)})^{1/4}
=
fRe^{1/4}
\simeq
\Phi^{(1)}(\Pi_1^*).
\end{aligned}
\end{equation}

The input coordinate is equivalent to the main-text form through the invertible power transformation $(\Pi_1^*)^{4/5}=Re^{3/5}(r/D)^{4/5}$.

\clearpage
\subsection{Case 9: Cess eddy-viscosity model in turbulent channel flow}

The Cess eddy-viscosity model provides a smooth representation of turbulent momentum transport across a channel. Near the wall, viscous scaling is naturally expressed through the inner coordinate $y^+=yu_\tau/\nu$, whereas the channel half-height $h$ and the friction Reynolds number $Re_\tau=hu_\tau/\nu$ become increasingly important toward the outer region. The Cess model therefore exhibits a continuous change in dominant wall-normal dependence, making it a useful test of whether physical-regime identification can recover distinct local similarity coordinates from a single smooth relation~\cite{zhang2024clustering}.

\noindent\textbf{Prompt.}
\begin{quote}\small\itshape
Perform dimensional analysis for the Cess eddy-viscosity model in turbulent channel flow to derive the dimensionless relationship governing the normalized eddy viscosity (denoted as nu\_t/nu).
This analysis uses ONLY the following 4 physical quantities as input parameters:
the channel half-height (h), the distance from the wall (y), the friction velocity (u\_tau), and the fluid kinematic viscosity (nu).
\end{quote}

\noindent\textbf{Variable definitions.}

\begin{table}[!htb]
\centering
\setlength{\tabcolsep}{4pt}
\small
\caption{Physical descriptions, dimensional assignments, and numerical ranges for the Cess model.}
\label{tab:vars-CESS}
\begin{tabular}{@{}
>{\raggedright\arraybackslash}p{16mm}
>{\raggedright\arraybackslash}p{44mm}
>{\raggedright\arraybackslash}p{25mm}
>{\centering\arraybackslash}p{31mm}
>{\raggedright\arraybackslash}p{32mm}
@{}}
\toprule
Symbol & Description & SI Unit &
\shortstack{Dimensional vector\\$(M,L,T)$} & Range\\
\midrule
$h$ & Channel half-height & $\mathrm m$ & $(0,1,0)$ & $0.5005$--\allowbreak $0.9998$\\
$y$ & Wall-normal distance & $\mathrm m$ & $(0,1,0)$ & $5.772\times10^{-7}$--\allowbreak $0.9673$\\
$u_\tau$ & Friction velocity & $\mathrm{m\,s^{-1}}$ & $(0,1,-1)$ & $0.0101$--\allowbreak $987$\\
$\nu$ & Kinematic viscosity & $\mathrm{m^2s^{-1}}$ & $(0,2,-1)$ & $1.017\times10^{-6}$--\allowbreak $9.995\times10^{-6}$\\
\midrule
$\nu_t/\nu$ & Normalized eddy viscosity & $1$ & $(0,0,0)$ & $4.996\times10^{-15}$--\allowbreak $3.245\times10^{7}$\\
\bottomrule
\end{tabular}
\end{table}

\noindent\textbf{Dimensional matrix and basis dimensionless parameters.}

\begin{equation}
\mathbf{D}=
\begin{array}{c|rrrr}
\! & h & y & u_\tau & \nu\\
\hline
L & 1 & 1 & 1 & 2\\
T & 0 & 0 & -1 & -1
\end{array}.
\end{equation}

The rank is 2 and the nullity is $n=2$. A basis is
\begin{equation}
\begin{aligned}
\Pi^{(1)}&=y/h,
&
\qquad
\Pi^{(2)}&=\nu/(hu_\tau).
\end{aligned}
\end{equation}

\begingroup
\small
\setlength{\tabcolsep}{2pt}
\setlength{\LTpre}{5pt}
\setlength{\LTpost}{5pt}
\setlength{\LTcapwidth}{\textwidth}
\renewcommand{\arraystretch}{1.00}
\begin{longtable}{@{}
>{\raggedright\arraybackslash}p{30mm}
>{\raggedleft\arraybackslash}p{30mm}
>{\raggedleft\arraybackslash}p{25mm}
>{\raggedright\arraybackslash}p{70mm}
@{}}
\caption{Numerical evidence and intermediate results for the Cess model.}
\label{tab:case-CESS}
\label{tab:partition-CESS}
\label{tab:metrics-CESS}
\label{tab:screen-CESS}
\\
\toprule
\endfirsthead
\caption[]{Numerical evidence and intermediate results for the Cess model (continued).}\\
\toprule
\endhead
\midrule
\endfoot
\bottomrule
\endlastfoot
\multicolumn{4}{@{}l}{\textbf{A. Regime assessment}}\\*
\multicolumn{4}{@{}l}{$N=1000$; $E_G=0.930863$.}\\*
$(K,I)$ & $N_I\ (N_I/N)$ & $\lambda_1^{(I)}/\lambda_n^{(I)}$ & $\mathbf{v}_1^{(I)}$\\*
\multicolumn{4}{@{}l}{$K=2$ (ratio $\times$, fraction $\checkmark$, sample count $\checkmark$)}\\*
$(2,A)$ & $849\ (0.849)$ & $140.552$ & $\begin{smallmatrix}(-0.37456,\ 0.92720)^T\end{smallmatrix}$\\*
$(2,B)$ & $151\ (0.151)$ & $25.183$ & $\begin{smallmatrix}(0.34583,\ 0.93830)^T\end{smallmatrix}$\\
\cmidrule(lr){1-4}
\multicolumn{4}{@{}l}{$K=3$ (ratio $\checkmark$, fraction $\checkmark$, sample count $\checkmark$)}\\*
$(3,A)$ & $107\ (0.107)$ & $66.133$ & $\begin{smallmatrix}(0.44247,\ 0.89678)^T\end{smallmatrix}$\\*
$(3,B)$ & $114\ (0.114)$ & $66.404$ & $\begin{smallmatrix}(-0.04390,\ 0.99904)^T\end{smallmatrix}$\\*
$(3,C)$ & $779\ (0.779)$ & $770.937$ & $\begin{smallmatrix}(-0.39469,\ 0.91881)^T\end{smallmatrix}$\\
\cmidrule(lr){1-4}
\multicolumn{4}{@{}l}{$K=4$ (ratio $\checkmark$, fraction $\checkmark$, sample count $\checkmark$)}\\*
$(4,A)$ & $82\ (0.082)$ & $204.469$ & $\begin{smallmatrix}(0.49311,\ 0.86996)^T\end{smallmatrix}$\\*
$(4,B)$ & $760\ (0.760)$ & $1373.671$ & $\begin{smallmatrix}(-0.39815,\ 0.91732)^T\end{smallmatrix}$\\*
$(4,C)$ & $67\ (0.067)$ & $109.484$ & $\begin{smallmatrix}(0.16055,\ 0.98703)^T\end{smallmatrix}$\\*
$(4,D)$ & $91\ (0.091)$ & $184.445$ & $\begin{smallmatrix}(-0.15193,\ 0.98839)^T\end{smallmatrix}$\\
\midrule
\multicolumn{4}{@{}l}{$K=2$ direction: $s_{\mathrm{dir}}=0.740462<\tau_{\mathrm{dir}}=0.90$ ($\checkmark$); selected $K=3$.}\\
\midrule
\multicolumn{4}{@{}l}{\textbf{B. Similarity-type screening}}\\*
Type and $\widehat I$ & \multicolumn{3}{l}{Optimized representation ($k=1$)}\\*
\multicolumn{4}{@{}l}{\textit{Regime 0}}\\*
Type~I\newline $\widehat I=2.247329$ & \multicolumn{3}{l}{$\begin{aligned}\Pi_{1,\mathrm{I}}&=\bigl(\Pi^{(1)}\bigr)^{-0.02275}\bigl(\Pi^{(2)}\bigr)^{-0.82361}\\\Pi_{y,\mathrm{I}}&=\Pi_y\end{aligned}$}\\*
Type~II\newline $\widehat I=2.333866$ & \multicolumn{3}{l}{$\begin{aligned}\Pi_{1,\mathrm{II}}&=\bigl(\Pi^{(1)}\bigr)^{0.27195}\bigl(\Pi^{(2)}\bigr)^{-0.30814}\\\widetilde\Pi_{y,\mathrm{II}}&=\Pi_y\bigl(\Pi^{(1)}\bigr)^{0.51848}\bigl(\Pi^{(2)}\bigr)^{0.5413}\end{aligned}$}\\*
\multicolumn{4}{@{}l}{$L_{\mathrm{out}}=0.037079\leq0.15$; Type~I.}\\
\cmidrule(lr){1-4}
\multicolumn{4}{@{}l}{\textit{Regime 1}}\\*
Type~I\newline $\widehat I=2.412678$ & \multicolumn{3}{l}{$\begin{aligned}\Pi_{1,\mathrm{I}}&=\bigl(\Pi^{(1)}\bigr)^{0.52948}\bigl(\Pi^{(2)}\bigr)^{-0.87452}\\\Pi_{y,\mathrm{I}}&=\Pi_y\end{aligned}$}\\*
Type~II\newline $\widehat I=2.564261$ & \multicolumn{3}{l}{$\begin{aligned}\Pi_{1,\mathrm{II}}&=\bigl(\Pi^{(1)}\bigr)^{-0.32198}\bigl(\Pi^{(2)}\bigr)^{-0.36327}\\\widetilde\Pi_{y,\mathrm{II}}&=\Pi_y\bigl(\Pi^{(1)}\bigr)^{-0.98367}\bigl(\Pi^{(2)}\bigr)^{0.5359}\end{aligned}$}\\*
\multicolumn{4}{@{}l}{$L_{\mathrm{out}}=0.059114\leq0.15$; Type~I.}\\
\cmidrule(lr){1-4}
\multicolumn{4}{@{}l}{\textit{Regime 2}}\\*
Type~I\newline $\widehat I=3.516019$ & \multicolumn{3}{l}{$\begin{aligned}\Pi_{1,\mathrm{I}}&=\bigl(\Pi^{(1)}\bigr)^{0.90976}\bigl(\Pi^{(2)}\bigr)^{-0.91303}\\\Pi_{y,\mathrm{I}}&=\Pi_y\end{aligned}$}\\*
Type~II\newline $\widehat I=3.529473$ & \multicolumn{3}{l}{$\begin{aligned}\Pi_{1,\mathrm{II}}&=\bigl(\Pi^{(1)}\bigr)^{0.89139}\bigl(\Pi^{(2)}\bigr)^{-0.90059}\\\widetilde\Pi_{y,\mathrm{II}}&=\Pi_y\bigl(\Pi^{(1)}\bigr)^{-0.01908}\bigl(\Pi^{(2)}\bigr)^{0.01187}\end{aligned}$}\\*
\multicolumn{4}{@{}l}{$L_{\mathrm{out}}=0.003812\leq0.15$; Type~I.}\\
\midrule
\multicolumn{4}{@{}l}{\textbf{C. Dimensionless learning}}\\*
Stage & $\widehat I$ & $\varepsilon$ & \multicolumn{1}{c}{$\overline{R^2}$}\\*
\multicolumn{4}{l}{\textit{Regime 0, $N=107$}}\\*[2pt]
$k=1$ & $2.247_{\pm 0.070}$ & $0.106_{\pm 0.017}$ & \multicolumn{1}{c}{$1.000$}\\*
\multicolumn{4}{@{}p{\textwidth}@{}}{\hspace{1em}
$\Pi^{\mathrm{cand}}_{1|1}
=\bigl(\Pi^{(1)}\bigr)^{-0.02275}
\bigl(\Pi^{(2)}\bigr)^{-0.82361}$}\\*
\addlinespace[1pt]
$k=2$ & $2.247_{\pm 0.070}$ & $0.106_{\pm 0.017}$ & \multicolumn{1}{c}{$1.000$}\\*
\multicolumn{4}{@{}p{\textwidth}@{}}{\hspace{1em}
$\Pi^{\mathrm{cand}}_{1|2}
=\bigl(\Pi^{(1)}\bigr)^{-0.02275}
\bigl(\Pi^{(2)}\bigr)^{-0.82359}$}\\*
\multicolumn{4}{@{}p{\textwidth}@{}}{\hspace{1em}
$\Pi^{\mathrm{cand}}_{2|2}
=\bigl(\Pi^{(1)}\bigr)^{0.9578}
\bigl(\Pi^{(2)}\bigr)^{0.0341}$}\\*
\addlinespace[1pt]
Sparsification & $2.236_{\pm 0.073}$ & $0.107_{\pm 0.018}$ & \multicolumn{1}{c}{--}\\*
Rationalization & $2.108_{\pm 0.057}$ & $0.121_{\pm 0.016}$ & \multicolumn{1}{c}{$1.000$}\\
\addlinespace[1pt]
\multicolumn{4}{l}{\textit{Regime 1, $N=114$}}\\*[2pt]
$k=1$ & $2.413_{\pm 0.063}$ & $0.090_{\pm 0.012}$ & \multicolumn{1}{c}{$1.000$}\\*
\multicolumn{4}{@{}p{\textwidth}@{}}{\hspace{1em}
$\Pi^{\mathrm{cand}}_{1|1}
=\bigl(\Pi^{(1)}\bigr)^{0.52948}
\bigl(\Pi^{(2)}\bigr)^{-0.87452}$}\\*
\addlinespace[1pt]
$k=2$ & $2.413_{\pm 0.062}$ & $0.090_{\pm 0.012}$ & \multicolumn{1}{c}{$1.000$}\\*
\multicolumn{4}{@{}p{\textwidth}@{}}{\hspace{1em}
$\Pi^{\mathrm{cand}}_{1|2}
=\bigl(\Pi^{(1)}\bigr)^{0.52388}
\bigl(\Pi^{(2)}\bigr)^{-0.87503}$}\\*
\multicolumn{4}{@{}p{\textwidth}@{}}{\hspace{1em}
$\Pi^{\mathrm{cand}}_{2|2}
=\bigl(\Pi^{(1)}\bigr)^{0.73066}
\bigl(\Pi^{(2)}\bigr)^{-0.03334}$}\\*
\addlinespace[1pt]
Rationalization & $2.396_{\pm 0.069}$ & $0.091_{\pm 0.013}$ & \multicolumn{1}{c}{$1.000$}\\
\addlinespace[1pt]
\multicolumn{4}{l}{\textit{Regime 2, $N=779$}}\\*[2pt]
$k=1$ & $3.516_{\pm 0.057}$ & $0.030_{\pm 0.003}$ & \multicolumn{1}{c}{$1.000$}\\*
\multicolumn{4}{@{}p{\textwidth}@{}}{\hspace{1em}
$\Pi^{\mathrm{cand}}_{1|1}
=\bigl(\Pi^{(1)}\bigr)^{0.90976}
\bigl(\Pi^{(2)}\bigr)^{-0.91303}$}\\*
\addlinespace[1pt]
$k=2$ & $3.516_{\pm 0.057}$ & $0.030_{\pm 0.003}$ & \multicolumn{1}{c}{$1.000$}\\*
\multicolumn{4}{@{}p{\textwidth}@{}}{\hspace{1em}
$\Pi^{\mathrm{cand}}_{1|2}
=\bigl(\Pi^{(1)}\bigr)^{0.90976}
\bigl(\Pi^{(2)}\bigr)^{-0.91303}$}\\*
\multicolumn{4}{@{}p{\textwidth}@{}}{\hspace{1em}
$\Pi^{\mathrm{cand}}_{2|2}
=\bigl(\Pi^{(1)}\bigr)^{0.93637}
\bigl(\Pi^{(2)}\bigr)^{0.63048}$}\\*
\addlinespace[1pt]
Sparsification & $3.479_{\pm 0.055}$ & $0.031_{\pm 0.003}$ & \multicolumn{1}{c}{--}\\*
Rationalization & $3.427_{\pm 0.051}$ & $0.032_{\pm 0.003}$ & \multicolumn{1}{c}{$1.000$}\\
\midrule
\multicolumn{4}{@{}l}{$k^*=1$ in regimes 0--2; $\Delta\varepsilon_2=(0,\,4.61\times10^{-6},\,0)$.}\\*
\multicolumn{4}{@{}l}{Relative MI losses after rationalization: $(0.057,\,0.007,\,0.015)$ (regimes 0--2).}\\
\end{longtable}
\endgroup

\noindent\textbf{Dominant dimensionless groups.}

\begin{equation}
\begin{aligned}
\Pi_{1,0}^*
&=
(\Pi^{(2)})^{-1}
=
Re_\tau,
\\
\Pi_{1,1}^*
&=
(\Pi^{(1)})^{0.57}
(\Pi^{(2)})^{-1}
=
Re_\tau(y/h)^{0.57},
\\
\Pi_{1,2}^*
&=
\Pi^{(1)}
(\Pi^{(2)})^{-1}
=
y^+,
\qquad
\Pi_y=\nu_t/\nu.
\end{aligned}
\end{equation}

The three retained groups correspond to the outer, intermediate, and near-wall dependences, respectively.

\clearpage
\subsection{Case 10: Viscous flow in a rough circular pipe}

Rough circular pipe flow provides a representative example in which different parts of parameter space are governed by qualitatively different dependencies. In laminar flow, the Darcy friction factor is determined solely by the Reynolds number through $f_D=64/Re$. In turbulent rough-pipe flow, wall roughness introduces the additional relative-roughness scale $\epsilon/D$, and the friction factor follows the implicit Colebrook relation. The piecewise construction of~\citet{zhang2024clustering} therefore provides a controlled test of whether physical-regime identification can separate these dependencies and recover a characteristic dimensionless group in each regime.

\noindent\textbf{Prompt.}
\begin{quote}\small\itshape
Perform dimensional analysis for viscous flow in a rough circular pipe to derive the dimensionless relationship governing the pipe friction factor (denoted as f\_D).
This analysis uses ONLY the following 5 physical quantities as input parameters:
the fluid density (rho), the dynamic viscosity (mu), the pipe diameter (D), the mean flow velocity (U), and the absolute wall roughness height (epsilon).
\end{quote}

\noindent\textbf{Variable definitions.}

\begin{table}[!htbp]
\centering
\setlength{\tabcolsep}{4pt}
\small
\caption{Physical descriptions, dimensional assignments, and numerical ranges for rough circular pipe flow.}
\label{tab:vars-PIPE}
\begin{tabular}{@{}
>{\raggedright\arraybackslash}p{16mm}
>{\raggedright\arraybackslash}p{44mm}
>{\raggedright\arraybackslash}p{25mm}
>{\centering\arraybackslash}p{31mm}
>{\raggedright\arraybackslash}p{32mm}
@{}}
\toprule
Symbol & Description & SI Unit &
\shortstack{Dimensional vector\\$(M,L,T)$} & Range\\
\midrule
$\rho$ & Mass density & $\mathrm{kg\,m^{-3}}$ & $(1,-3,0)$ & $0.1003$--\allowbreak $0.1799$\\
$\mu$ & Dynamic viscosity & $\mathrm{kg\,m^{-1}s^{-1}}$ & $(1,-1,-1)$ & $1.004\times10^{-6}$--\allowbreak $9.997\times10^{-6}$\\
$D$ & Pipe diameter & $\mathrm m$ & $(0,1,0)$ & $0.3013$--\allowbreak $1$\\
$U$ & Mean flow velocity & $\mathrm{m\,s^{-1}}$ & $(0,1,-1)$ & $0.02517$--\allowbreak $795.1$\\
$\epsilon$ & Absolute roughness height & $\mathrm m$ & $(0,1,0)$ & $0.005001$--\allowbreak $0.09991$\\
\midrule
$f_D$ & Darcy friction factor & $1$ & $(0,0,0)$ & $0.02138$--\allowbreak $0.2301$\\
\bottomrule
\end{tabular}
\end{table}

\noindent\textbf{Dimensional matrix and basis dimensionless parameters.}

\begin{equation}
\mathbf{D}=
\begin{array}{c|rrrrr}
\! & \rho & \mu & D & U & \epsilon\\
\hline
M & 1 & 1 & 0 & 0 & 0\\
L & -3 & -1 & 1 & 1 & 1\\
T & 0 & -1 & 0 & -1 & 0
\end{array}.
\end{equation}

The rank is 3 and the nullity is $n=2$. A basis is
\begin{equation}
\begin{aligned}
\Pi^{(1)}&=\rho DU/\mu,
&
\qquad
\Pi^{(2)}&=\epsilon/D.
\end{aligned}
\end{equation}

\begingroup
\small
\setlength{\tabcolsep}{2pt}
\setlength{\LTpre}{5pt}
\setlength{\LTpost}{5pt}
\setlength{\LTcapwidth}{\textwidth}
\renewcommand{\arraystretch}{1.00}
\begin{longtable}{@{}
>{\raggedright\arraybackslash}p{30mm}
>{\raggedleft\arraybackslash}p{30mm}
>{\raggedleft\arraybackslash}p{25mm}
>{\raggedright\arraybackslash}p{70mm}
@{}}
\caption{Numerical evidence and intermediate results for rough circular pipe flow.}
\label{tab:case-PIPE}
\label{tab:partition-PIPE}
\label{tab:metrics-PIPE}
\label{tab:screen-PIPE}
\\
\toprule
\endfirsthead
\caption[]{Numerical evidence and intermediate results for rough circular pipe flow (continued).}\\
\toprule
\endhead
\midrule
\endfoot
\bottomrule
\endlastfoot
\multicolumn{4}{@{}l}{\textbf{A. Regime assessment}}\\*
\multicolumn{4}{@{}l}{$N=1200$; $E_G=0.845428$.}\\*
$(K,I)$ & $N_I\ (N_I/N)$ & $\lambda_1^{(I)}/\lambda_n^{(I)}$ & $\mathbf{v}_1^{(I)}$\\*
\multicolumn{4}{@{}l}{$K=2$ (ratio $\checkmark$, fraction $\checkmark$, sample count $\checkmark$)}\\*
$(2,A)$ & $975\ (0.812)$ & $382.039$ & $\begin{smallmatrix}(-0.37590,\ 0.92666)^T\end{smallmatrix}$\\*
$(2,B)$ & $225\ (0.188)$ & $943.661$ & $\begin{smallmatrix}(-0.99990,\ 0.01398)^T\end{smallmatrix}$\\
\cmidrule(lr){1-4}
\multicolumn{4}{@{}l}{$K=3$ (ratio $\checkmark$, fraction $\checkmark$, sample count $\checkmark$)}\\*
$(3,A)$ & $539\ (0.449)$ & $722.197$ & $\begin{smallmatrix}(-0.34408,\ 0.93894)^T\end{smallmatrix}$\\*
$(3,B)$ & $436\ (0.363)$ & $1083.199$ & $\begin{smallmatrix}(-0.41461,\ 0.91000)^T\end{smallmatrix}$\\*
$(3,C)$ & $225\ (0.188)$ & $943.661$ & $\begin{smallmatrix}(-0.99990,\ 0.01398)^T\end{smallmatrix}$\\
\cmidrule(lr){1-4}
\multicolumn{4}{@{}l}{$K=4$ (ratio $\checkmark$, fraction $\times$, sample count $\times$)}\\*
$(4,A)$ & $225\ (0.188)$ & $943.661$ & $\begin{smallmatrix}(-0.99990,\ 0.01398)^T\end{smallmatrix}$\\*
$(4,B)$ & $434\ (0.362)$ & $1085.873$ & $\begin{smallmatrix}(-0.41477,\ 0.90993)^T\end{smallmatrix}$\\*
$(4,C)$ & $2\ (0.002)$ & $218.416$ & $\begin{smallmatrix}(0.15546,\ 0.98784)^T\end{smallmatrix}$\\*
$(4,D)$ & $539\ (0.449)$ & $1996.759$ & $\begin{smallmatrix}(-0.34568,\ 0.93835)^T\end{smallmatrix}$\\
\midrule
\multicolumn{4}{@{}l}{$K=2$ direction: $s_{\mathrm{dir}}=0.388814<\tau_{\mathrm{dir}}=0.90$ ($\checkmark$); selected $K=2$.}\\
\midrule
\multicolumn{4}{@{}l}{\textbf{B. Similarity-type screening}}\\*
Type and $\widehat I$ & \multicolumn{3}{l}{Optimized representation ($k=1$)}\\*
\multicolumn{4}{@{}l}{\textit{Regime 0}}\\*
Type~I\newline $\widehat I=2.354549$ & \multicolumn{3}{l}{$\begin{aligned}\Pi_{1,\mathrm{I}}&=\bigl(\Pi^{(1)}\bigr)^{-0.00573}\bigl(\Pi^{(2)}\bigr)^{0.67663}\\\Pi_{y,\mathrm{I}}&=\Pi_y\end{aligned}$}\\*
Type~II\newline $\widehat I=2.313032$ & \multicolumn{3}{l}{$\begin{aligned}\Pi_{1,\mathrm{II}}&=\bigl(\Pi^{(1)}\bigr)^{-0.26935}\bigl(\Pi^{(2)}\bigr)^{0.09025}\\\widetilde\Pi_{y,\mathrm{II}}&=\Pi_y\bigl(\Pi^{(1)}\bigr)^{-0.249}\bigl(\Pi^{(2)}\bigr)^{-0.35993}\end{aligned}$}\\*
\multicolumn{4}{@{}l}{$L_{\mathrm{out}}=0.000000\leq0.15$; Type~I.}\\
\cmidrule(lr){1-4}
\multicolumn{4}{@{}l}{\textit{Regime 1}}\\*
Type~I\newline $\widehat I=3.101890$ & \multicolumn{3}{l}{$\begin{aligned}\Pi_{1,\mathrm{I}}&=\bigl(\Pi^{(1)}\bigr)^{-0.23523}\bigl(\Pi^{(2)}\bigr)^{0.0005}\\\Pi_{y,\mathrm{I}}&=\Pi_y\end{aligned}$}\\*
Type~II\newline $\widehat I=3.093622$ & \multicolumn{3}{l}{$\begin{aligned}\Pi_{1,\mathrm{II}}&=\bigl(\Pi^{(1)}\bigr)^{0.42071}\bigl(\Pi^{(2)}\bigr)^{0.26594}\\\widetilde\Pi_{y,\mathrm{II}}&=\Pi_y\bigl(\Pi^{(1)}\bigr)^{0.72699}\bigl(\Pi^{(2)}\bigr)^{-0.17229}\end{aligned}$}\\*
\multicolumn{4}{@{}l}{$L_{\mathrm{out}}=0.000000\leq0.15$; Type~I.}\\
\midrule
\multicolumn{4}{@{}l}{\textbf{C. Dimensionless learning}}\\*
Stage & $\widehat I$ & $\varepsilon$ & \multicolumn{1}{c}{$\overline{R^2}$}\\*
\multicolumn{4}{l}{\textit{Regime 0, $N=975$}}\\*[2pt]
$k=1$ & $2.355_{\pm 0.051}$ & $0.095_{\pm 0.005}$ & \multicolumn{1}{c}{$0.987$}\\*
\multicolumn{4}{@{}p{\textwidth}@{}}{\hspace{1em}
$\Pi^{\mathrm{cand}}_{1|1}
=\bigl(\Pi^{(1)}\bigr)^{-0.00573}
\bigl(\Pi^{(2)}\bigr)^{0.67663}$}\\*
\addlinespace[1pt]
$k=2$ & $2.725_{\pm 0.025}$ & $0.066_{\pm 0.002}$ & \multicolumn{1}{c}{$1.000$}\\*
\multicolumn{4}{@{}p{\textwidth}@{}}{\hspace{1em}
$\Pi^{\mathrm{cand}}_{1|2}
=\bigl(\Pi^{(1)}\bigr)^{-0.10625}
\bigl(\Pi^{(2)}\bigr)^{-0.02106}$}\\*
\multicolumn{4}{@{}p{\textwidth}@{}}{\hspace{1em}
$\Pi^{\mathrm{cand}}_{2|2}
=\bigl(\Pi^{(1)}\bigr)^{-0.00708}
\bigl(\Pi^{(2)}\bigr)^{0.99843}$}\\*
\addlinespace[1pt]
Sparsification & $2.263_{\pm 0.029}$ & $0.104_{\pm 0.003}$ & \multicolumn{1}{c}{--}\\*
Rationalization & $2.198_{\pm 0.030}$ & $0.111_{\pm 0.003}$ & \multicolumn{1}{c}{$0.984$}\\
\addlinespace[1pt]
\multicolumn{4}{l}{\textit{Regime 1, $N=225$}}\\*[2pt]
$k=1$ & $3.102_{\pm 0.053}$ & $0.045_{\pm 0.005}$ & \multicolumn{1}{c}{$0.974$}\\*
\multicolumn{4}{@{}p{\textwidth}@{}}{\hspace{1em}
$\Pi^{\mathrm{cand}}_{1|1}
=\bigl(\Pi^{(1)}\bigr)^{-0.23523}
\bigl(\Pi^{(2)}\bigr)^{0.0005}$}\\*
\addlinespace[1pt]
$k=2$ & $3.103_{\pm 0.054}$ & $0.045_{\pm 0.005}$ & \multicolumn{1}{c}{$0.973$}\\*
\multicolumn{4}{@{}p{\textwidth}@{}}{\hspace{1em}
$\Pi^{\mathrm{cand}}_{1|2}
=\bigl(\Pi^{(1)}\bigr)^{-0.70705}
\bigl(\Pi^{(2)}\bigr)^{0.91814}$}\\*
\multicolumn{4}{@{}p{\textwidth}@{}}{\hspace{1em}
$\Pi^{\mathrm{cand}}_{2|2}
=\bigl(\Pi^{(1)}\bigr)^{-0.23866}
\bigl(\Pi^{(2)}\bigr)^{0.00053}$}\\*
\addlinespace[1pt]
Sparsification & $3.094_{\pm 0.053}$ & $0.045_{\pm 0.005}$ & \multicolumn{1}{c}{--}\\*
Rationalization & $3.085_{\pm 0.053}$ & $0.046_{\pm 0.005}$ & \multicolumn{1}{c}{$0.974$}\\
\midrule
\multicolumn{4}{@{}l}{$k^*=1$ in regimes 0 and 1; $\Delta\varepsilon_2=(0.02937,\,6.908\times10^{-5})$.}\\*
\multicolumn{4}{@{}l}{Relative MI losses after rationalization: $(0.029,\,0.003)$ (regimes 0 and 1).}\\
\end{longtable}
\endgroup

\noindent\textbf{Dominant dimensionless groups.}

\begin{equation}
\Pi_{1,0}^*
=
\Pi^{(2)}
=
\epsilon/D,
\qquad
\Pi_{1,1}^*
=
(\Pi^{(1)})^{-1}
=
Re^{-1},
\qquad
\Pi_y=f_D.
\end{equation}

The two retained groups recover the distinct dependencies built into the piecewise relation. Regime~0 is predominantly controlled by relative roughness, consistent with the fully rough limit, whereas regime~1 recovers the inverse-Reynolds-number dependence of the laminar branch.

\section{Agentic workflow evaluation and comparison}
\label{sec:discussion}

\subsection{Comparison with existing methods in functionality and capabilities}
\label{sec:discussion-agent}

Existing data-driven methods for dimensional analysis target a specific discovery task, leaving the broader workflow to the user (Supplementary Table~\ref{tab:agent-capabilities}). Some seek a dominant dimensionless relation for the full dataset; others also identify multiple physical regimes governed by different dominant mechanisms or infer a rescaling of the dimensionless dependent variable that enables data collapse under incomplete similarity. AI Buckingham integrates these capabilities within a unified orchestration. It plans the analysis route and configures numerical settings according to the physical problem and natural-language instructions. When automatic analysis is requested, its numerical tools apply explicit criteria to determine the regime description and similarity type; the corresponding analysis policies can also be prescribed by the user. The assigned physical dimensions are checked against retrieved knowledge-base entries to identify inconsistencies and potential LLM hallucinations before numerical inference. In addition, AI Buckingham simplifies the inferred representations, yielding compact dimensionless scalings with clearer physical interpretation.

\begin{table}[!htbp]
\centering
\small
\setlength{\tabcolsep}{1.5pt}
\renewcommand{\arraystretch}{1.16}
\caption{
Functional comparison of dimensionless-learning methods. Similarity coverage refers to the principal tasks addressed. Simplification includes sparsity, symbolic-complexity control or rationalization. Workflow autonomy comprises workflow planning, parameter configuration and analysis decisions based on numerical criteria.}
\label{tab:agent-capabilities}
\begin{tabular}{@{}>{\raggedright\arraybackslash}m{0.20\linewidth}>{\centering\arraybackslash}m{0.13\linewidth}>{\centering\arraybackslash}m{0.05\linewidth}>{\centering\arraybackslash}m{0.05\linewidth}>{\centering\arraybackslash}m{0.1\linewidth}>{\centering\arraybackslash}m{0.15\linewidth}>{\centering\arraybackslash}m{0.165\linewidth}>{\centering\arraybackslash}m{0.1\linewidth}@{}}
\toprule
\multicolumn{1}{l}{Method} & \shortstack{Dimensional\\homogeneity} &
\multicolumn{2}{c}{\begingroup%
\renewcommand{\arraystretch}{0.88}%
\begin{tabular}[c]{@{}>{\centering\arraybackslash}m{0.05\linewidth}@{\hspace{3pt}}>{\centering\arraybackslash}m{0.05\linewidth}@{}}
\multicolumn{2}{@{}c@{}}{Similarity}\\[-1.5pt]
I & II
\end{tabular}%
\endgroup} & \shortstack{Regime\\analysis} & \shortstack{Representation\\simplification} & \shortstack{RAG-based\\dimension check} & \shortstack{Workflow\\autonomy}\\
\midrule
Active subspaces~\cite{constantine2017data} & Comp. & $\checkmark$ & $\times$ & $\times$ & $\times$ & $\times$ & $\times$\\
ANN--AS~\cite{xu2022artificial} & Comp. & $\checkmark$ & $\times$ & $\times$ & $\times$ & $\times$ & $\times$\\
PyDimension~\cite{xie2022data} & Comp. & $\checkmark$ & $\times$ & $\times$ & $\checkmark$ & $\times$ & $\times$\\
BuckiNet~\cite{bakarji2022dimensionally} & Penalty & $\checkmark$ & $\times$ & $\times$ & $\checkmark$ & $\times$ & $\times$\\
CDL~\cite{zhang2024clustering} & Comp. & $\checkmark$ & $\times$ & $\checkmark$ & $\times$ & $\times$ & $\times$\\
MIDL~\cite{zhang2025mutual} & Comp. & $\checkmark$ & $\times$ & $\times$ & $\times$ & $\times$ & $\times$\\
IT--$\pi$~\cite{yuan2025dimensionless} & Comp. & $\checkmark$ & $\checkmark$ & $\checkmark$ & $\checkmark$ & $\times$ & $\times$\\
Barenet~\cite{sanfins2025similarity} & Prov. & $\times$ & $\checkmark$ & $\times$ & $\times$ & $\times$ & $\times$\\
Watanabe et al.~\cite{watanabe2025datadriven} & Prov. & $\times$ & $\checkmark$ & $\times$ & $\times$ & $\times$ & $\times$\\
BSM~\cite{bempedelis2025extracting} & Penalty & $\checkmark$ & $\checkmark$ & $\times$ & $\checkmark$ & $\times$ & $\times$\\
AI Feynman~\cite{udrescu2020ai} & Comp. & -- & -- & $\times$ & $\checkmark$ & $\times$ & $\times$\\
Hi--$\pi$~\cite{gan2025hierarchical} & Comp. & $\checkmark$ & $\times$ & $\times$ & $\checkmark$ & $\times$ & $\times$\\
Gan et al.~\cite{gan2021universal} & Comp. & $\checkmark$ & $\times$ & $\times$ & $\times$ & $\times$ & $\times$\\
\midrule
\textbf{AI Buckingham} & Comp. & $\checkmark$ & $\checkmark$ & $\checkmark$ & $\checkmark$ & $\checkmark$ & $\checkmark$\\
\bottomrule
\end{tabular}
\par\vspace{3pt}
\begin{minipage}{\linewidth}
\footnotesize
ANN--AS, neural-network-assisted active subspaces; CDL, clustering dimensionless learning; MIDL, mutual-information-based dimensional learning. Comp. (computed): dimensionless groups derived from the dimensions of the physical variables; Prov. (provided): dimensionless inputs specified a priori; Penalty: an added loss term penalizing dimensional mismatch. A dash denotes no explicit Type~I/II formulation.
\end{minipage}
\end{table}

\subsection{LLM ablation and API usage in dimensional checking}
\label{sec:discussion-RAG}

To compare the performance of different LLMs in semantic parsing and quantify the gains from RAG-based dimensional checking, the present ablation analysis evaluated ten cases spanning laminar and turbulent flows, magnetohydrodynamics, thermal convection, and laser processing. Each original prompt was paired with a rephrased prompt in which the name of one governing quantity was replaced by a physical description or defining relation to make its semantic identification more challenging. Three LLMs (GPT-4o mini, GPT-4.1 mini, and GPT-5 mini) were employed to identify the governing quantities and dependent variable, and to assign candidate SI exponents to each governing quantity. In all configurations, GPT-5 mini was tasked with reviewing these assignments against retrieved physical-quantity records. When revision was requested, the parsing LLM revised its initial assignments once, followed by fresh retrieval and a second review. Success required agreement with the reference governing quantities, their SI exponents and the dependent variable, a passed review, and construction of a Buckingham--$\Pi$ basis. Ten repetitions per case, prompt version, and LLM yielded 600 trials for comparing success rates, token usage by stage, and API time.

\begingroup
\small
\setlength{\tabcolsep}{1.5pt}
\setlength{\LTcapwidth}{\linewidth}
\renewcommand{\arraystretch}{1.12}
\definecolor{DCStripe}{HTML}{ECF9F7}
\newcommand{\DCSep}[1]{{\color{black!#1}\vrule width 0.3pt}}
\newcommand{\DCTransition}[3]{%
\makebox[8mm][c]{#1}%
\makebox[\dimexpr2\tabcolsep\relax][c]{%
\kern1mm\makebox[0pt][c]{%
\ensuremath{\ifnum#3>0\relax\overset{\scriptscriptstyle+#3}{\longrightarrow}\else\longrightarrow\fi}}%
\kern-1mm}%
\makebox[12mm][c]{#2}}
\newcommand{\DCOutcomeHead}{%
\toprule
Case & \multicolumn{4}{c!{\DCSep{32}}}{GPT-4.1 mini} & \multicolumn{4}{c!{\DCSep{32}}}{GPT-5 mini} & \multicolumn{4}{c@{}}{GPT-4o mini} \\
\cmidrule(lr){2-5}\cmidrule(lr){6-9}\cmidrule(lr){10-13}
 & \multicolumn{2}{c!{\DCSep{18}}}{Orig.} & \multicolumn{2}{c!{\DCSep{32}}}{Rephr.} & \multicolumn{2}{c!{\DCSep{18}}}{Orig.} & \multicolumn{2}{c!{\DCSep{32}}}{Rephr.} & \multicolumn{2}{c!{\DCSep{18}}}{Orig.} & \multicolumn{2}{c@{}}{Rephr.} \\
\cmidrule(lr){2-3}\cmidrule(lr){4-5}\cmidrule(lr){6-7}\cmidrule(lr){8-9}\cmidrule(lr){10-11}\cmidrule(lr){12-13}
 & Init. & Final & Init. & Final & Init. & Final & Init. & Final & Init. & Final & Init. & Final \\
\midrule}
\begin{longtable}{@{}>{\centering\arraybackslash}m{20mm}!{\DCSep{20}}
*{2}{>{\centering\arraybackslash}m{8mm}>{\centering\arraybackslash}m{12mm}!{\DCSep{18}}>{\centering\arraybackslash}m{8mm}>{\centering\arraybackslash}m{12mm}!{\DCSep{32}}}
>{\centering\arraybackslash}m{8mm}>{\centering\arraybackslash}m{12mm}!{\DCSep{18}}
>{\centering\arraybackslash}m{8mm}>{\centering\arraybackslash}m{12mm}@{}}
\caption{LLM ablation in semantic parsing. Orig.\ and Rephr.\ denote original and rephrased prompts. Init.\ counts successes after the initial response and dimensional check. Final also includes successes after revision and rechecking.}
\label{tab:dimensional-checking-outcomes}\\
\DCOutcomeHead
\endfirsthead
\caption[]{(continued.)}\\
\DCOutcomeHead
\endhead
\midrule
\multicolumn{13}{r@{}}{\footnotesize Continued on next page}\\
\endfoot
\bottomrule
\endlastfoot
\rowcolor{DCStripe} MHD & \multicolumn{2}{c!{\DCSep{18}}}{\DCTransition{1}{9}{8}} & \multicolumn{2}{c!{\DCSep{32}}}{\DCTransition{7}{8}{1}} & \multicolumn{2}{c!{\DCSep{18}}}{\DCTransition{9}{10}{1}} & \multicolumn{2}{c!{\DCSep{32}}}{\DCTransition{9}{9}{0}} & \multicolumn{2}{c!{\DCSep{18}}}{\DCTransition{0}{7}{7}} & \multicolumn{2}{c@{}}{\DCTransition{0}{10}{10}} \\
PresGrad & \multicolumn{2}{c!{\DCSep{18}}}{\DCTransition{0}{10}{10}} & \multicolumn{2}{c!{\DCSep{32}}}{\DCTransition{0}{10}{10}} & \multicolumn{2}{c!{\DCSep{18}}}{\DCTransition{10}{10}{0}} & \multicolumn{2}{c!{\DCSep{32}}}{\DCTransition{9}{9}{0}} & \multicolumn{2}{c!{\DCSep{18}}}{\DCTransition{0}{10}{10}} & \multicolumn{2}{c@{}}{\DCTransition{0}{10}{10}} \\
\rowcolor{DCStripe} Rayleigh & \multicolumn{2}{c!{\DCSep{18}}}{\DCTransition{10}{10}{0}} & \multicolumn{2}{c!{\DCSep{32}}}{\DCTransition{10}{10}{0}} & \multicolumn{2}{c!{\DCSep{18}}}{\DCTransition{9}{10}{1}} & \multicolumn{2}{c!{\DCSep{32}}}{\DCTransition{10}{10}{0}} & \multicolumn{2}{c!{\DCSep{18}}}{\DCTransition{0}{9}{9}} & \multicolumn{2}{c@{}}{\DCTransition{0}{6}{6}} \\
Keyhole & \multicolumn{2}{c!{\DCSep{18}}}{\DCTransition{10}{10}{0}} & \multicolumn{2}{c!{\DCSep{32}}}{\DCTransition{10}{10}{0}} & \multicolumn{2}{c!{\DCSep{18}}}{\DCTransition{10}{10}{0}} & \multicolumn{2}{c!{\DCSep{32}}}{\DCTransition{9}{10}{1}} & \multicolumn{2}{c!{\DCSep{18}}}{\DCTransition{0}{9}{9}} & \multicolumn{2}{c@{}}{\DCTransition{0}{9}{9}} \\
\rowcolor{DCStripe} Blasius & \multicolumn{2}{c!{\DCSep{18}}}{\DCTransition{10}{10}{0}} & \multicolumn{2}{c!{\DCSep{32}}}{\DCTransition{10}{10}{0}} & \multicolumn{2}{c!{\DCSep{18}}}{\DCTransition{9}{9}{0}} & \multicolumn{2}{c!{\DCSep{32}}}{\DCTransition{10}{10}{0}} & \multicolumn{2}{c!{\DCSep{18}}}{\DCTransition{4}{9}{5}} & \multicolumn{2}{c@{}}{\DCTransition{3}{9}{6}} \\
SandGrain & \multicolumn{2}{c!{\DCSep{18}}}{\DCTransition{8}{8}{0}} & \multicolumn{2}{c!{\DCSep{32}}}{\DCTransition{9}{9}{0}} & \multicolumn{2}{c!{\DCSep{18}}}{\DCTransition{10}{10}{0}} & \multicolumn{2}{c!{\DCSep{32}}}{\DCTransition{10}{10}{0}} & \multicolumn{2}{c!{\DCSep{18}}}{\DCTransition{9}{9}{0}} & \multicolumn{2}{c@{}}{\DCTransition{8}{8}{0}} \\
\rowcolor{DCStripe} VeloTrans & \multicolumn{2}{c!{\DCSep{18}}}{\DCTransition{10}{10}{0}} & \multicolumn{2}{c!{\DCSep{32}}}{\DCTransition{10}{10}{0}} & \multicolumn{2}{c!{\DCSep{18}}}{\DCTransition{10}{10}{0}} & \multicolumn{2}{c!{\DCSep{32}}}{\DCTransition{10}{10}{0}} & \multicolumn{2}{c!{\DCSep{18}}}{\DCTransition{9}{10}{1}} & \multicolumn{2}{c@{}}{\DCTransition{10}{10}{0}} \\
Goldenfeld & \multicolumn{2}{c!{\DCSep{18}}}{\DCTransition{10}{10}{0}} & \multicolumn{2}{c!{\DCSep{32}}}{\DCTransition{10}{10}{0}} & \multicolumn{2}{c!{\DCSep{18}}}{\DCTransition{10}{10}{0}} & \multicolumn{2}{c!{\DCSep{32}}}{\DCTransition{9}{10}{1}} & \multicolumn{2}{c!{\DCSep{18}}}{\DCTransition{9}{10}{1}} & \multicolumn{2}{c@{}}{\DCTransition{1}{10}{9}} \\
\rowcolor{DCStripe} CessVisc & \multicolumn{2}{c!{\DCSep{18}}}{\DCTransition{10}{10}{0}} & \multicolumn{2}{c!{\DCSep{32}}}{\DCTransition{9}{10}{1}} & \multicolumn{2}{c!{\DCSep{18}}}{\DCTransition{10}{10}{0}} & \multicolumn{2}{c!{\DCSep{32}}}{\DCTransition{10}{10}{0}} & \multicolumn{2}{c!{\DCSep{18}}}{\DCTransition{2}{6}{4}} & \multicolumn{2}{c@{}}{\DCTransition{4}{8}{4}} \\
CirclePipe & \multicolumn{2}{c!{\DCSep{18}}}{\DCTransition{10}{10}{0}} & \multicolumn{2}{c!{\DCSep{32}}}{\DCTransition{10}{10}{0}} & \multicolumn{2}{c!{\DCSep{18}}}{\DCTransition{10}{10}{0}} & \multicolumn{2}{c!{\DCSep{32}}}{\DCTransition{10}{10}{0}} & \multicolumn{2}{c!{\DCSep{18}}}{\DCTransition{7}{9}{2}} & \multicolumn{2}{c@{}}{\DCTransition{7}{10}{3}} \\
\midrule
\textbf{Total} & \multicolumn{2}{c!{\DCSep{18}}}{\DCTransition{79}{97}{18}} & \multicolumn{2}{c!{\DCSep{32}}}{\DCTransition{85}{97}{12}} & \multicolumn{2}{c!{\DCSep{18}}}{\DCTransition{97}{99}{2}} & \multicolumn{2}{c!{\DCSep{32}}}{\DCTransition{96}{98}{2}} & \multicolumn{2}{c!{\DCSep{18}}}{\DCTransition{40}{88}{48}} & \multicolumn{2}{c@{}}{\DCTransition{33}{90}{57}} \\
\midrule
\textbf{Rate (\%)} & \multicolumn{2}{c!{\DCSep{18}}}{\DCTransition{79}{97}{18}} & \multicolumn{2}{c!{\DCSep{32}}}{\DCTransition{85}{97}{12}} & \multicolumn{2}{c!{\DCSep{18}}}{\DCTransition{97}{99}{2}} & \multicolumn{2}{c!{\DCSep{32}}}{\DCTransition{96}{98}{2}} & \multicolumn{2}{c!{\DCSep{18}}}{\DCTransition{40}{88}{48}} & \multicolumn{2}{c@{}}{\DCTransition{33}{90}{57}} \\
\end{longtable}
\endgroup

\begingroup
\small
\setlength{\tabcolsep}{3.5pt}
\setlength{\LTcapwidth}{\linewidth}
\renewcommand{\arraystretch}{1.12}
\definecolor{DCStripe}{HTML}{ECF9F7}
\newcommand{\DCAPISep}{{\color{black!24}\vrule width 0.3pt}}
\newcommand{\DCUsageHead}{%
\toprule
 \multirow{2}{*}{Case}& \multirow{2}{*}{LLMs} & \multicolumn{7}{c!{\DCAPISep}}{Token usage (k tokens/trial)} & \multirow{2}{*}{\shortstack{API time\\(s/trial)}} \\
\cmidrule(lr){3-9}
& & \multicolumn{1}{c}{Init.} & \multicolumn{1}{c}{Emb.\,1} & \multicolumn{1}{c!{\DCAPISep}}{Chk.\,1} & \multicolumn{1}{c}{Rev.} & \multicolumn{1}{c}{Emb.\,2} & \multicolumn{1}{c!{\DCAPISep}}{Chk.\,2} & \multicolumn{1}{c!{\DCAPISep}}{Sum} & \\
\midrule}
\begin{longtable}{@{}>{\centering\arraybackslash}m{17.5mm}!{\DCAPISep}
>{\centering\arraybackslash}m{10mm}!{\DCAPISep}
rrr!{\DCAPISep}rrr!{\DCAPISep}r!{\DCAPISep}c@{}}
\caption{API usage per scheduled trial, averaged over ten original and ten rephrased trials per row. The LLMs column lists the GPT mini variants used for semantic parsing. Init.\ and Rev.\ denote the initial response and the revision based on feedback. Emb.\ and Chk.\ report query-embedding tokens and tokens used by GPT-5 mini for checking, respectively. Sum adds token usage across the four LLM stages, excluding embedding tokens. A dash indicates a stage with no calls. API time sums the durations of all recorded requests, including retries and output-column verification.}
\label{tab:dimensional-checking-api-usage}\\
\DCUsageHead
\endfirsthead
\caption[]{(continued.)}\\
\DCUsageHead
\endhead
\midrule
\multicolumn{10}{r@{}}{\footnotesize Continued on next page}\\
\endfoot
\bottomrule
\endlastfoot
\rowcolor{DCStripe} MHD & 4o & 1.20 & 0.71 & 7.30 & 3.99 & 0.72 & 7.34 & 19.83 & 146.0 \\*
 & 4.1 & 1.27 & 0.84 & 8.06 & 2.80 & 0.46 & 5.26 & 17.39 & 144.2 \\*
\rowcolor{DCStripe} & 5 & 3.26 & 0.87 & 7.76 & 0.32 & 0.04 & 0.37 & 11.71 & 101.5 \\
\specialrule{0.3pt}{0pt}{0pt}
PresGrad & 4o & 1.27 & 1.12 & 8.13 & 4.57 & 1.11 & 7.44 & 21.41 & 144.5 \\*
\rowcolor{DCStripe} & 4.1 & 1.33 & 1.15 & 7.89 & 4.54 & 1.15 & 7.54 & 21.30 & 138.6 \\*
 & 5 & 2.80 & 1.15 & 7.35 & -- & -- & -- & 10.14 & 72.8 \\
\specialrule{0.3pt}{0pt}{0pt}
\rowcolor{DCStripe} Rayleigh & 4o & 1.34 & 1.19 & 9.49 & 3.94 & 0.90 & 6.56 & 21.34 & 144.4 \\*
 & 4.1 & 1.42 & 1.20 & 8.65 & -- & -- & -- & 10.07 & 95.0 \\*
\rowcolor{DCStripe} & 5 & 2.98 & 1.24 & 8.15 & 0.26 & 0.06 & 0.31 & 11.69 & 90.9 \\
\specialrule{0.3pt}{0pt}{0pt}
Keyhole & 4o & 1.45 & 0.97 & 8.26 & 4.91 & 0.93 & 7.72 & 22.34 & 155.7 \\*
\rowcolor{DCStripe} & 4.1 & 1.51 & 1.05 & 9.45 & -- & -- & -- & 10.96 & 99.6 \\*
 & 5 & 2.97 & 1.07 & 8.05 & 0.28 & 0.05 & 0.42 & 11.71 & 99.7 \\
\specialrule{0.3pt}{0pt}{0pt}
\rowcolor{DCStripe} Blasius & 4o & 0.99 & 0.56 & 5.57 & 2.13 & 0.35 & 3.13 & 11.82 & 110.8 \\*
 & 4.1 & 1.05 & 0.59 & 5.96 & -- & -- & -- & 7.01 & 71.1 \\*
\rowcolor{DCStripe} & 5 & 2.35 & 0.59 & 5.24 & -- & -- & -- & 7.59 & 60.1 \\
\specialrule{0.3pt}{0pt}{0pt}
SandGrain & 4o & 1.43 & 0.79 & 7.76 & -- & -- & -- & 9.20 & 49.4 \\*
\rowcolor{DCStripe} & 4.1 & 1.49 & 0.79 & 7.97 & -- & -- & -- & 9.46 & 52.6 \\*
 & 5 & 2.96 & 0.84 & 8.11 & -- & -- & -- & 11.07 & 46.5 \\
\specialrule{0.3pt}{0pt}{0pt}
\rowcolor{DCStripe} VeloTrans & 4o & 1.26 & 0.78 & 7.85 & 0.22 & 0.04 & 0.49 & 9.82 & 87.0 \\*
 & 4.1 & 1.31 & 0.76 & 8.01 & -- & -- & -- & 9.32 & 80.3 \\*
\rowcolor{DCStripe} & 5 & 2.72 & 0.80 & 7.79 & -- & -- & -- & 10.51 & 79.4 \\
\specialrule{0.3pt}{0pt}{0pt}
Goldenfeld & 4o & 1.14 & 0.65 & 7.35 & 1.94 & 0.33 & 3.52 & 13.94 & 105.8 \\*
\rowcolor{DCStripe} & 4.1 & 1.19 & 0.62 & 6.73 & -- & -- & -- & 7.91 & 76.3 \\*
 & 5 & 2.53 & 0.64 & 6.08 & 0.22 & 0.03 & 0.30 & 9.14 & 66.8 \\
\specialrule{0.3pt}{0pt}{0pt}
\rowcolor{DCStripe} CessVisc & 4o & 1.02 & 0.44 & 4.60 & 1.36 & 0.25 & 2.43 & 9.41 & 73.6 \\*
 & 4.1 & 1.09 & 0.61 & 6.12 & 0.18 & 0.03 & 0.29 & 7.68 & 66.2 \\*
\rowcolor{DCStripe} & 5 & 2.56 & 0.63 & 5.85 & -- & -- & -- & 8.41 & 62.6 \\
\specialrule{0.3pt}{0pt}{0pt}
CirclePipe & 4o & 1.07 & 0.65 & 6.26 & 0.92 & 0.17 & 1.59 & 9.83 & 74.8 \\*
\rowcolor{DCStripe} & 4.1 & 1.18 & 0.70 & 6.37 & -- & -- & -- & 7.55 & 62.0 \\*
 & 5 & 2.39 & 0.71 & 6.10 & -- & -- & -- & 8.49 & 64.0 \\
\end{longtable}
\endgroup

With RAG-based dimensional checking, the permitted revision increased success rates by 52.5, 15.0, and 2.0 percentage points for 4o mini, 4.1 mini, and 5 mini, respectively (Supplementary Table~\ref{tab:dimensional-checking-outcomes}). GPT-4o mini showed the weakest dimensional reasoning and least reliable structured responses in these trials, producing incorrect SI exponents and malformed assignments. GPT-4.1 mini's errors were concentrated in derived quantities and electromagnetic dimensions, whereas GPT-5 mini made isolated errors in quantity identification or exponent assignment. Specifically, the magnetohydrodynamic trial records show how retrieved definitions supported correction of GPT-5 mini's interpretation of $\eta$ as magnetic diffusivity ($L^2T^{-1}$) to electrical resistivity ($ML^3T^{-3}I^{-2}$). Resistivity and magnetic-flux-density records also supported corrections to GPT-4o mini's electromagnetic exponents. In the laser keyhole case, 4o mini assigned a temperature difference to the time dimension and omitted the inverse-temperature factor in specific heat capacity. In the pressure-gradient boundary layer, 4o mini and 4.1 mini misinterpreted $\int U^2\,\mathrm{d}y$, omitting the length factor introduced by wall-normal integration or introducing a spurious mass dimension. In the sand-grain case, all initial dimensional assignments were correct, but six trials failed the consistency checks on the Checker's corrections and cited records.

In terms of API usage (Supplementary Table~\ref{tab:dimensional-checking-api-usage}), GPT-5 mini used more than twice as many tokens as 4o mini for the initial response. Its higher success rate after the initial review limited the need for revision and rechecking. These subsequent calls (Rev. and Chk. 2) accounted for 2.5\% of its reported LLM token usage (Sum), compared with 43.1\% for 4o mini. With this lower revision overhead, GPT-5 mini achieved the highest final success count with the lowest overall reported LLM token usage and mean cumulative API time of 74.4~s per trial, supporting its selection for semantic parsing in AI Buckingham.

\subsection{Relation to the Trettel--Larsson transformation}
\label{sec:discussion-TL}

\begin{figure}[!t]
\centering
\includegraphics[width=\linewidth]{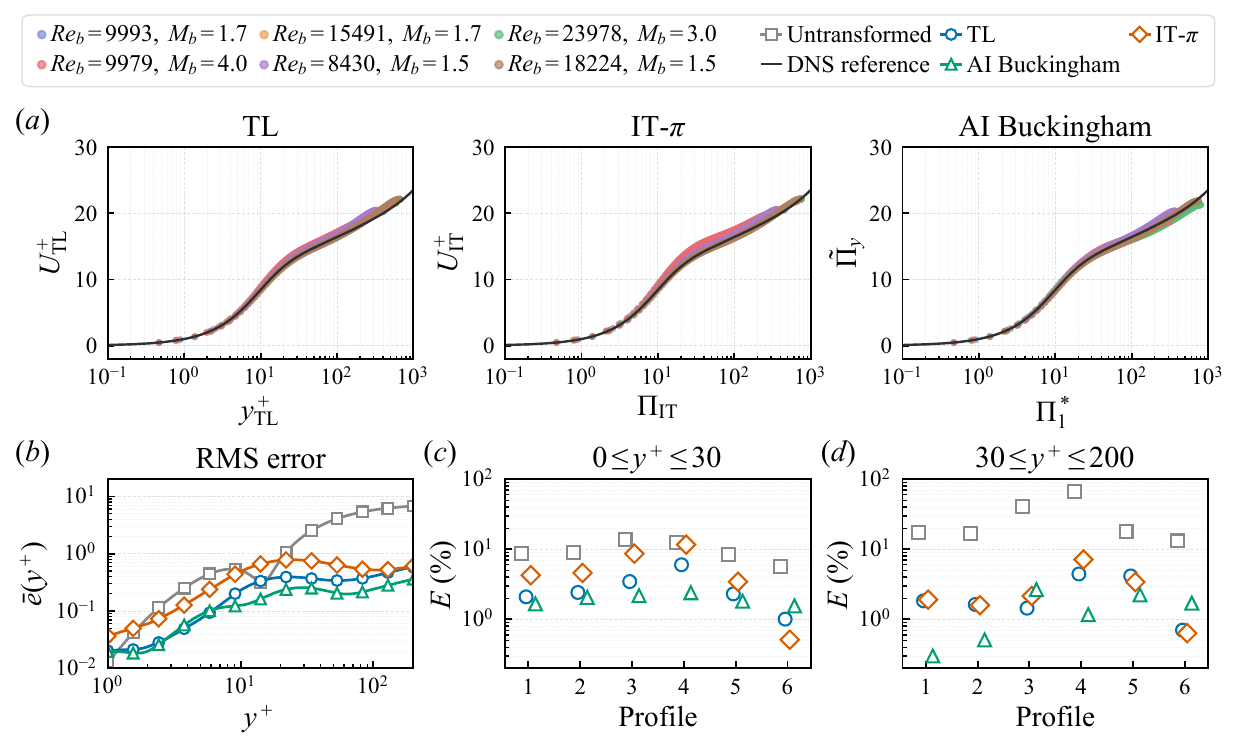}
\caption{Comparison of velocity transformations. (a) Transformed profiles and the incompressible DNS reference~\cite{sillero2013one}. (b) Six-profile RMS error $\bar e$. (c,d) Normalized errors $E_j^{[0,30]}$ and $E_j^{[30,200]}$, respectively. ``Untransformed'' denotes the original $(y^+,U^+)$ profiles. Profiles 1--6 follow the legend from left to right across its first and then second row. Symbols in (c,d) are horizontally offset to avoid overlap.}
\label{fig:TL-comparison}
\end{figure}

The scaling discovered by AI Buckingham and the Trettel--Larsson (TL) transformation~\cite{trettel2016mean} both account for mean-property variations through the local-to-wall density and viscosity ratios, $R=\rho/\rho_w$ and $M=\mu/\mu_w$. Their wall-normal coordinates are
\begin{equation}
y_{\mathrm{TL}}^+=y^+R^{1/2}M^{-1},\quad
\Pi_1^*=y^+R^{2/3}M^{-1/3}
=y_{\mathrm{TL}}^+R^{1/6}M^{2/3}.
\label{eq:velotrans-TL-coordinate}
\end{equation}

For the mean velocity, AI Buckingham uses the local rescaling $\widetilde\Pi_y=U^+R^{0.86}M^{0.40}$, whereas TL defines the transformed mean velocity by the profile integral
\begin{equation}
U_{\mathrm{TL}}^+
=\int_0^{U^+}\!\sqrt R
\left(1+\frac{y^+}{2R}\frac{\mathrm dR}{\mathrm dy^+}
-\frac{y^+}{M}\frac{\mathrm dM}{\mathrm dy^+}\right)
\,\mathrm dU^+,
\label{eq:velotrans-TL-integral}
\end{equation}

Trettel and Larsson~\cite{trettel2016mean} obtain this transformation by requiring the transformed flow to recover the incompressible logarithmic velocity gradient and preserve the inner-layer shear-stress balance. In the logarithmic region, the velocity-gradient scaling uses the wall distance and the local velocity scale $\sqrt{\tau_w/\rho}$, where $\tau_w$ is the wall shear stress. This assumes that mean-property variations do not alter turbulent length scales. Across the inner layer, the total shear stress, comprising viscous and turbulent contributions, is matched between the compressible and transformed flows. The density-weighted turbulent shear stress is also assumed to match throughout this layer, extending Morkovin scaling into the viscous sublayer and buffer layer; the viscous stresses must then agree as well. These requirements determine the TL coordinate and velocity mapping. AI Buckingham discovers the scaling exponents from data without imposing these TL assumptions. The resulting algebraic transformation requires neither profile differentiation nor integration.

The comparison uses four compressible channel-flow DNS profiles~\cite{trettel2016mean} and two pipe-flow DNS profiles~\cite{modesti2019direct}. The incompressible mean-velocity reference $U_{\mathrm{ref}}^+(y^+)$ is taken from DNS of a zero-pressure-gradient turbulent boundary layer at $Re_\theta=6500$~\cite{sillero2013one}. For quantitative comparison with existing benchmarks, the velocity scaling proposed in IT--$\pi$~\cite{yuan2025dimensionless} is also evaluated:
\begin{equation}
\Pi_{\mathrm{IT}}=y^+R^{-0.06}M^{-1.38},
\quad
U_{\mathrm{IT}}^+=U^+R^{1.44}M^{1.23}.
\label{eq:velotrans-ITPI}
\end{equation}
The profiles are interpolated with piecewise cubic Hermite interpolating polynomials, and the TL integral is evaluated from the no-slip wall using composite Simpson integration.

The transformed velocities are compared with the incompressible reference at matched wall distances $y^+$, with the untransformed profiles as a baseline. For each comparison, let $\Delta U_j^+=U_j^+-U_{\mathrm{ref}}^+$ denote the velocity difference for DNS profile $j=1,\ldots,6$. We report the root-mean-square (RMS) velocity error $\bar e$ over the six profiles and the normalized RMS velocity-profile error employed by~\citet{ying2025general} for each profile:
\begin{equation}
\bar e(y^+)=\left[\frac{1}{6}\sum_{j=1}^{6}
\left(\Delta U_j^+(y^+)\right)^2\right]^{1/2},
\quad
E_j^{[a,b]}=
\frac{\left[\dfrac{1}{b-a}\int_a^b
\left(\Delta U_j^+(y^+)\right)^2\,\mathrm dy^+\right]^{1/2}}
{\dfrac{1}{b-a}\int_a^b U_{\mathrm{ref}}^+(y^+)\,\mathrm dy^+}.
\label{eq:TL-RMS-errors}
\end{equation}
The RMS error $\bar e$ is evaluated over $1\leq y^+\leq200$. We evaluate $E_j^{[a,b]}$ separately over $0\leq y^+\leq30$, covering the viscous sublayer and buffer layer, and over $30\leq y^+\leq200$, within the log-law region of the incompressible reference~\cite{pope2000turbulent}. The six-profile mean of $E$ is $1.94\%$ and $1.43\%$ for AI Buckingham on $[0,30]$ and $[30,200]$, respectively, compared with $2.88\%$ and $2.37\%$ for TL and $5.51\%$ and $2.81\%$ for IT--$\pi$ (Supplementary Fig.~\ref{fig:TL-comparison}). These errors quantify collapse to the chosen incompressible reference.

\subsection{Physical-regime detection in IT--$\pi$ and AI Buckingham}
\label{sec:discussion-ITPI}

IT--$\pi$~\cite{yuan2025dimensionless} identifies physical regimes by assessing the local predictive relevance of dimensionless groups. Let $\boldsymbol{\Pi}^{\mathrm{IT}}=(\Pi_1^{\mathrm{IT}},\ldots,\Pi_k^{\mathrm{IT}})$ denote the groups derived from the full dataset. Following the implementation of~\citet{yuan2025dimensionless}, the group coordinates are logarithmically transformed and standardized, then partitioned into a prescribed number $M$ of regions using $k$-means. Thus, $M$ controls the resolution of the geometric partition rather than selecting the number of distinct physical regimes. Within each region $r$, the prediction score compares the information carried by an individual group with that carried by the complete representation:
\begin{equation}
\mathcal R_i(r)=\frac{\widehat I_r(\Pi_i^{\mathrm{IT}};\Pi_y)}
{\widehat I_r(\boldsymbol{\Pi}^{\mathrm{IT}};\Pi_y)},\qquad i=1,\ldots,k,
\label{eq:ITPI-regional-score}
\end{equation}
where $\widehat I_r$ denotes MI estimated from samples in region $r$. A score close to unity indicates that one group retains most of the locally available predictive information. The Cess dimensional matrix admits two independent Buckingham--$\Pi$ groups. The comparison therefore uses a two-group representation, with regions labelled ``$\Pi_i^{\mathrm{IT}}$ dominant'' or ``Both $\Pi_1^{\mathrm{IT}}$ and $\Pi_2^{\mathrm{IT}}$'' according to their relevance scores.

\begin{figure}[!t]
\centering
\includegraphics[width=\linewidth]{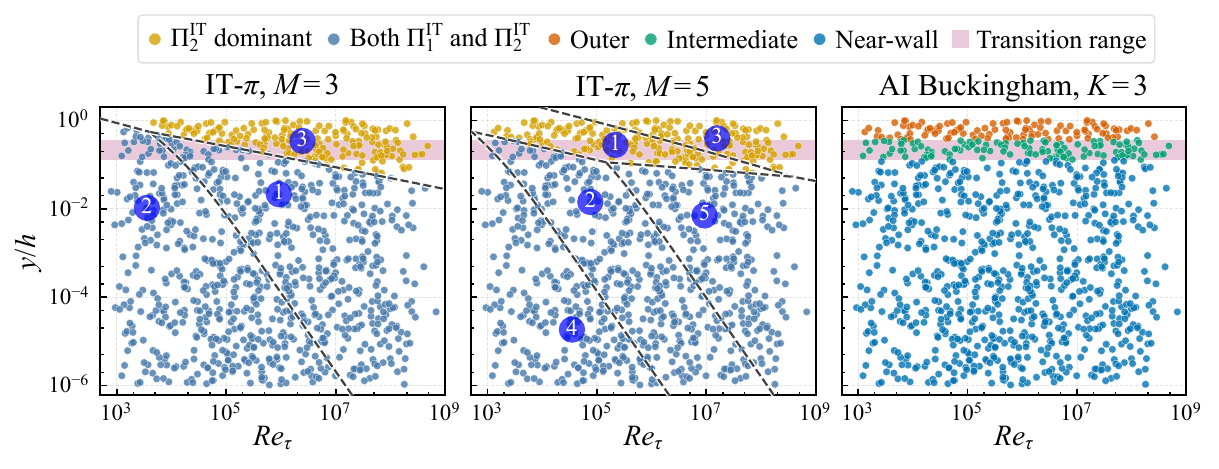}
\caption{Physical-regime analysis of the Cess model. Left and centre: regional relevance categories for the IT--$\pi$ groups with $M=3,5$, respectively. Numbers and dashed curves mark geometric $k$-means regions. Right: AI Buckingham's gradient-based partition. Shading indicates the wall-normal range spanned by its intermediate samples.}
\label{fig:ITPI-Cess}
\end{figure}

Both methods are evaluated on the same 1,000 samples of the Cess eddy-viscosity model~\cite{zhang2024clustering}. IT--$\pi$ first searches the full dataset for two dimensionless groups, yielding
\begin{equation}
\Pi_1^{\mathrm{IT}}=(y^+)Re_\tau^{-0.83},\quad
\Pi_2^{\mathrm{IT}}=(y^+)^{0.56}Re_\tau^{0.44}.
\label{eq:ITPI-Cess-inputs}
\end{equation}
Local MI is estimated from the logarithms of the groups and dependent variable. For $M=3$ and $M=5$, both the $\Pi_2^{\mathrm{IT}}$-dominant and joint-group categories extend inside and outside the shaded intermediate band; neither partition isolates this transition regime as a distinct relevance category (Supplementary Fig.~\ref{fig:ITPI-Cess}). AI Buckingham clusters samples according to the similarity of local GPR gradient directions and selects the smallest admissible cluster count from the candidate partitions. The resulting $K=3$ partition identifies an intermediate transition regime characterized by the dominant dimensionless group $Re_\tau^{0.43}(y^+)^{0.57}$, which retains contributions from both limiting dependences, $Re_\tau$ and $y^+$.

IT--$\pi$ relevance scores depend on the prescribed geometric partition: varying $M$ modifies the sample subsets used for MI estimation and can alter the resulting relevance categories. These categories are therefore conditional on a geometric resolution specified a priori. Within each region, Eq.~\eqref{eq:ITPI-regional-score} requires MI estimation over the two-dimensional joint spaces $(\Pi_i^{\mathrm{IT}},\Pi_y)$ and the $(k+1)$-dimensional joint space $(\boldsymbol{\Pi}^{\mathrm{IT}},\Pi_y)$. At a fixed total sample count, refining the partition reduces the average number of samples per region. Reliable MI estimation may therefore require more data points as $M$ or $k$ increases, since finite-sample estimation errors propagate to the relevance scores~\cite{kraskov2004estimating}.

AI Buckingham instead evaluates regional MI only after gradient-based partitioning, to discover the dominant dimensionless groups within each regime. Regime detection therefore avoids repeated local MI estimation and relies on the global GPR surrogate to resolve local gradient directions accurately.

\bibliographystyle{unsrtnat}
\bibliography{sn-bibliography}